\documentclass[preprint,12pt,authoryear]{elsarticle}
\usepackage{amssymb}
\usepackage{amsmath}
\usepackage{url}
\usepackage{hyperref}
\usepackage{amsmath,amssymb,amsfonts}%
\usepackage{amsthm}%
\usepackage{amsfonts}
\usepackage{multirow}
\usepackage{booktabs}
\usepackage{adjustbox}
\usepackage{xcolor}
\usepackage[ruled,vlined,linesnumbered]{algorithm2e}
\newtheorem{theorem}{Theorem}
\newtheorem{assumption}{Assumption}

\newtheorem{corollary}{Corollary}
\newtheorem{lemma}{Lemma}
\newtheorem{remark}{Remark}

\journal{Spatial Statistics}

\begin{document}

\begin{frontmatter}



\title{A Generalized Spatial Correlated Mean Component Model: Theory and Application to U.S. County Voting Data}

\author[label1]{N.A. Cruz\corref{cor1}}
\affiliation[label1]{organization={Profesor Visitante, Universitat de les Illes Balears, Departament de Matemàtiques i Informàtica, Phone: +34 637 54 6888, Palma de Mallorca, España},
ead={nelson-alirio.cruz@uib.es}, ead={https://orcid.org/0000-0002-7370-5111}}
\author[label3]{J.D. Toloza-Delgado}
\affiliation[label3]{organization={Department of Statistics, Faculty of Sciences,  Universidad Nacional de Colombia, Bogota, Colombia},
ead={jdtolozad@unal.edu.co}, ead={https://orcid.org/0000-0001-7523-7625}}

\author[label2]{O.O. Melo}
\affiliation[label2]{organization={Department of Statistics, Faculty of Sciences,  Universidad Nacional de Colombia, Bogota, Colombia},
ead={oomelo@unal.edu.co}, ead={https://orcid.org/0000-0002-0296-4511}}

\begin{abstract}
This paper introduces the generalized spatial correlated inverse mean component (GSCIMC), a flexible and computationally efficient framework for modeling spatial dependence in non-Gaussian data from the exponential family. 
We establish asymptotic properties of the proposed estimators, including consistency, asymptotic normality, and efficiency, and derive robust variance expressions that enable valid inference under heteroskedasticity or partial model misspecification.  Through extensive simulation studies, we show that the GSCIMC estimator provides stable estimation, accurate inference, and reliable coverage across a wide range of spatial dependence structures and response distributions. An empirical analysis of U.S. county-level voting patterns in the 2020 presidential election illustrates the model’s ability to identify spatial spillover effects while maintaining an interpretable regression framework, the proposed framework provides a practical and theoretically grounded tool for analyzing binary, count, or skewed continuous spatial responses. An open-source implementation is available in the \texttt{spatemR} package, facilitating the application of GSCIMC across disciplines such as environmental science, epidemiology, and regional economics.
\end{abstract}

\begin{keyword}
Asymptotic properties\sep spatial econometrics\sep categorical data\sep voting behavior
\end{keyword}

\end{frontmatter}
\section{Introduction}
Voting behavior in the United States exhibits strong geographic structure, with neighboring counties often displaying remarkably similar electoral outcomes. These spatial patterns have been extensively documented in the political science literature and are commonly associated with persistent regional polarization and urban–rural divides \citep{gelman2009red,rodden2019cities,campbell2018polarized}. Several mechanisms have been proposed to explain this phenomenon, including social interaction and peer effects \citep{mutz2013reflections}, as well as the influence of geographically structured media markets and migration flows \citep{green2019get}. As a result, electoral outcomes tend to exhibit significant spatial dependence, whereby the behavior observed in one county is systematically related to that of its neighbors \citep{lacombe2018use}.
From a statistical perspective, this dependence violates the assumption of independent observations underlying classical regression models and necessitates the use of spatial econometric techniques \citep{anselin1988spatial,lesage2009introduction}.

A variety of approaches have been proposed to model spatial dependence, including spatial autoregressive (SAR) models, {spatial error (SEM) models}, conditional autoregressive (CAR) models, intrinsic autoregressive (IAR) models, and Gaussian Markov random fields (GMRFs) \citep{anselin1988spatial,cressie1993statistics}. These model classes differ in how spatial dependence is represented and interpreted. {In SAR and SEM models, the autoregressive structure is imposed on a stochastic component of the data-generating process, either through the response variable itself or through the error term, generating the familiar spatial multiplier representation \citep{anselin1988spatial}. By contrast,} CAR, IAR, and GMRF models are typically formulated as latent random effects embedded within a hierarchical or GLM-based structure \citep{lee2017data,wood2017generalized,de2018gaussian}. Each class has found applications in economics, epidemiology, environmental sciences, and political analysis \citep{plant2018spatial,minguez2019alternative,toloza2021determinantes}.

Recent developments in SAR modeling have explored several extensions. Bayesian estimation frameworks have been proposed to incorporate prior information and deal with parameter uncertainty \citep{lesage1997bayesian,cepeda2022spatial}. Other contributions have focused on modeling both the mean and variance jointly \citep{toloza2024joint,toloza2026spatial, toloza2026SAR}, or introducing nonparametric terms to capture complex spatial effects \citep{montero2012sar}. Despite these developments, SAR-type models still lack a general and unified framework for integrating spatial dependence into the broader class of generalized linear models (GLMs), particularly for responses arising from non-Gaussian members of the exponential family.

A key limitation of SAR models in this regard is that the conditional distributions of the spatially autocorrelated errors are not easily generalized beyond the Gaussian case \citep{banerjee2003hierarchical}. While some efforts have been made to adapt SAR models to other distributions---such as logistic or probit models for binary outcomes, and Poisson models for count data \citep{GRIFFITH2009396}, or negative binomial models \citep{MUTISO2022100703}---these approaches are often developed in isolation and lack a unified estimation strategy applicable across different members of the exponential family.

A separate strand of the literature addresses non-Gaussian spatial dependence through fully model-based hierarchical frameworks. Hierarchical spatial GLMs \citep{banerjee2003hierarchical} embed a latent spatial process---typically a Gaussian Markov random field---within a generalized linear model, and are commonly estimated via Markov chain Monte Carlo or, more efficiently, through the Integrated Nested Laplace Approximation (INLA) \citep{rue2009approximate,lindgren2015bayesian}. Similarly, generalized additive models (GAMs) offer a flexible, nonparametric representation of spatial dependence within the GLM framework, either through smooth bivariate spline terms or through Markov random field smoothers defined directly over an areal neighborhood structure \citep{wood2017generalized}. Both approaches are well established, extensively validated, and often highly competitive in terms of predictive accuracy. However, they represent spatial dependence through a latent random effect or a smooth surface rather than through an explicit, interpretable autoregressive parameter, which complicates the direct quantification of spatial spillovers of the type routinely computed in the SAR literature \citep{lesage2009introduction}. 

This limitation motivates the development of more flexible frameworks that can jointly accommodate spatial dependence and the distributional properties of data arising from the exponential family, while retaining the interpretability of an explicit spatial dependence parameter that hierarchical and nonparametric approaches do not directly provide. An alternative line of research models spatial dependence through marginal mean structures and estimating equations. Examples include the spatial generalized linear modeling approaches of \citet{albert1995generalized}, \citet{gotway1997generalized}, \citet{liu2005}, and more recently \citet{wang2025using}. These methods avoid the specification of a full joint distribution and focus on obtaining consistent inference on regression effects under spatial dependence, and are particularly attractive when the primary goal is inference rather than the construction of a complete probabilistic spatial model. Although these approaches share the same general philosophy, they differ from the present proposal in important respects. \citet{liu2005} develop quasi-likelihood estimating equations for spatially correlated \emph{binary} data specifically, without extending the framework to other members of the exponential family or providing an explicit mechanism for estimating the spatial dependence parameter itself. \citet{wang2025using} propose a general Generalized Estimating Equations (GEE) based approach for nonlinear models with spatial data, but treat spatial dependence primarily as a nuisance captured through the working correlation structure, without deriving a companion quasi-likelihood procedure that yields a consistent, asymptotically normal estimator of the autoregressive parameter $\rho$ together with an interpretable spatial-multiplier representation.

Motivated by these developments, we propose the Generalized Spatial Correlated Inverse Mean Component (GSCIMC) model. The proposed framework combines a marginal spatial mean specification with log quasi-likelihood estimation and GEE, allowing spatially correlated responses from several members of the exponential family to be analyzed within a common inferential framework. GSCIMC is best understood not as a generalization of the classical SAR model but as a working alternative within the GEE paradigm: it coincides asymptotically with the SAR estimator of \citet{anselin1988spatial} in the Gaussian case, while extending naturally to non-Gaussian responses. Relative to the hierarchical and nonparametric approaches discussed above, GSCIMC is proposed for settings in which an interpretable, low-dimensional measure of spatial dependence in the linear predictor---together with direct and indirect effect decompositions analogous to those of \citet{lesage2009introduction}---is of primary interest, while still providing predictive performance that remains competitive with GAM and INLA.

The proposed estimation algorithm is implemented in the \texttt{GSCIMC()} function of the package spatemR in \cite{Rmanual}. This function is used both in the simulation study and in the real-data application presented in this paper, ensuring full reproducibility of the results.
The remainder of the paper is organized as follows. Section \ref{sec2} introduces the GSCIMC and the proposed estimation procedure, along with key theoretical properties. Section \ref{sec3} presents simulation studies assessing the model's performance under various distributional assumptions. Section \ref{sec4} applies the GSCIMC to a real dataset, illustrating its empirical relevance. Finally, Section \ref{sec5} discusses the main conclusions and outlines directions for future research.
\section{Generalized Spatial Correlated Inverse Mean Component}\label{sec2}
\subsection{Model Specification}

Let $\mathbf{Y}_n = (Y_1, \ldots, Y_n)^\top$ denote a vector of $n$ spatially indexed random variables, each $Y_i$ following a distribution from the exponential family with mean $\mu_i = \mathbb{E}(Y_i)$ and variance $\mbox{Var}(Y_i) = \phi_i V(\mu_i)$, where $V(\cdot)$ is the variance function associated with the exponential family. The linear predictor is defined via a monotonic and differentiable link function $g(\cdot)$, such that $g(\mu_i) = \eta_i$, where $\eta_i$ denotes the linear predictor for unit $i$. 
To incorporate spatial dependence, the model assumes that $\eta_i$ follows an autoregressive structure:
\begin{equation}
    \eta_i = \rho \sum_{j=1}^n w_{ij} \eta_j + \mathbf{x}_i^\top \pmb{\beta}, \label{eq:eta_scalar}
\end{equation}
where $\mathbf{x}_i$ is a $p$-dimensional vector of covariates, $\pmb{\beta}$ is a $p$-vector of regression coefficients, $\rho \in (-1,1)$ is the spatial autoregressive parameter, and $w_{ij}$ are the elements of $\mathbf{W} = \{w_{ij}\}_{n \times n}$. 
In order to guarantee the existence of the spatial process, identifiability of the parameters, and the validity of the asymptotic results derived later, we introduce the following assumptions.
\begin{assumption}\label{as:mom}
There exists $\delta>0$ such that
\[
\sup_{1\leq i\leq n} E\left(|Y_i|^{2+\delta}\right)<\infty.
\]
Moreover, the variance function $V(\mu_i)$ is bounded away from zero and infinity on compact subsets of the parameter space.
\end{assumption}

\begin{assumption}\label{as:gmu}
Let $\mu_i = \mathbb{E}(Y_i)$ denote the marginal mean of $Y_i$. The conditional mean structure satisfies
\[
g(\mu_i) =  \rho \sum_{j=1}^n w_{ij} g(\mu_j) + \mathbf{x}_i^\top \pmb{\beta},
\]
where $g(\cdot)$ is twice continuously differentiable and strictly monotone, with $g'(\mu) \neq 0$ for all $\mu$ in the interior of the parameter space.
\end{assumption}

\begin{assumption}\label{as::w1}
 Let $\mathbf{W}=\{w_{ij}\}_{n \times n}$ be an $n \times n$ spatial weights matrix known and non-stochastic, and $\rho \in \mathbb{R}$ a scalar parameter with $w_{ij}\geq 0$. We assume that the matrix power series
\[
\sum_{j=0}^{\infty} \rho^j \mathbf{W}^j
\]
is convergent. 
\end{assumption}
A sufficient condition for convergence is $-1/\lambda_{\max}(\mathbf{W})<|\rho| < 1/\lambda_{\max}(\mathbf{W})$, where $\lambda_{\max}(\mathbf{W})$ denotes the spectral radius of $\mathbf{W}$. In practice, a common special case is when $\mathbf{W}$ is row-standardized, which implies $\lambda_{\max}(\mathbf{W})=1$ and therefore $|\rho|<1$. 
\begin{assumption}\label{as:lee}
The sequence of spatial weights matrices $\{\mathbf{W}_n\}$ satisfies
\[
\sup_{n\in \mathbb{N}} \max_{1\leq i \leq n}\sum_{j=1}^n |w_{ij}| < \infty,
\qquad
\sup_{n\in \mathbb{N}} \max_{1\leq j \leq n}\sum_{i=1}^n |w_{ij}| < \infty.
\]

\end{assumption}
Assumption~\ref{as:lee} is commonly used in the asymptotic analysis of SAR \citep{lee2004}. {It imposes uniformly bounded row and column sums on the spatial weights matrices, which ensures that the magnitude of spatial interactions remains controlled as the sample size increases.}  Throughout this paper, we consider row-standardized spatial weight matrices, for which the row sums are equal to one and the condition is automatically satisfied.

\begin{assumption}\label{as::xp}
Let $\mathbf{X}$ be the $n \times p$ matrix of covariates. We assume that $\mbox{rank}(\mathbf{X})=p$ for all $n$ and that
\[
\lim_{n\to \infty} \frac{1}{n}\mathbf{X}^\top \mathbf{X}
\]
exists and is finite.
\end{assumption}

Under Assumptions~\ref{as:gmu}-\ref{as::xp}, the model is well defined. In matrix form, Equation~\eqref{eq:eta_scalar} can be rewritten as:
\begin{equation}
    \pmb{\eta} = \rho \mathbf{W} \pmb{\eta} + \mathbf{X} \pmb{\beta}=\mathbf{A}^{-1} \mathbf{X} \pmb{\beta}, \label{eq:eta_vector}
\end{equation}
where $\pmb{\eta} = (\eta_1, \ldots, \eta_n)^\top$ and $\mathbf{X} = (\mathbf{x}_1^\top, \ldots, \mathbf{x}_n^\top)^\top$ is the $n \times p$ design matrix, and $\mathbf{A} = \mathbb{I}_n - \rho \mathbf{W}$. Under Assumption~\ref{as::w1}, we obtain that
\[
\mathbf{A}^{-1}=(\mathbb{I}_n - \rho \mathbf{W})^{-1} = \sum_{j=0}^{\infty} \rho^j \mathbf{W}^j .
\] 
Consequently, the mean vector is obtained by applying the inverse of the link function:
\begin{equation}
    \pmb{\mu} = g^{-1}(\pmb{\eta}) = g^{-1}(\mathbf{A}^{-1} \mathbf{X} \pmb{\beta}). \label{eq:mu}
\end{equation}
In the standard GEE framework, the covariance matrix of the response vector is specified as
\[
\mbox{Cov}(\mathbf{Y}_n) = \phi \mathbf{V}(\pmb{\mu})^{\frac{1}{2}} \mathbf{R}(\alpha) \mathbf{V}(\pmb{\mu})^{\frac{1}{2}},
\]
where $\phi$ is a scale parameter, $\mathbf{V}=\mbox{diag}(V(\mu_1),\ldots,V(\mu_n))$ contains the variance function evaluated at the marginal means, and $\mathbf{R}(\alpha)$ is a working correlation matrix that approximates the dependence structure of the data. 
The autoregressive parameter $\rho$ is defined through the marginal mean structure in Equation~\eqref{eq:eta_vector}, and therefore, it has a direct interpretation as a measure of spatial dependence in the linear predictor. Consequently, the mean specification in Equation~\eqref{eq:eta_scalar} constitutes the primary modeling assumption and captures the systematic spatial dependence of the response variable.

By contrast, the covariance matrix in Equation~\eqref{eq:covariance} is not derived from a fully specified joint distribution. Instead, it is introduced as a working covariance structure in the spirit of GEE, providing a tractable approximation to the dependence induced by the autoregressive mean structure. Its role is to improve the efficiency of parameter estimation and statistical inference rather than to represent the true covariance structure of the data-generating process.

In the spatial setting, we approximate the dependence induced by the autoregressive structure through a spatial working correlation matrix. In particular, we specify
\begin{equation}
    \mbox{Cov}(\mathbf{Y}_n) = \phi \mathbf{V}(\pmb{\mu})^{\frac{1}{2}} \left( \mathbf{A}^\top\mathbf {A}\right)^{-1}  \mathbf{V}(\pmb{\mu})^{\frac{1}{2}}, \label{eq:covariance}
\end{equation}
where the matrix $(\mathbf{A}^\top\mathbf{A})^{-1}$ captures the spatial dependence implied by the autoregressive operator. This formulation follows the GEE philosophy of approximating the true covariance structure through a working correlation matrix while preserving robustness of the estimating equations for the mean parameters.

Importantly, the spatial parameter $\rho$ is identified through the mean structure, while the working covariance matrix provides additional information that improves the efficiency of the estimating equations and the resulting inference without requiring repeated measurements on the same observational units, as it is typically needed in classical GEE applications. 
The spatial working correlation induced by $\left(\mathbf{A}^\top\mathbf {A}\right)^{-1}$ effectively captures the dependence between units, providing additional information for $\rho$ while preserving consistency of the estimator for the mean parameters.
From Equation~\ref{eq:covariance}, the dispersion parameter of each unit to vary according to:
\begin{equation}
    \phi_i = \phi \omega_i, \label{eq:phi1}
\end{equation}
where $\omega_i=\left(\mathbf{A}^\top\mathbf {A}\right)^{-1}_{ii}$ is the $i$-th diagonal element of $\left(\mathbf{A}^\top\mathbf{A} \right)^{-1}$. 
Equations~\eqref{eq:eta_scalar}–\eqref{eq:covariance} define the GSCIMC model. 
This specification constitutes a spatially structured generalized mean model, estimated within a GEE-type framework, that extends the GLM framework to accommodate spatial dependence in the linear predictor. The resulting model enables the analysis of spatially correlated outcomes with flexible non-Gaussian marginal distributions and spatial heterogeneity in mean.

The spatial convergence condition guarantees the existence of $(\mathbb{I}_n-\rho \mathbf{W})^{-1}$, and therefore, the spatial multiplier representation. The full column rank of $\mathbf{X}$ ensures the identifiability of the regression parameters $\pmb{\beta}$, while the regularity conditions on the regressors guarantee finite moments and the stability of the estimating equations. Together with the monotonicity and differentiability of the link function $g(\cdot)$, these conditions ensure that the mean structure is properly defined and that the GSCIMC estimating equations admit a well–defined solution.
\subsection{Parameter estimation}
The proposed estimation procedure adopts a hybrid strategy in which the spatial dependence parameter $\rho$ plays a dual role. On the one hand, $\rho$ enters the mean structure through the spatially adjusted covariates, ensuring its identifiability and, as shown in the asymptotic results, the consistency of its estimator. Also, $\rho$ defines the correlation structure, allowing the GEE framework to improve the estimation of the covariance matrix of $\hat{\pmb{\beta}}$ and $\hat{\rho}$, leading to more efficient statistical inference.

For proposing an algorithm to estimate the parameters in a GSCIMC, the log quasi-likelihood function associated with the exponential family and parameter $\rho$ is constructed as
\begin{equation}
    \ell(\rho, \pmb{\beta}, \mathbf{Y}_n, \mathbf{X}) = \sum_{i=1}^n \ell_i(\rho, Y_i, \mathbf{x}_i, \pmb{\beta}), \label{eq:quasilikelihood}
\end{equation}
where the contribution $\ell_i(\rho, Y_i, \mathbf{x}_i, \pmb{\beta}) $ depends on the assumed distribution of $Y_i$ \citep{hardin2002generalized,dobson2018introduction}, and it is defined by:
\begin{equation}
\ell_i(\cdot) = \begin{cases}
-\dfrac{(Y_i - \tilde{\mathbf{x}}_i^\top \pmb{\beta})^2}{2 \phi_i} & \text{if } Y_i \sim \mathcal{N}(\mu_i, \phi_i),\\
Y_i \ln(\mu_i) + (m_i - Y_i) \ln(1 - \mu_i) & \text{if } Y_i \sim \text{Binomial}(\mu_i, m_i),\\
Y_i \ln(\mu_i) - \mu_i & \text{if } Y_i \sim \text{Poisson}(\mu_i),\\
-\dfrac{Y_i}{\mu_i} - \ln(\mu_i) & \text{if } Y_i \sim \text{Gamma}(\mu_i, \phi_i),\\
\dfrac{Y_i \mu_i}{\mu_i + V(\mu_i)} + V(\mu_i) \ln\left( \dfrac{V(\mu_i)}{V(\mu_i) + \mu_i} \right) & \text{if } Y_i \sim \text{Neg Binom}(\mu_i, \phi_i).
\end{cases} \label{eq:qlfunction}
\end{equation}
or any other chosen distribution. For instance, in our examples we consider $Y_i$ from the Gaussian, Poisson, or Binomial families, but any member of the exponential family could be accommodated within this framework.
Here, $\mu_i = g^{-1}(\tilde{\mathbf{x}}_i^\top \pmb{\beta})$ with the spatially adjusted design matrix given by $\tilde{\mathbf{X}} = \mathbf{A}^{-1} \mathbf{X} = (\mathbb{I}_n - \rho \mathbf{W})^{-1} \mathbf{X}. \label{eq:tildeX}$
The estimation of the model parameters $(\pmb{\beta}, \rho)$, as defined in Equations~\eqref{eq:eta_scalar}–\eqref{eq:covariance}, is performed through an iterative log quasi-likelihood procedure that combines the spatial estimation framework of \cite{anselin1998introduction} with the GEE methodology of \cite{liang1986longitudinal}. The full procedure is outlined in Algorithm~\ref{algorithm}. 

\begin{algorithm}[!ht]\label{algorithm}
\footnotesize
\caption{Quasi-likelihood estimation for GSCIMC model}
\KwIn{Response vector $\mathbf{Y}_n$, design matrix $\mathbf{X}$, spatial weights matrix $\mathbf{W}$, convergence thresholds $\epsilon_\beta$, and $\epsilon_\rho$.}
\KwOut{Estimates $\hat{\pmb{\beta}}$, $\hat{\rho}$, $\hat{\phi}$.}
Construct $\tilde{\mathbf{Y}}=g(\mathbf{Y}+1)$ and fit a robust spatial lag model to obtain the initial value $\rho^{(0)}$.
Estimate $\pmb{\beta}^{(0)}$ using \texttt{glm} with design matrix $\tilde{\mathbf{X}}$.

\Repeat{$|\rho^{(k+1)} - \rho^{(k)}| \leq \epsilon_\rho$}{
    
    Construct the log quasi-likelihood function:
    \[
    \ell(\rho, \pmb{\beta}^{(k)}, \mathbf{Y}_n, \mathbf{X}) = \sum_{i=1}^n \ell_i\left(\rho, Y_i, \mathbf{x}_i, \pmb{\beta}^{(k)}\right)
    \]
    
    Maximize $\ell$ with respect to $\rho$ using the \texttt{optim} function in \texttt{R} with the L-BFGS-B algorithm, subject to the constraint $\rho\in(-0.999,0.999)$, to obtain $\rho^{(k+1)}$.

    Update $\tilde{\mathbf{X}}^{(k+1)} \gets (\mathbb{I}_n - \rho^{(k+1)}\mathbf{W})^{-1}\mathbf{X}$.

    Estimate initial $\pmb{\beta}^{(t_0)}$ using \texttt{glm} with $\tilde{\mathbf{X}}^{(k+1)}$.

    \Repeat{$\|\pmb{\beta}^{(t_{m+1})} - \pmb{\beta}^{(t_m)}\| \leq \epsilon_\beta$}{
        Compute working response:
        \[
        \mathbf{z}^{(t_m)} = \tilde{\mathbf{X}}^{(k+1)} \pmb{\beta}^{(t_m)} - \mathbf{N}^{(t_m)}\left(\mathbf{Y}_n - \pmb{\mu}^{(t_m)}\right)
        \]
        \[
        \mathbf{N}^{(t_m)} =\mbox{diag}\left\{\frac{\partial \mu_1^{(t_m)}}{\partial \eta_1^{(t_m)}}, \ldots, \frac{\partial \mu_n^{(t_m)}}{\partial \eta_n^{(t_m)}} \right\}
        \]
        Update $\pmb{\beta}^{(t_{m+1})}$ using (Apply numerical stabilization rules to prevent underflow, overflow, divisions by zero.):
        \[
        \pmb{\beta}^{(t_{m+1})} = \left(\tilde{\mathbf{X}}^{(k+1)\top} \mbox{Cov}(\mathbf{Y})^{-1} \tilde{\mathbf{X}}^{(k+1)}\right)^{-1} \tilde{\mathbf{X}}^{(k+1)\top} \mbox{Cov}(\mathbf{Y})^{-1} \mathbf{z}^{(t_m)}
        \]
    }

    Set $\pmb{\beta}^{(k+1)} \gets \pmb{\beta}^{(t_{m+1})}$.
}

Set $\hat{\pmb{\beta}} \gets \pmb{\beta}^{(k+1)}$, $\hat{\rho} \gets \rho^{(k+1)}$, $ \hat{\phi}=  \frac{1}{n-p-1}\sum_{i=1}^n \frac{(Y_i-\hat{\mu }_i)^2}{\hat{\omega}_iV(\hat{\mu}_i)}$.
\KwRet{$\hat{\pmb{\beta}}, \hat{\rho}, \hat{\phi}$}.
\end{algorithm}
\noindent To improve numerical stability when highly heterogeneous weights are present, all quantities involved in products, divisions, and likelihood evaluations are constrained to remain within prespecified numerical bounds ($10^{-6}$). This prevents underflow, overflow, and numerical instabilities associated with extremely small or large intermediate values. 
\begin{remark} 
Computational efficiency of Algorithm \ref{algorithm} is achieved through sparse matrix representations, avoiding the storage of unnecessary intermediate quantities and exploiting matrix operations implemented in the \texttt{Matrix} package. In our empirical application ($n\approx 3000$), the complete estimation procedure required less than 60 seconds on a standard desktop-class computer. Additional implementation details
are available in the package documentation.
\end{remark}
 Once the algorithm \ref{algorithm} obtained convergence, $\phi$ can be estimated as:
\begin{equation}\label{eq:phi}
    \hat{\phi}=  \frac{1}{n-p-1}\sum_{i=1}^n \frac{(Y_i-\hat{\mu }_i)^2}{\hat{\omega}_iV(\hat{\mu}_i)}
 \end{equation}
where $\hat{\omega}_i$ is the $i$-th diagonal element of $\left(\hat{\mathbf{A}}^\top\hat{\mathbf{A}}\right)^{-1}$. The variance of the estimators for $\pmb{\beta}$ is obtained by adapting the sandwich variance estimator proposed by \cite{pan2001robust}, which is defined as:
\begin{equation}\label{eq:pan}
 \widehat{Var}\left(\hat{\pmb{\beta}}\right) = \hat{\mathbf{J}}^{-1} \hat{\mathbf{B}} \hat{\mathbf{J}}^{-1}
\end{equation}
where
\begin{align}
    \mathbf{J}&=\frac{1}{\phi}\tilde{\mathbf{X}}^{\top}\mathbf{N}^{\top} \mathbf{V}(\pmb{\mu})^{-\frac{1}{2}}\left( \mathbf{A}^\top\mathbf{A}\right) \mathbf{V}(\pmb{\mu})^{-\frac{1}{2}}\mathbf{N}\tilde{\mathbf{X}},\nonumber\\
    \mathbf{B} &= \tilde{\mathbf{X}}^{\top}\mathbf{N}^{\top}\mathbf{V}(\pmb{\mu})^{-\frac{1}{2}}\left( \mathbf{Y}-\hat{\pmb{\mu}}\right) \left( \mathbf{Y}-\hat{\pmb{\mu}}\right)^\top\mathbf{V}(\pmb{\mu})^{-\frac{1}{2}} \mathbf{N}\tilde{\mathbf{X}},  \nonumber  \\
     \mathbf{N} &=\mbox{diag}\left\{\frac{\partial \mu_1}{\partial \eta_1}, \ldots, \frac{\partial \mu_n}{\partial \eta_n} \right\},\label{eq:jb}
\end{align}
The variance of the estimator $\hat{\rho}$ is given by the following expression:
\begin{align}
\widehat{Var}(\hat{\rho})^{-1} &= -\mathbb{E}\left[\frac{\partial^2}{\partial \rho^2} \ell(\rho, \hat{\pmb{\beta}}, \mathbf{Y}_n, \mathbf{X}, \mathbf{W})\right]\nonumber\\
&=-\sum_{i=1}^{n} \mathbb{E}\left[\frac{\partial^2}{\partial \rho^2} \ell_i(\rho, \hat{\pmb{\beta}}, Y_i, \mathbf{X}, \mathbf{W})\right]\label{eq:varrho}
\end{align}
Note that the mean is well specified, and therefore its scoring function $\frac{\partial \ell_i}{\partial \mu_i}$ has expectation 0. Using the chain rule in Equation \eqref{eq:varrho} and taking into account that $\ell_i(\cdot)$ is a function of $\mu_i=g^{-1}(\eta_i)$, which in turn depends of $\rho$, it is obtained that:
\begin{align}
    \widehat{Var}(\hat{\rho})^{-1} &=-\sum_{i=1}^{n}\left(\frac{\partial\mu_i}{\partial\rho}\right)^2 \mathbb{E}\left[\frac{\partial^2\ell_i(\rho, \hat{\pmb{\beta}}, Y_i, \mathbf{X}, \mathbf{W})}{\partial \mu_i^2} \right] \nonumber\\
    &-\sum_{i=1}^{n}\left(\frac{\partial^2\mu_i}{\partial\rho^2}\right) \mathbb{E}\left[\frac{\partial \ell_i(\rho, \hat{\pmb{\beta}}, Y_i, \mathbf{X}, \mathbf{W})}{\partial \mu_i} \right]\nonumber\\
    &=-\sum_{i=1}^{n}\left(\frac{\partial\mu_i}{\partial\rho}\right)^2 \mathbb{E}\left[\frac{\partial^2\ell_i(\rho, \hat{\pmb{\beta}}, Y_i, \mathbf{X}, \mathbf{W})}{\partial \mu_i^2} \right] \label{eq:varrho1}
\end{align}

The shape of each of the variances of $\hat{\rho}$ will depend on the functional form of the function $\ell_i(\cdot)$ defined in Equation \eqref{eq:qlfunction}. As shown in Table~\ref{tab:var_rho}, the expressions for the inverse variance $\mbox{Var}(\hat{\rho})^{-1}$ vary according to the distribution and the log quasi-likelihood derivatives. 
\begin{table}[!ht]
\centering
\caption{Expressions for $\mbox{Var}(\hat{\rho})^{-1}$, under different distributions, derived using the log quasi-likelihood functions defined in Equation~\eqref{eq:qlfunction} and their derivatives given in Equations~\eqref{deta}--\eqref{deta3}.}
\begin{tabular}{ll}
\hline
\textbf{Distribution} & $\mbox{Var}(\hat{\rho})^{-1}$ \\
\hline
$\mathcal{N}(\mu_i, \phi)$ &
$\displaystyle \frac{1}{\phi} \sum_{i=1}^n \left( \frac{\partial \mu_i}{\partial \rho} \right)^2 $ \\
 $\text{Poisson}(\mu_i)$ &
$\displaystyle \sum_{i=1}^n \left( \frac{\partial \mu_i}{\partial \rho} \right)^2  \frac{1}{\mu_i}$ \\
 $\text{Binomial}(\mu_i, m_i)$ &
$\displaystyle
\sum_{i=1}^n \left( \frac{\partial \mu_i}{\partial \rho} \right)^2 
\left( -\frac{m_i}{\mu_i^2} + \frac{m_i - 1}{(1 - \mu_i)^2} \right)
$ \\
 $\text{Gamma}(\mu_i, \phi)$ &
$\displaystyle \sum_{i=1}^n \left( \frac{\partial \mu_i}{\partial \rho} \right)^2  \left( \frac{1}{\mu_i^2} - \frac{1}{\mu_i \phi} \right)$ \\
$\text{Neg Binom}(\mu_i, \phi)$ &
$\displaystyle \sum_{i=1}^n \left( \frac{\partial \mu_i}{\partial \rho} \right)^2  \left( \mu_i + \frac{\mu_i^2}{\phi} \right)^{-1}$ \\
\hline
\end{tabular}
\label{tab:var_rho}
\end{table}
For any other distribution, the form $Var(\hat{\rho})^{-1}$ is obtained in the same way, or it can be approximated by numerical methods, taking advantage of the fact that the function \eqref{eq:qlfunction} under Assumptions~\ref{as::w1} and \ref{as::xp} is continuous and it has real-valued for all $\rho\in \left(\frac{-1}{\lambda_{max}},\frac{1}{\lambda_{max}}\right)$. However, the following properties will allow the variances of $\hat{\rho}$ to be calculated:
\begin{align}
    \frac{\partial g(\mu_i)}{\partial\rho}&= \left(\mathbf{A}^{-1}\mathbf{WA}^{-1}\mathbf{X}\pmb{\beta}\right)_i, \label{deta}\\
    \frac{\partial g^{-1}(\eta_i)}{\partial\rho}&=  \frac{\partial \mu_i}{\partial\rho}= \frac{\partial g^{-1}(\eta_i)}{\partial\eta_i}\left(\mathbf{A}^{-1}\mathbf{WA}^{-1}\mathbf{X}\pmb{\beta}\right)_i\label{deta3}
\end{align}
where $\left(\mathbf{A}^{-1}\mathbf{WA}^{-1}\mathbf{X}\pmb{\beta}\right)_i$ is the $i$-th component of the vector $\mathbf{A}^{-1}\mathbf{WA}^{-1}\mathbf{X}\pmb{\beta}$.
Although the inverse Fisher information derived in Equation~\eqref{eq:varrho1} provides an analytical approximation for the variance of $\hat{\rho}$; this expression relies on the correct specification of the working covariance structure assumed in Equation~\eqref{eq:covariance}. In practice, the spatial working correlation matrix $(\mathbf{A}^{\top}\mathbf{A})^{-1}$ is only an approximation to the true dependence structure of the data. Consequently, the variance obtained from the expected Hessian may underestimate the true sampling variability of $\hat{\rho}$.
To account for possible misspecification of the covariance structure, we additionally consider a robust variance estimator of the sandwich type, following the GEE framework \citep{liang1986longitudinal}. The sandwich estimator of the variance of $\hat{\rho}$ is then defined as
\begin{equation}\label{eq:varrhosand}
\widehat{\mbox{Var}}_{\text{sand}}(\hat{\rho}) =
 \widehat{Var}(\hat{\rho})^{-1}\widehat{B}_{\rho} \widehat{Var}(\hat{\rho})^{-1},    
\end{equation}
where $ \widehat{Var}(\hat{\rho})$ is defined in Equation~\eqref{eq:varrho1} and
\[
B_{\rho} =
\sum_{i=1}^{n} \left(\frac{\partial \ell_i(\rho,Y_i,\mathbf{x}_i,\pmb{\beta})}{\partial \rho}\right)^2
\]
is the empirical variance of the score contributions. 
Assumption~\ref{as:mom} is a standard moment condition ensuring the existence of higher-order moments required for the asymptotic normality of the estimating equations and the consistency of the sandwich variance estimator.

This estimator remains consistent even if the working covariance matrix is misspecified, while preserving the efficiency gains obtained when the assumed spatial structure approximates the true dependence \cite{yin2006asymptotic}. Therefore, the sandwich estimator provides a robust alternative for inference on the spatial autoregressive parameter $\rho$ within the proposed GSCIMC framework. Note that the variance expression in Equation~\eqref{eq:varrho1} is conditionally on 
$\hat{\beta}$ because the estimator $\hat{\rho}$ is obtained given the 
estimated mean parameters. Since $\hat{\beta}$ is a consistent estimator, 
the conditional and unconditional variances coincide asymptotically.
\subsection{Asymptotic inference}
With the estimation method above, the following results are obtained:
\begin{theorem}\label{te1}
Under Assumptions~\ref{as:mom}-\ref{as::xp} and model defined in Equations~\eqref{eq:eta_scalar}–\eqref{eq:covariance}  then:
\begin{align}
\hat{\rho} & \xrightarrow[n\to \infty]{d} \mathcal{N}(\rho, Var_{\text{sand}}(\hat{\rho})) \\
\hat{\pmb{\beta}} & \xrightarrow[n\to \infty]{d} N_{p}\left(\pmb{\beta}, \mathbf{J}^{-1} \mathbf{B} \mathbf{J}^{-1} \right)
 \end{align}
\end{theorem}
\begin{proof}
See \ref{proofte1}
\end{proof}
\begin{corollary}\label{cor1}
If $\rho=0$ is specified in Equations~\eqref{eq:eta_scalar}–\eqref{eq:covariance}, then the proposed estimator $\hat{\pmb{\beta}}$ reduces exactly to the usual GLM estimator.
\end{corollary}
\begin{proof}
See \ref{proofcor1}
\end{proof}
\begin{lemma}\label{lem1}
If the true value of $\rho$ is zero in Equations~\eqref{eq:eta_scalar}–\eqref{eq:covariance}, then the estimator $\hat{\pmb{\beta}}$ obtained with the proposed methodology converges in probability to the estimator from a usual GLM.
\end{lemma}
\begin{proof}
See \ref{prooflema1}
\end{proof}
{
\begin{corollary}\label{cor2} 
If $Y_i\sim \mathcal{N}(\mu_i,\sigma_i^2)$ in Equations~\eqref{eq:eta_scalar}--\eqref{eq:covariance}, then the proposed model and the Gaussian SAR model of \cite{anselin1988spatial} share the same marginal mean and covariance structure. Under Assumptions~\ref{as:mom} and \ref{as:lee}, the estimators $\hat{\pmb{\beta}}$ and $\hat{\rho}$ converge in probability to the same true parameter values as the corresponding estimators of the Gaussian SAR model.
\end{corollary}
}
\begin{proof}
See \ref{proofcor2}
\end{proof}

\begin{corollary}\label{cor3}

{If $Y_i\sim Po(\lambda_i)$ in Equations~\eqref{eq:eta_scalar}--\eqref{eq:covariance}, then the proposed estimator and the estimator of \cite{lambert2010two} target the same marginal mean structure. Under Assumptions~\ref{as:mom} and \ref{as:lee}, both estimators converge asymptotically to the same values of $\pmb{\beta}$ and $\rho$.}
\end{corollary}

\begin{proof}
See \ref{proofcor3}
\end{proof}

\begin{corollary}\label{cor4}
{If $Y_i\sim Bin(\mu_i,m_i)$ in Equations~\eqref{eq:eta_scalar}--\eqref{eq:covariance}, then the proposed estimator and the estimator of \cite{song2022variable} target the same marginal mean structure. Under Assumptions~\ref{as:mom} and \ref{as:lee}, both estimators converge asymptotically to the same values of $\pmb{\beta}$ and $\rho$.}
\end{corollary}

\begin{proof}
See \ref{proofcor4}
\end{proof}

The variance dominance follows from the classical GEE theory of \cite{liang1986longitudinal}. When the working covariance structure incorporates information on the true dependence pattern, the resulting estimating equations are asymptotically at least as efficient as those obtained under an independence working covariance matrix. Since the estimators of \cite{lambert2010two} and \cite{song2022variable} rely primarily on the marginal mean specification, the proposed estimator exploits additional information contained in the spatial dependence structure through Equation~\eqref{eq:covariance}. 
These theoretical results establish connections between the proposed methodology and several existing spatial estimators under particular distributional assumptions. They also suggest potential efficiency gains arising from the use of the proposed working covariance structure.
Moreover, it enables extensions to other distribution families such as the Gamma, the negative binomial, among others \citep{dobson2018introduction}.

When analyzing the effect of covariates, it is insufficient to rely solely on the estimated coefficients as marginal effects. As in the SAR model, spatial interactions between different areas must be considered. Consequently, it is necessary to compute the average direct, indirect (spillover), and total effects.

The direct effect is defined as the average of the diagonal elements of the partial derivative matrix, $\pmb{S}_k$, which is given by:
\begin{equation}\label{sk_dit}
\pmb{S}_k = \beta_k(\mathbb{I}-\rho\pmb{W})^{-1}
\end{equation}

The indirect effect corresponds to the average of the off-diagonal elements in each row (or column) of the matrix $\pmb{S}_k$. The total effect is computed as the sum of the direct and indirect effects. These estimators can be interpreted following the classical GLM framework \citep{dobson2018introduction}, while also generalizing the approach introduced by \cite{lesage2009introduction}.
The Corrected Information Criterion (CIC) is an extension of the Akaike Information Criterion (AIC) for quasi-likelihood models with correlated or overdispersed data. It is used for model selection, balancing goodness-of-fit and model complexity. CIC is computed as
\[
\mbox{CIC} = \mbox{tr}\left(\mathbf{J}^{-1} \mathbf{B}\right),
\]
where $\mathbf{J}$ defined in Equation~\eqref{eq:pan} is the expected information matrix and $\mathbf{B}$ is the variance of the score. Intuitively, $\mathbf{J}$ captures the sensitivity of the likelihood to the parameters, while $\mathbf{B}$ accounts for the variability of the score under the log quasi-likelihood. Lower CIC values indicate models that achieve a better trade-off between fit and parsimony, and it is particularly useful when comparing models that include correlated observations or different dispersion assumptions.
As with other models estimated through GEE, model adequacy can be assessed using residual diagnostics and influence measures. In particular, Pearson residuals,
\[
r_{Pi}=\frac{Y_i-\hat{\mu}_i}
{\sqrt{\hat{\phi}\,\omega_i\,V(\hat{\mu}_i)}},
\]
and deviance residuals,
\[
r_{Di}=\text{sign}(Y_i-\hat{\mu}_i)
\sqrt{2\left\{\ell(Y_i;Y_i)-\ell(Y_i;\hat{\mu}_i)\right\}},
\]
can be used to identify outlying observations, departures from distributional assumptions, and systematic lack of fit. Since the proposed model explicitly incorporates spatial dependence, it is also advisable to examine residual spatial autocorrelation, for example through Moran's $I$ computed from Pearson or deviance residuals. Values close to zero indicate that the model has adequately captured the underlying spatial structure. In addition, leverage measures and influence diagnostics analogous to those developed for GEE models may be employed to identify observations exerting an undue impact on parameter estimates and standard errors \citep{hardin2002generalized, cruz2023semiparametric}. Together, these diagnostics provide a practical framework for evaluating model adequacy, identifying influential spatial units, and assessing the robustness of inferential conclusions.
\section{Simulation}\label{sec3}
\subsection{Simulation under correct model specification}
Three simulation studies were conducted to evaluate the proposed methodology under different distributional assumptions: Poisson, Gamma and Binomial. For each scenario, data were generated on a regular spatial grid with sample sizes $n = 49, 81, 144, 400$, and each setting was replicated 500 times. The spatial dependence was modeled through a Rook's contiguity matrix $W = (w_{ij})$ \citep{bivand2022r}. Covariates $x_{1i} \sim \mathcal{N}(0,1)$, $x_{2i} \sim \mathcal{N}(2,1)$, and $x_{3i} \sim U(0,1)$ were included, with fixed regression coefficients $\beta_0=0.5$, $\beta_1=-0.5$, and $\beta_2=1$. The spatial autoregressive parameter $\rho$ varied over $\{-0.75, -0.5, -0.25, 0, 0.25, 0.5, 0.75\}$.  The data-generating processes for each distribution were as follows. In all cases, the linear predictor was defined as
\[
\eta_i = \rho \sum_{j=1}^n w_{ij} \eta_j + \beta_0 + \beta_1 x_{1i} + \beta_2 x_{2i}.
\]
 For each simulation setting, the vector of linear predictors was obtained by solving the spatial system $\pmb{\eta}=(\mathbf{I}-\rho\mathbf{W})^{-1}\mathbf{X}\pmb{\beta}$, from which the mean vector was computed as
$\pmb{\mu}=g^{-1}(\pmb{\eta})$. The response variable was then generated from the corresponding distribution conditional on $\pmb{\mu}$.
\begin{itemize}
    \item \textbf{Poisson:} $y_i \sim \text{Poisson}(\mu_i)$, with $\log(\mu_i) = \eta_i$, and $\phi=1$. 
    \item \textbf{Gamma:} $ y_i \sim \text{Gamma}(\mu_i, \phi)$, with $\log(\mu_i) = \eta_i$ and $\phi\sim Unif(1,5)$.
    \item \textbf{Binomial:} $y_i \sim \text{Binomial}(\mu_i, M_i)$, with $\text{logit}(\mu_i) = \eta_i$, and $M_i \sim U\{1, 100\}$ is a discrete uniform random variable.
\end{itemize}
The gamma dispersion parameter will allow us to  evaluate scenarios where $\phi$ is not constant.  {Although the log link is used throughout this study, the canonical Gamma link is the inverse link.}

\begin{table}[ht]
\centering
\caption{Summary of models fitted in the simulation studies. For the SAR, SARAR, and Ro-SAR models, the transformed response was used: $\log(Y)$ for the Poisson and Gamma simulations, and $\mathrm{logit}(Y)$ for the Binomial simulations. These models were therefore fitted as Gaussian SAR on the transformed response.}
\label{tab:model_summary}
\begin{tabular}{|l|c|c|l|}
\hline
\textbf{Model} & \textbf{Working response} & \textbf{Link} & \textbf{R package::function} \\
\hline
Proposed    & $Y$ & Log / Logit & \textit{spatemR::GSCIMC} \\
GLM         & $Y$ & Log / Logit & \textit{base::glm} \\
Pois-SAR    & $Y$ & Log & \textit{sppois::sarpoisson} \\
SAR         & $\log(Y)$ / $\mathrm{logit}(Y)$ & Identity & \textit{spatialreg::sacsarlm} \\
SARAR       & $\log(Y)$ / $\mathrm{logit}(Y)$ & Identity & \textit{spatialreg::sacsarlm} \\
Ro-SAR      & $\log(Y)$ / $\mathrm{logit}(Y)$ & Identity & \textit{sphet::spreg} \\
Probit-SAR  & $1(Y\geq0.5)$ & Probit & \textit{spatialprobit::sarprobit} \\
GAM         & $Y$ & Log / Logit & \textit{mgcv::gam} \\
INLA        & $Y$ & Log / Logit & \textit{INLA::inla} \\
\hline
\end{tabular}
\end{table}

For each simulated dataset, parameters were estimated using the \textbf{Proposed} model given in Equation~\eqref{eq:eta_scalar}, as well as several comparative approaches (summarized in Table~\ref{tab:model_summary}):
\begin{enumerate}
    \item \textbf{GLM:} A classical GLM with linear predictor
    \[
    \eta_i = \beta_0 + \beta_1 x_{1i} + \beta_2 x_{2i} +\beta_3 \tilde{y}_{j},
    \]
    where $\tilde{y}_{j}=\sum_{j=1}^n w_{ij} y_j$ denotes the spatially lagged response variable, and the coefficient $\beta_3$ may capture spatial autocorrelation.  This formulation can be seen as a simple extension of a standard GLM, 
where we include the spatial lag, $\tilde{y}_i$, as an additional covariate. 
Consequently, one can estimate the model using the usual \texttt{glm} function in \textsf{R}, 
treating $\tilde{y}_i$ just like any other explanatory variable.
    \item \textbf{SAR and SARAR models:}  SAR and SARAR maximum likelihood models assuming normality on $\log(y_i)$ for Poisson and Gamma  distribution or $logit(y_i)$ for Binomial distribution \citep{SpatialReg}, estimated with the \textit{spatialreg} package \citep{SpatialReg}.
    \item \textbf{Ro-SAR:} Heteroscedasticity-robust SAR model \citep{arraiz2010spatial} on $\log(y_i)$ for Poisson and Gamma  distribution or $logit(y_i)$ for Binomial distribution.
    \item \textbf{Pois-SAR:} For the Poisson case, a SAR model with log-link \citep{lambert2010two} estimated via the \textit{sppois} package \citep{psar}.
    \item \textbf{Probit-SAR:} For the Binomial case, a SAR model with probit link estimated using the \textit{spatialprobit} package \citep{SARprobit} and response indicator $y_i^* = 1(y_i \geq 0.5)$ \citep{lacombe2018use}. To compare the estimates between the proposed model and the Probit-SAR model, the approximation of $\beta_{i-logit} \approx 1.6\beta_{i-probit}$ explained in \cite[pag 35]{hosmer2013applied} will be used.
    \item \textbf{GAM:} A GAM is fitted using the \textit{mgcv} package \citep{wood2017generalized}, incorporating a smooth spatial component through a bidimensional spline $s(i,j)$ to capture spatial dependence. The model is specified with the appropriate distribution and link function (log for Poisson and Gamma, logit for Binomial), allowing for flexible nonparametric modeling of spatial effects without imposing a parametric spatial structure.
\item \textbf{INLA:} A Bayesian spatial model is estimated using the INLA framework \citep{rue2009approximate} implemented in the \textit{INLA} package \citep{lindgren2015bayesian}. Spatial dependence is modeled through a latent Gaussian field using a Besag (intrinsic CAR) structure defined over a neighborhood graph. The model includes fixed effects for covariates and a spatial random effect $f(i)$, with appropriate likelihoods (Binomial, Poisson, Gamma) and corresponding link functions (logit or log), enabling efficient approximate Bayesian inference for spatially structured data.
\end{enumerate}
These comparative models allow assessing the performance of the proposed methodology against both classical and spatial alternatives under each distributional setting. In each scenario of the simulation, the following metrics were calculated: 
\begin{align*}
    RMSE(\hat{\theta})&=\sqrt{\frac{1}{500}\sum_{s=1}^{500} {(\hat{\theta}_s-\theta_s)^2}}\\
    \mbox{Coverage}(\%) &=\frac{100\%}{500}\sum_{s=1}^{500} \mathbf{1} \left(\theta\in\left\{\hat{\theta}_s
    \pm 1.96\sqrt{\widehat{Var}(\hat{\theta}_s)}\right\}\right)\\
   R^2
&=
\frac{1}{500}
\sum_{s=1}^{500}\left(1-\frac{(n^{(s)}-1)\sum_{i=1
}^{n^{(s)}}(Y_{i}^{(s)}-\hat{Y}_{i}^{(s)})^2}{(n^{(s)}-p^{(s)})\sum_{i=1
}^{n^{(s)}}(Y_{i}^{(s)}-\bar{Y}^{(s)})^2}\right)\\
    RMSE(Y) & =\frac{1}{500}
\sum_{s=1}^{500}\sqrt{\frac{1}{n^{(s)}-p^{(s)}}\sum_{i=1}^{n^{(s)}} (Y_i^{(s)}-\hat{Y_i}^{(s)})^2}
\end{align*}
{where $p^{(s)}$ and $n^{(s)}$ are the number of estimated parameters and the number of spatial units in each simulation scenario, respectively}. We report the response-based pseudo-$R^2$ because it provides a unified {in-sample} goodness-of-fit measure that can be consistently compared across the different probabilistic models considered, regardless of the underlying response distribution.

Confidence intervals were computed using the theoretical variance formula proposed in Equation~\eqref{eq:varrhosand}.
\begin{figure}[!ht]
\centering
\includegraphics[width=14cm]{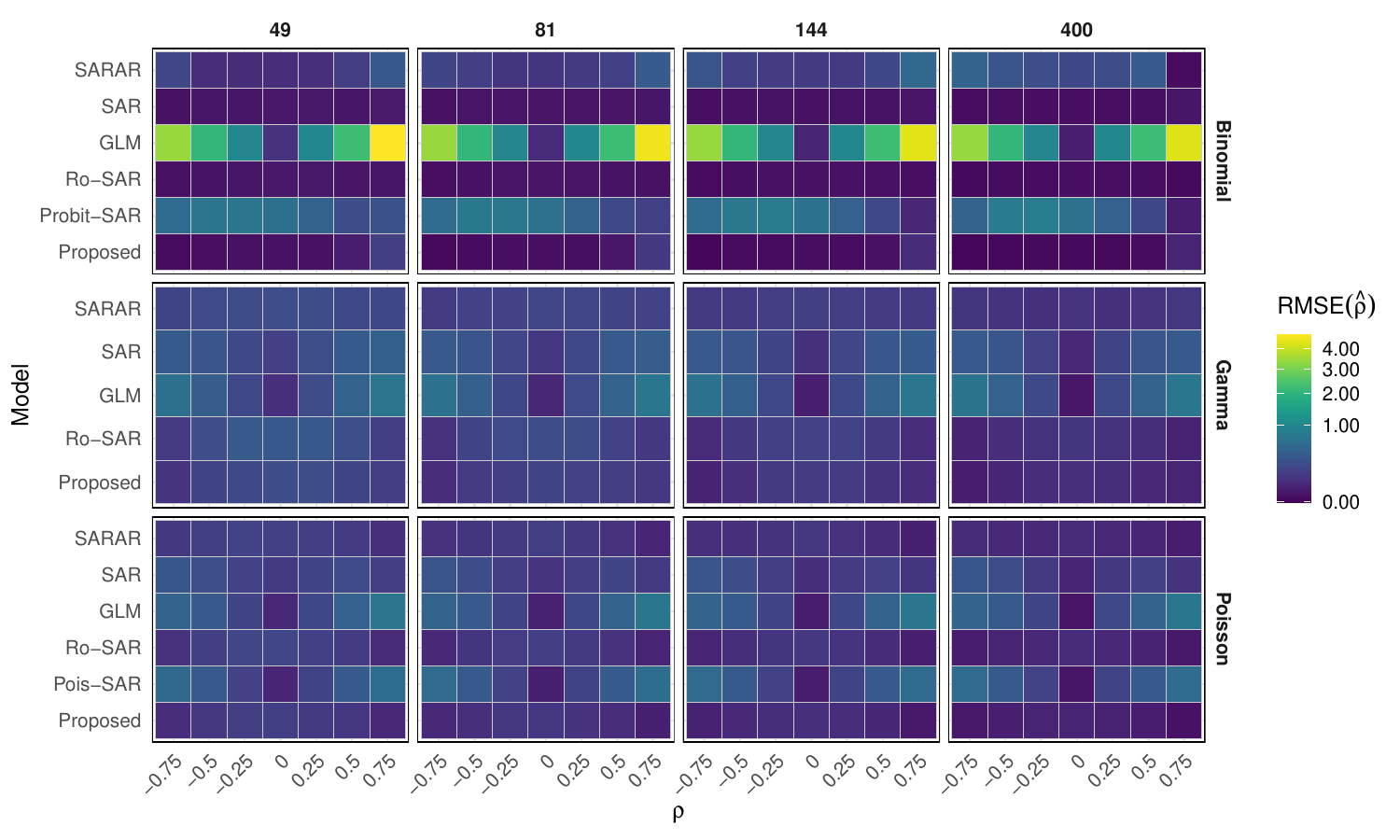}
\caption{Heatmap of RMSE (log scale) for the estimated spatial parameter $\hat{\rho}$ at different true $\rho$ values (x-axis), by model (y-axis), distribution (rows), and sample size (columns). }
\label{rmserho}
\end{figure}
Figure~\ref{rmserho} displays the Root Mean Squared Error (RMSE) for the estimator of the spatial autoregressive parameter $\rho$ across all simulation scenarios. Under the Gamma distribution, only the proposed model exhibits optimal performance. The Ro-SAR models exhibit similar performance to the proposed model. This can be attributed to the logarithmic transformations typically applied to Gamma-distributed data, which stabilize the variance and mean of the data. The SARAR model performs well, but worse than the previous two. Conversely, the SAR and GLM models produce poor results except for $\rho=0$, where they behave similarly to the proposed model.

\begin{figure}[!ht]
\centering
\includegraphics[width=14cm]{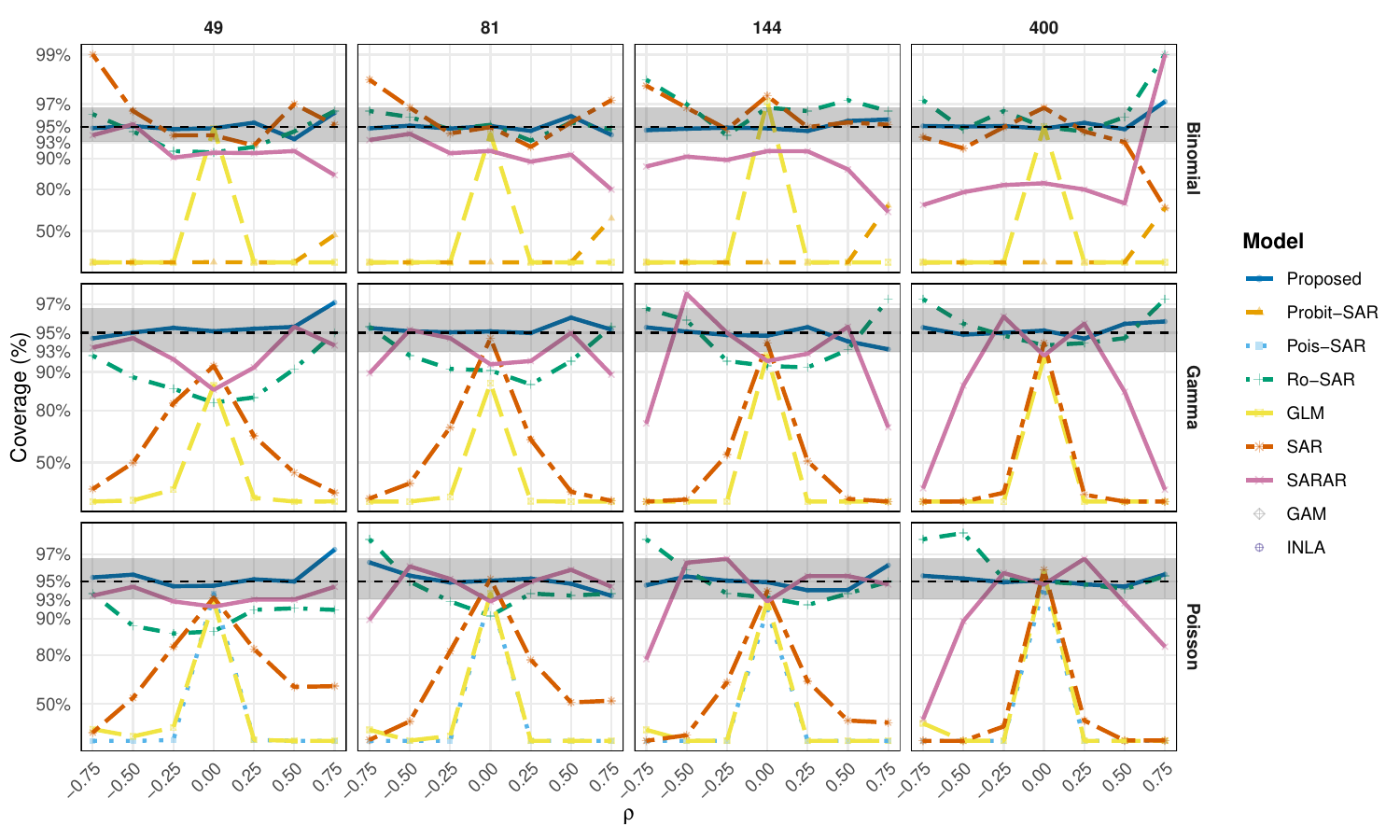}
\caption{Coverage rates of 95\% confidence intervals for the estimated spatial parameter $\hat{\rho}$ (y-axis), across true $\rho$ values (x-axis), by model, distribution (rows), and sample size (columns). The dashed horizontal line indicates the nominal 95\% level. Shaded bands represent the approximate confidence interval for a binomial proportion with $0.95$ and $500$ replications. Additionally, the y-axis is transformed using a log-scale of the complement, $-\log(1 - y)$, to enhance the visualization of deviations from the nominal level. Models based on GAM and INLA do not estimate a spatial autoregressive parameter $\rho$, and therefore, they are not included in this evaluation.}
\label{coverrho}
\end{figure}

Under the binomial distribution, the Probit-SAR and GLM models approaches generally exhibit larger RMSE values than the proposed method under the simulation settings considered.
The SAR model also shows inferior performance. Notably, the proposed model only exhibits elevated RMSE when $\rho = 0.75$. This behavior is explained by the fact that the Pearson correlation between binomial random variables is bounded by the odds ratio, as discussed by \cite{lipsitz1991generalized}. Additionally, due to the accurate normal approximation of the binomial distribution when $M_i$ is large, the Ro-SAR and SAR models perform well across all values of $\rho$. In the Poisson case, the implementation of the Poisson-SAR model considered in this study generally produced larger RMSE values than several competing approaches. The proposed model achieves very good performance, comparable to the SARAR and Ro-SAR model. These RMSE results indicate that the estimates obtained by the proposed model are consistent, as the RMSE decreases with increasing sample size. This behavior is theoretically supported by Theorem~\ref{te1}. The GAM and INLA models are omitted from this comparison since they do not estimate an explicit spatial autoregressive parameter $\rho$.

As shown in Figure~\ref{coverrho}, the proposed model consistently achieves coverage rates close to the nominal 95\% level across all distributions and scenarios. Slight deviations are observed when $\rho$ approaches the boundaries $-1$ or $1$, which can be attributed to high spatial correlation, potentially affecting variance stability.

Several competing methods exhibit departures from the nominal coverage level in at least some simulation settings. Only the Ro-SAR and SAR models demonstrate performance comparable to the proposed method under the binomial distribution, except when $\rho=0.75$, when exibit poor performance. Under the Gamma distribution, the Ro-SAR model approximates the proposed model’s coverage only when $n = 144$ or $n = 400$. In the Poisson case, Ro-SAR displays competitive behavior when $n > 49$ and $\rho \neq -0.75$. This is likely because the log-linear transformation used in the Ro-SAR model compensates for heteroscedasticity in Poisson data.
Poisson SAR model is competitive when $\rho$ is near to zero and $n>49$
\begin{figure}[!ht]
\centering
\includegraphics[width=14cm]{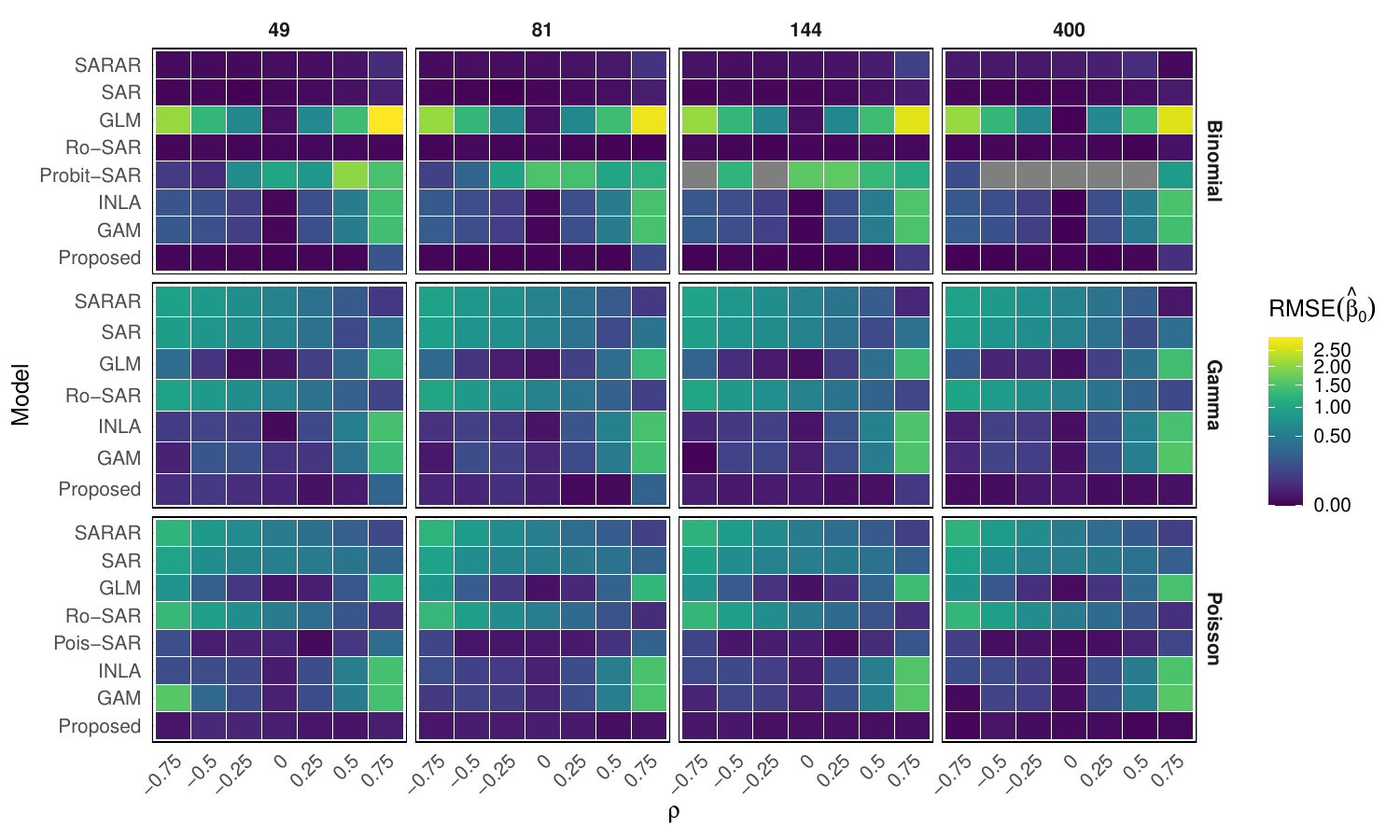}
\caption{Heatmap of RMSE (log scale) for the estimated parameter $\hat{\beta}_0$ at different true $\rho$ values (x-axis), by model (y-axis), distribution (rows), and sample size (columns). }
\label{rmsebeta0}
\end{figure}

The numerical results presented in Figures~\ref{rmserho}, \ref{rmsebeta0}, \ref{rmsebeta1}, and \ref{rmsebeta2} are reported in detail in Tables~\ref{RMSERhoAA}, \ref{RMSEBetaAA}, \ref{RMSEBeta1AA}, and \ref{RMSEBeta2AA}, respectively (\ref{apen00}). Consistent with the visual patterns observed in the heatmaps and coverage curves, the proposed model consistently exhibits the lowest RMSE values in most configurations, specially as the sample size increases. Differences between models become more pronounced for high values of $|\rho|$, where several methods show clear losses in precision or stability. Distribution-specific models {(Probit-SAR and Pois-SAR)} improve performance in their natural contexts but do not achieve the overall robustness of the proposed method.

\begin{figure}[!ht]
\centering
\includegraphics[width=14cm]{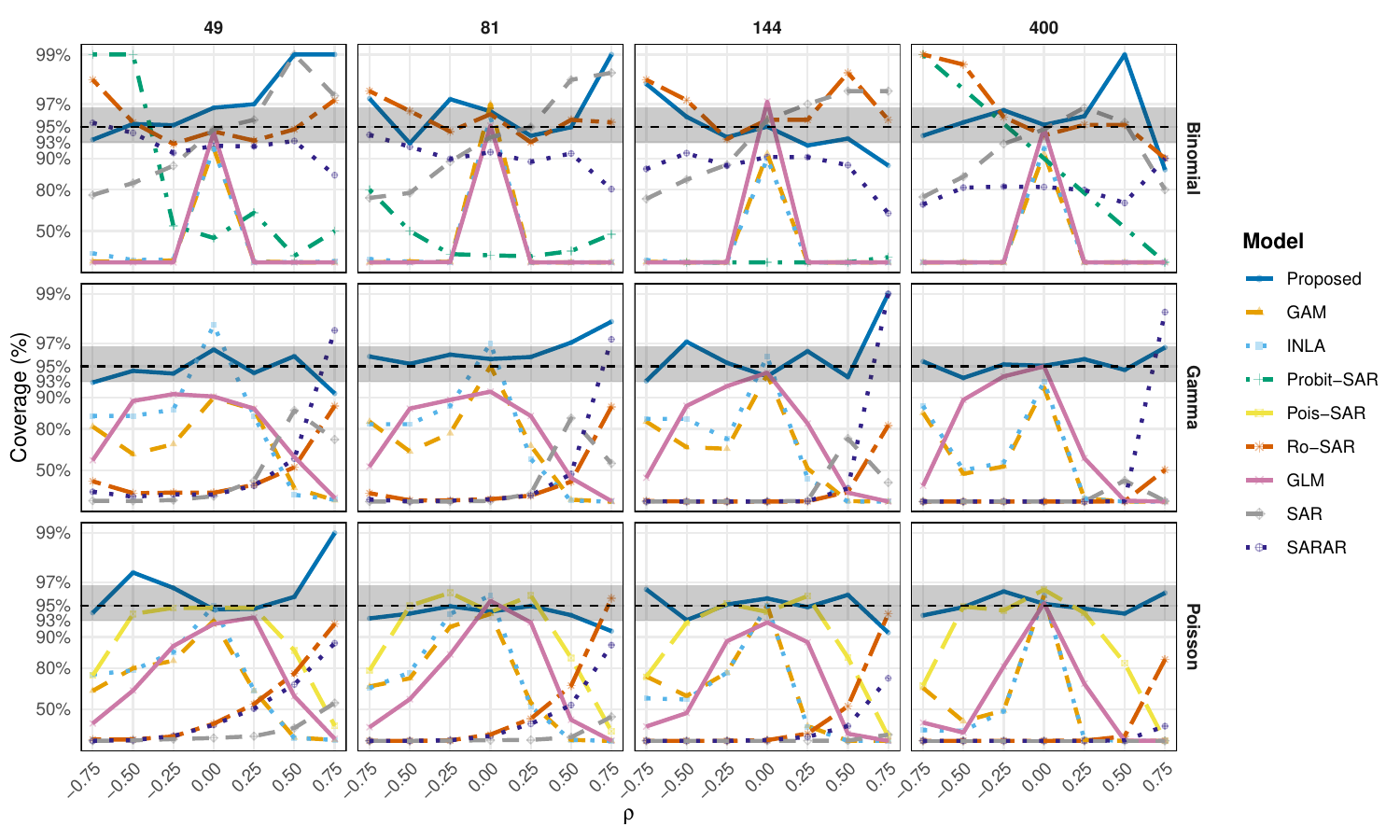}
\caption{Coverage rates of 95\% confidence intervals for the estimated parameter $\hat{\beta}_0$ (y-axis), across true $\rho$ values (x-axis), by model, distribution (rows), and sample size (columns). The dashed horizontal line indicates the nominal 95\% level. Shaded bands represent the approximate confidence interval for a binomial proportion with $0.95$ and $500$ replications. Additionally, the y-axis is transformed using a log-scale of the complement, $-\log(1 - y)$, to enhance the visualization of deviations from the nominal level.}
\label{coverbeta0}
\end{figure}

The Probit-SAR and Pois-SAR implementations frequently exhibited convergence difficulties, which affected the corresponding coverage estimates and complicate a direct comparison with the remaining methods. The GLM and SAR models achieve near-nominal coverage when there is no spatial correlation, which is consistent with Corollary~\ref{cor1} and Lemma~\ref{lem1}.

Figure~\ref{rmsebeta0} shows the RMSE for $\hat{\beta}_0$, with behavior similar to that seen in Figure~\ref{rmserho}. The Probit-SAR model stands out with gray cells indicating convergence failures in all 500 simulations for some scenarios, resulting in \texttt{NaN} values for either the estimator, its standard error, or both. The Pois-SAR model produced larger RMSE values than the proposed approach in most scenarios, although its performance remained competitive with some of the alternative specifications. This indicates that the Pois-SAR model is an intermediate between the GLM and the proposed model. Similar patterns are observed for $\beta_1$ and $\beta_2$ in Figures~\ref{rmsebeta1} and~\ref{rmsebeta2}, which are shown in~\ref{apen0}.

The GAM and INLA models also provide accurate estimates of the intercept, with RMSE decreasing as the sample size increases. Their performance is generally comparable to that of the proposed model, although both methods deteriorate as $\rho$ approaches 1, where strong spatial dependence becomes more difficult to capture without an explicit autoregressive parameter.

\begin{figure}[!ht]
\centering
\includegraphics[width=14cm]{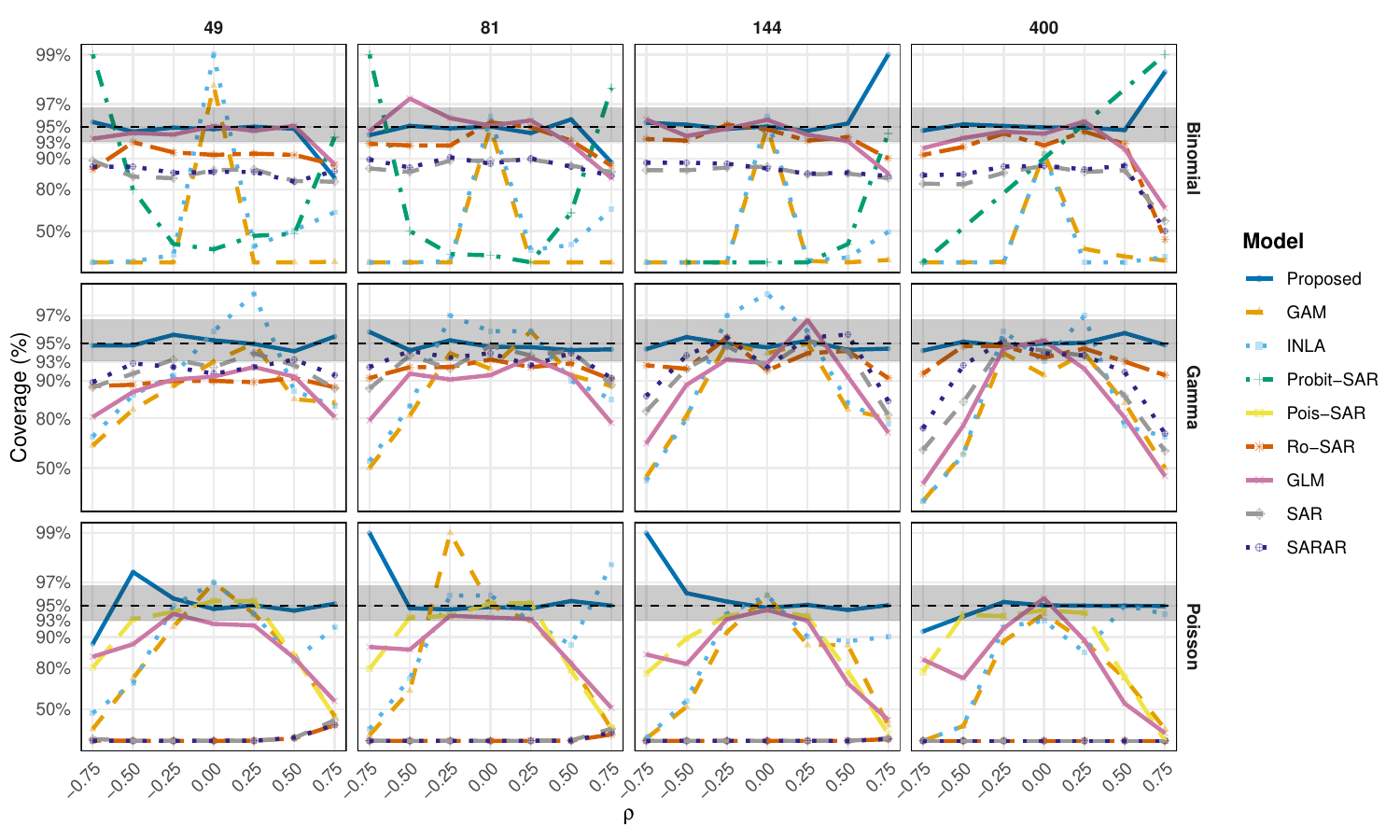}
\caption{Coverage rates of 95\% confidence intervals for the estimated parameter $\hat{\beta}_2$ (y-axis), across true $\rho$ values (x-axis), by model, distribution (rows), and sample size (columns). The dashed horizontal line indicates the nominal 95\% level. Shaded bands represent the approximate confidence interval for a binomial proportion with $0.95$ and $500$ replications. Additionally, the y-axis is transformed using a log-scale of the complement, $-\log(1 - y)$, to enhance the visualization of deviations from the nominal level.}
\label{coverbeta2}
\end{figure}

As shown in Figure~\ref{coverbeta0}, the proposed model maintains coverage close to the nominal 95\% for $\beta_0$, with reduced variability as $n$ increases. The Ro-SAR model approximates this performance under the binomial distribution, while other models perform poorly, except for the GLM when $n$ is large and $\rho$ is close to zero. This supports findings from the literature showing that intercept variance is often misestimated under normal approximations. The Pois-SAR model achieves good coverage when the autoregressive parameter is near zero. For $\rho$ values of -0.75, 0.5, or 0.75, coverage decreases. This is explained by Corollary \ref{cor3}.

Figure~\ref{coverbeta2} shows similar results for $\beta_2$, with the proposed model consistently performing well across all distributions and scenarios. The GLM shows slight improvements over previous cases, likely due to the inclusion of the spatial lag term $\mathbf{W}\mathbf{y}$, which helps stabilize the estimation of non-intercept parameters. However, this effect is limited to the binomial distribution, under Gamma and Poisson, performance deteriorates.

The GAM and INLA models achieve competitive RMSE values, particularly for moderate and large sample sizes, reflecting their ability to adequately account for spatial dependence through smooth or latent spatial effects. However, unlike the proposed model, these approaches do not estimate an explicit spatial autoregressive parameter, making them suitable benchmarks for prediction and fixed-effect estimation rather than for quantifying the strength of spatial dependence.

Robust models show slight undercoverage, generally around 90\%, except in Poisson settings, where it worsens. Similar results are found for $\beta_1$ in Figure~\ref{coverbeta1} (~\ref{apen0}). These results confirm that the proposed model is robust across all values of $\rho$, providing reliable inference. Theoretically, Corollaries~\ref{cor3} and~\ref{cor4} support this, indicating that the model not only produces accurate estimates but also well-calibrated variances.

Overall, the simulation study highlights the numerical stability and favorable inferential performance of the proposed \texttt{spatemR::GSCIMC} algorithm across the scenarios considered. Across all scenarios (regardless of response distribution, sample size, or spatial dependence level), the proposed method consistently achieved 100\% convergence without numerical warnings or estimation failures. The software implementations considered for the Probit-SAR and Poisson-SAR models occasionally exhibited convergence difficulties, particularly in scenarios involving strong spatial dependence or complex likelihood surfaces. Notably, even in simulations where these alternative routines did converge, their estimation accuracy was often comparable to or worse than that of the proposed method in terms of bias and efficiency. Because the competing methods were evaluated through available software implementations, part of the observed differences may reflect numerical and computational aspects of those implementations in addition to the underlying modeling assumptions. By providing robust, stable, and accurate estimation for a broader class of spatial non-Gaussian models, the proposed approach offers a practical and reliable alternative for real-world spatial data analysis.

In contrast, the standard SAR and SARAR models, as well as their robust variants, converged consistently in all scenarios. This outcome is expected, as their estimation relies on least squares and the normal likelihood, which are generally well-behaved and numerically stable. However, their applicability is limited to settings where the normality assumption holds. In addition, GAM and INLA are based on highly developed and extensively tested modeling frameworks with a long methodological and applied track record, while allowing for flexible distributional assumptions.

\begin{table}[ht]
\footnotesize
\centering
\begin{adjustbox}{width=1.2\textwidth,center}
\begin{tabular}{lllllllll}
    \hline
Dist & $\rho$ & Proposed & GAM & INLA & Ro-SAR & GLM & SAR & SARAR \\ 
   \hline
\multirow{4}{*}{Binomial} & \multirow{1}{*}{0} & \textbf{.012 (.008)} & .013 (.009) & .017 (.011) & .278 (.014) & .013 (.008) & .279 (.018) & .279 (.017) \\ 
   & \multirow{1}{*}{.25} & .012 (.006) & .025 (.008) & \textbf{.011 (.009)} & .282 (.018) & .013 (.006) & .285 (.025) & .285 (.020) \\ 
   & \multirow{1}{*}{.5} & .012 (.007) & .049 (.021) & \textbf{.009 (.010)} & .298 (.032) & .018 (.006) & .316 (.035) & .313 (.026) \\ 
   & \multirow{1}{*}{.75} & .010 (.005) & .090 (.058) & \textbf{.006 (.009)} & .340 (.058) & .028 (.006) & .392 (.039) & .378 (.031) \\ 
\multirow{4}{*}{Gamma} & \multirow{1}{*}{0} & \textbf{5.143 (2.782)} & 5.497 (2.910) & 7.264 (4.167) & 32.293 (31.163) & 5.290 (2.933) & 5.612 (3.154) & 5.603 (3.158) \\ 
   & \multirow{1}{*}{.25} & \textbf{5.608 (3.213)} & 6.079 (3.616) & 7.977 (4.812) & 35.065 (31.654) & 5.806 (3.407) & 6.158 (3.691) & 6.118 (3.671) \\ 
   & \multirow{1}{*}{.5} & \textbf{8.271 (5.732)} & 9.140 (6.296) & 12.093 (8.423) & {60.219} (94.831) & 9.045 (6.502) & 9.817 (7.076) & 9.580 (6.817) \\ 
   & \multirow{1}{*}{.75} & \textbf{39.249 (61.373)} & 44.876 (89.502) & 47.875 (64.054) & 276.054 (543.591) & 185.237 ($1.40\times10^{3}$) & 53.105 (75.908) & 51.853 (74.585) \\ 
\multirow{4}{*}{Poisson} & \multirow{1}{*}{0} & \textbf{1.689 (.225)} & 1.734 (.229) & 2.354 (.314) & $2.68\times10^{4}$ ($9.62\times10^{4}$) & 1.733 (.230) & $1.53\times10^{3}$ ($1.10\times10^{4}$) & $1.50\times10^{3}$ ($1.05\times10^{4}$) \\ 
   & \multirow{1}{*}{.25} & \textbf{1.741 (.285)} & 1.839 (.307) & 2.607 (.452) & $4.78\times10^{4}$ ($6.49\times10^{5}$) & 1.877 (.354) & $2.02\times10^{3}$ ($1.50\times10^{4}$) & 191.866 ($1.28\times10^{4}$) \\ 
   & \multirow{1}{*}{.5} & \textbf{1.970 (.464)} & 2.168 (.642) & 2.636 (.660) & $3.27\times10^{4}$ ($1.25\times10^{5}$) & 2.698 (.979) & $1.32\times10^{3}$ ($9.95\times10^{3}$) & $1.25\times10^{3}$ ($8.14\times10^{3}$) \\ 
   & \multirow{1}{*}{.75} & 3.283 (1.689) & 5.983 (6.495) & \textbf{1.900 (.425)} & $8.02\times10^{4}$ ($3.74\times10^{4}$) & 14.647 (23.491) & 301.721 ($1.49\times10^{3}$) & 381.790 ($1.65\times10^{3}$) \\
   \hline
\end{tabular}
\end{adjustbox}
\caption{$RMSE$ across simulation scenarios, reported as mean RMSE with its standard deviation in parentheses. Results are averaged over all sample sizes for each combination of response distribution (Dist) and absolute spatial dependence level ($|\rho|$).  The distribution-specific spatial models (Probit-SAR for Binomial data and Poisson-SAR for Poisson data) were excluded from this summary because frequent convergence failures substantially reduced the number of valid replications in several simulation settings.}
\label{RMSEYYs}
\end{table}
Table \ref{RMSEYYs} summarizes the predictive performance of the competing methods. The proposed approach consistently achieved the best or nearly the best predictive performance across the considered scenarios. Although GAM and INLA occasionally attained similar RMSE values, these methods rely on substantially more flexible model structures and a larger number of parameters. Despite its comparatively parsimonious formulation, the proposed approach achieved predictive accuracy comparable to or better than these alternatives. In contrast, SAR, SARAR, and Ro-SAR generally produced substantially larger prediction errors, particularly as the strength of spatial dependence increased. For the Binomial distribution, several methods performed similarly. This result is expected because, as the cluster size $M_i$ increases, the Binomial distribution approaches a Gaussian distribution through the Central Limit Theorem, making the normality assumptions underlying these models less restrictive.

\begin{table}[ht]
\footnotesize
\centering
\begin{adjustbox}{width=1.2\textwidth,center}
\begin{tabular}{lllllllll}
  \hline
Dist & $\rho$ & Proposed & GAM & INLA & Ro-SAR & GLM & SAR & SARAR \\ 
 \hline
\multirow{4}{*}{Binomial} & \multirow{1}{*}{0} & \textbf{.996 (.016)} & .995 (.020) & .991 (.033) & .037 (.273) & \textbf{.996 (.015)} & .052 (.301) & .032 (.299) \\ 
   & \multirow{1}{*}{.25} & \textbf{.996 (.007)} & .986 (.010) & \textbf{.996 (.013)} & .075 (.383) & \textbf{.996 (.007)} & .023 (.350) & .024 (.333) \\ 
   & \multirow{1}{*}{.5} & \textbf{.996 (.010)} & .948 (.035) & \textbf{.996 (.017)} & .050 (.480) & .993 (.009) & .028 (.456) & .038 (.446) \\ 
   & \multirow{1}{*}{.75} & .997 (.005) & .844 (.125) & \textbf{.998 (.010)} & .090 (.426) & .984 (.012) & .065 (.324) & .029 (.459) \\ 
  \multirow{4}{*}{Gamma} & \multirow{1}{*}{0} & \textbf{.333 (.313)} & .220 (.432) & .083 (.352) & .018 (.402) & .331 (.335) & .264 (.164) & .267 (.164) \\ 
   & \multirow{1}{*}{.25} & \textbf{.264 (.381)} & .121 (.438) & .054 (.445) & .046 (.430) & .254 (.376) & .244 (.214) & .252 (.214) \\ 
   & \multirow{1}{*}{.5} & \textbf{.297 (.435)} & .166 (.486) & .080 (.456) & .055 (.420) & .244 (.423) & .174 (.181) & .204 (.181) \\ 
   & \multirow{1}{*}{.75} & \textbf{.162 (.309)} & .028 (.453) & .088 (.358) & .023 (.366) & .032 (.393) & .034 (.145) & .064 (.144) \\ 
  \multirow{4}{*}{Poisson} & \multirow{1}{*}{0} & \textbf{.852 (.074)} & .842 (.076) & .708 (.150) & .088 (.465) & \textbf{.852 (.073)} & .079 (.414) & .064 (.417) \\ 
   & \multirow{1}{*}{.25} & \textbf{.861 (.079)} & .845 (.085) & .689 (.173) & .061 (.330) & .851 (.079) & .047 (.375) & .099 (.416) \\ 
   & \multirow{1}{*}{.5} & \textbf{.909 (.065)} & .888 (.082) & .795 (.199) & .013 (.424) & .861 (.076) & .046 (.413) & .092 (.462) \\ 
   & \multirow{1}{*}{.75} & \textbf{.969 (.038)} & .937 (.070) & .966 (.098) & .046 (.327) & .814 (.113) & .055 (.470) & .081 (.341) \\ 
   \hline
\end{tabular}
\end{adjustbox}
\caption{Average $R^2$ across simulation scenarios, reported as the mean $R^2$ with its standard deviation in parentheses. Results are averaged over all sample sizes for each combination of response distribution (Dist) and absolute spatial dependence level ($|\rho|$). The distribution-specific spatial models (Probit-SAR for Binomial data and Poisson-SAR for Poisson data) were excluded from this summary because frequent convergence failures substantially reduced the number of valid replications in several simulation settings.}
\label{R2YYs}
\end{table}

Table \ref{R2YYs} summarizes the explanatory performance of the competing methods. For the Binomial distribution, the proposed model, GAM, INLA, and the classical GLM all achieved nearly perfect fits, with only negligible differences among them. 

For the Gamma distribution, the overall $R^2$ values were substantially lower for all methods. This is a direct consequence of allowing the dispersion parameter to vary across observations, with values sampled from a Uniform$(1,5)$ distribution. The resulting increase in residual variability reduces the proportion of variance that can be explained by the conditional mean, thereby lowering the attainable $R^2$ for every competing method. Nevertheless, the proposed approach consistently achieved the highest $R^2$ values across all levels of spatial dependence, with GAM and INLA providing similar, although slightly lower, explanatory performance.

For the Poisson distribution, the proposed approach again attained the highest $R^2$ values in nearly all scenarios, with GAM and INLA producing comparable results. In contrast, the Gaussian SAR-based models generally exhibited poor explanatory performance, particularly under moderate and strong spatial dependence. Overall, these results indicate that the proposed method achieves predictive and explanatory performance comparable to, and often better than, more flexible alternatives such as GAM and INLA, while simultaneously providing direct inference on the spatial dependence parameter through a considerably more parsimonious model.

Overall, the proposed method demonstrated stable convergence and reliable estimation across all simulation settings. The \texttt{GSCIMC} implementation consistently produced low RMSE values, high explanatory power, and coverage rates close to their nominal levels, even under strong spatial dependence and non-Gaussian outcomes. GAM and INLA achieved similar predictive performance in many scenarios, reflecting their greater modeling flexibility. Nevertheless, the proposed approach, despite requiring a considerably more parsimonious specification, matched or exceeded their predictive accuracy while maintaining reliable estimation of both regression and spatial dependence parameters. These findings support the use of log quasi-likelihood and estimating equations as a practical framework for inference in spatially correlated non-Gaussian data.

\subsection{Simulation under model misspecification}
To further assess the robustness of the proposed methodology, we consider an scenario in which the data-generating process does not match the assumed model specification. While the previous experiments focus on correctly specified models—where spatial dependence is explicitly incorporated in the mean structure—we now analyze alternative setting in which spatial dependence arises from different mechanisms.  In this case, spatial autocorrelation is present in the residual component, violating the structural assumption of the proposed model.
Second, a \textit{spatial covariance model} is considered, where dependence is induced through a latent Gaussian process:
\begin{equation}\label{eq:miss}
  \mathbf{u} \sim \mathcal{N}(0, \pmb{\Sigma}), \quad \Sigma_{ij} = \exp\left(-\rho\, d_{ij}\right),
\quad \pmb{\eta} = \mathbf{X}\pmb{\beta} + \mathbf{u}, \; \rho>0,  
\end{equation}
with $d_{ij}$ denoting normalized distances. This specification generates smooth spatial correlation driven by distance rather than a discrete autoregressive structure as \cite{wang2025using}
For both settings, responses are generated under the same distributional assumptions (Poisson, Gamma, and Binomial) and link functions described previously. All models listed in Table~\ref{tab:model_summary} are fitted to these datasets, enabling a direct comparison of their robustness under misspecification.
\begin{figure}[!ht]
\centering
\includegraphics[width=14cm]{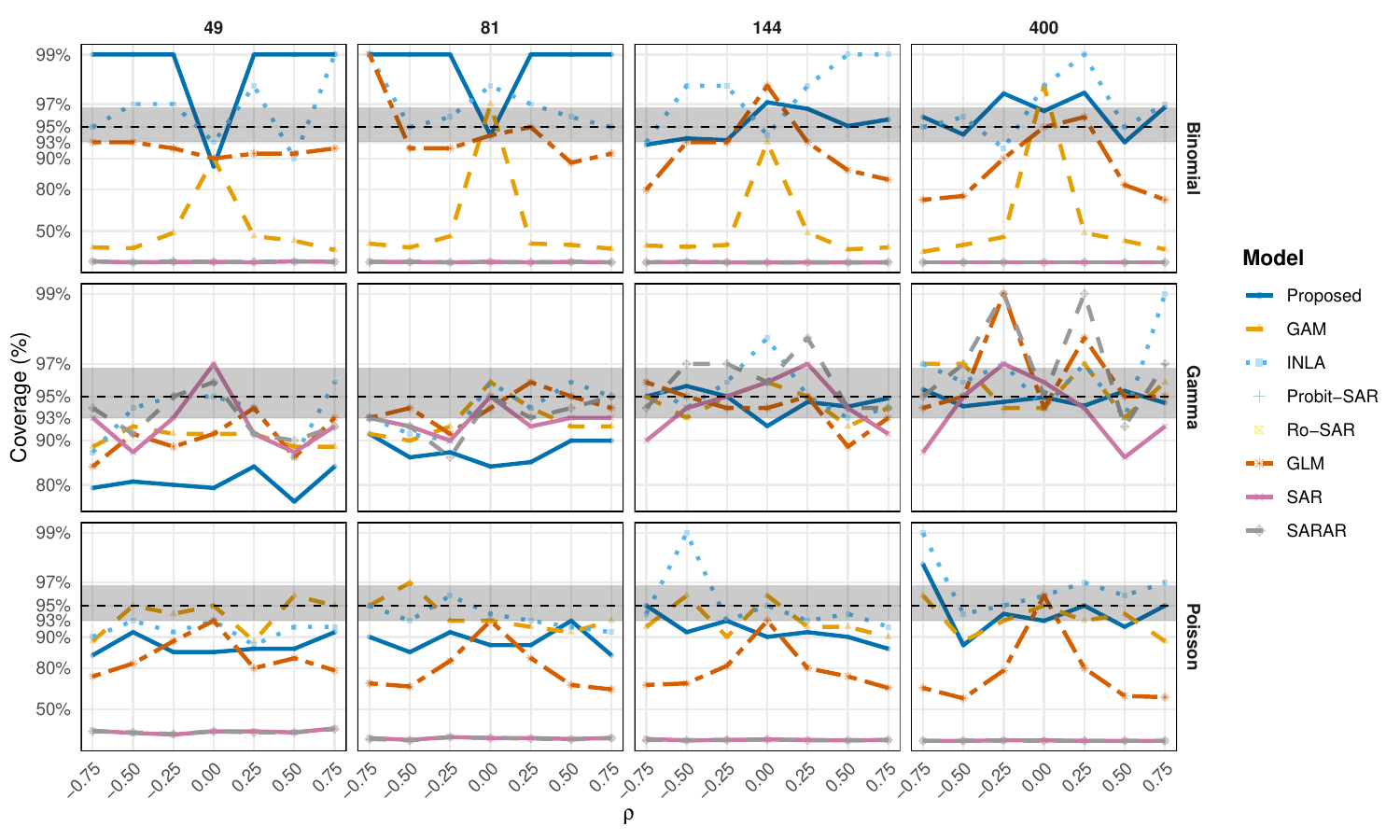}
\caption{Coverage rates of 95\% confidence intervals for the estimated parameter $\hat{\beta}_2$ (y-axis) under model misspecification (Equation \eqref{eq:miss}), across true $\rho$ values (x-axis), by model, distribution (rows), and sample size (columns). The dashed horizontal line indicates the nominal 95\% level. Shaded bands represent the approximate confidence interval for a binomial proportion with $0.95$ and $500$ replications. Additionally, the y-axis is transformed using a log-scale of the complement, $-\log(1 - y)$, to enhance the visualization of deviations from the nominal level.}
\label{coverbeta2Miss}
\end{figure}

\begin{table}[ht]
\footnotesize
\centering
\begin{adjustbox}{width=1.2\textwidth,center}
\begin{tabular}{lllllllll}
\hline
 Dist & $\rho$ & Proposed & GAM & INLA & Ro-SAR & GLM & SAR & SARAR \\ 
\hline
\multirow{4}{*}{Binomial} & \multirow{1}{*}{0} & \textbf{.011 (.007)} & .012 (.007) & .016 (.009) & .273 (.088) & .012 (.007) & .270 (.097) & .270 (.097) \\ 
   & \multirow{1}{*}{.25} & .057 (.017) & .023 (.007) & \textbf{.011 (.009)} & .278 (.081) & .054 (.014) & .285 (.085) & .266 (.092) \\ 
   & \multirow{1}{*}{.5} & .078 (.021) & .031 (.008) & \textbf{.010 (.010)} & .272 (.072) & .070 (.015) & .288 (.075) & .258 (.082) \\ 
   & \multirow{1}{*}{.75} & .092 (.025) & .036 (.008) & \textbf{.009 (.008)} & .274 (.077) & .079 (.016) & .299 (.078) & .261 (.086) \\ 
\multirow{4}{*}{Gamma} & \multirow{1}{*}{0} & \textbf{8.764 (13.531)} & 9.391 (14.822) & 12.356 (19.413) & 59.617 (118.832) & 9.033 (13.987) & 9.617 (15.206) & 9.536 (15.368) \\ 
   & \multirow{1}{*}{.25} & \textbf{9.337 (14.558)} & 9.875 (15.884) & 12.635 (19.789) & 74.228 (156.540) & 9.550 (14.532) & 10.456 (2.553) & 10.369 (19.812) \\ 
   & \multirow{1}{*}{.5} & \textbf{8.617 (11.896)} & 8.798 (12.624) & 11.164 (15.587) & 73.549 (128.623) & 8.965 (12.500) & 9.336 (13.770) & 9.301 (13.756) \\ 
   & \multirow{1}{*}{.75} & 9.662 (11.930) & \textbf{9.641 (11.832)} & 11.852 (14.523) & 96.424 (237.852) & 10.511 (17.094) & 10.560 (13.483) & 10.570 (13.830) \\ 
\multirow{4}{*}{Poisson} & \multirow{1}{*}{0} & \textbf{1.891 (1.195)} & 1.934 (1.161) & 2.657 (1.664) & $8.78\times10^{13}$ ($1.76\times10^{15}$) & 1.952 (1.241) & 598.731 ($4.67\times10^{3}$) & 827.981 ($9.53\times10^{3}$) \\ 
   & \multirow{1}{*}{.25} & 3.108 (3.619) & \textbf{2.019 (1.261)} & 2.403 (.994) & $1.10\times10^{5}$ ($7.71\times10^{5}$) & 3.186 (4.056) & 82.231 ($7.10\times10^{3}$) & 899.685 ($7.16\times10^{3}$) \\ 
   & \multirow{1}{*}{.5} & 3.423 (3.535) & \textbf{1.976 (1.243)} & 2.143 (.699) & $2.19\times10^{8}$ ($6.18\times10^{9}$) & 3.421 (3.436) & 484.247 ($2.19\times10^{3}$) & 562.674 ($2.50\times10^{3}$) \\ 
   & \multirow{1}{*}{.75} & 4.177 (3.985) & 2.118 (1.325) & \textbf{2.045 (.677)} & $1.32\times10^{5}$ ($1.98\times10^{6}$) & 4.167 (3.953) & 432.064 ($2.05\times10^{3}$) & 708.560 ($6.23\times10^{3}$) \\ 
\hline
\end{tabular}
\end{adjustbox}
\caption{$RMSE$ across simulation scenarios under model misspecification, reported as mean RMSE with its standard deviation in parentheses. Results are averaged over all sample sizes for each combination of response distribution (Dist) and absolute spatial dependence level ($|\rho|$).  The distribution-specific spatial models (Probit-SAR for Binomial data and Poisson-SAR for Poisson data) were excluded from this summary because frequent convergence failures substantially reduced the number of valid replications in several simulation settings.}
\label{RMSEYYsMis}
\end{table}
Figure \ref{coverbeta2Miss} displays the empirical coverage rates for $\beta_2$ under model misspecification. As expected, distributional misspecification deteriorates inferential performance for all methods, with the largest departures from the nominal level occurring in the Gamma and Poisson settings and for the smallest sample sizes. Nevertheless, the coverage of the proposed estimator improves steadily as the sample size increases, approaching the nominal level in most scenarios. This pattern is consistent with the asymptotic properties of the estimating-equations framework and indicates that reliable inference is progressively recovered as more information becomes available.
\newline
In contrast, several Gaussian SAR-type models exhibit substantial undercoverage together with markedly larger prediction errors across a wide range of scenarios, particularly under moderate and strong spatial dependence. Overall, these findings indicate that, although model misspecification inevitably degrades both prediction and inference, the proposed estimator remains inferentially robust, retains competitive predictive performance, and benefits substantially from increasing sample size.
\newline
Because the misspecification scenarios involve different data-generating mechanisms, the spatial dependence parameter $\rho$ is not directly comparable across methods. In particular, the competing models are fitted under different covariance structures and, in several cases, these do not target the same notion of spatial dependence. Therefore, the comparison is restricted to the regression parameters and predictive performance. Complete results for all estimated parameters, including $\rho$, are reported in~\ref{apen0}.
\begin{table}[ht]
\footnotesize
\centering
\begin{adjustbox}{width=1.2\textwidth,center}
\begin{tabular}{lllllllll}
 \hline
Dist & $\rho$ & Proposed & GAM & INLA & Ro-SAR & GLM & SAR & SARAR \\ 
  \hline
\multirow{4}{*}{Binomial} & \multirow{1}{*}{0} & \textbf{.995 (.012)} & .994 (.014) & .991 (.024) & .059 (.377) & \textbf{.995 (.012)} & .039 (.307) & .099 (.366) \\ 
   & \multirow{1}{*}{.25} & .917 (.044) & .987 (.009) & \textbf{.995 (.014)} & .010 (.372) & .927 (.034) & .057 (.425) & .030 (.467) \\ 
   & \multirow{1}{*}{.5} & .859 (.069) & .977 (.015) & \textbf{.995 (.019)} & .050 (.434) & .888 (.044) & .032 (.434) & .038 (.355) \\ 
   & \multirow{1}{*}{.75} & .809 (.085) & .971 (.011) & \textbf{.997 (.007)} & .014 (.311) & .859 (.051) & .091 (.320) & .012 (.463) \\ 
  \multirow{4}{*}{Gamma} & \multirow{1}{*}{0} & \textbf{.271 (.476)} & .222 (.310) & .082 (.335) & .096 (.445) & .233 (.490) & .239 (.199) & .249 (.195) \\ 
   & \multirow{1}{*}{.25} & \textbf{.264 (.429)} & .195 (.339) & .016 (.343) & .071 (.424) & .226 (.441) & .205 (.297) & .219 (.242) \\ 
   & \multirow{1}{*}{.5} & \textbf{.228 (.413)} & .169 (.489) & .087 (.341) & .040 (.346) & .161 (.479) & .169 (.184) & .184 (.168) \\ 
   & \multirow{1}{*}{.75} & \textbf{.212 (.427)} & .193 (.308) & .084 (.332) & .074 (.412) & .027 (.323) & .160 (.252) & .187 (.162) \\ 
  \multirow{4}{*}{Poisson} & \multirow{1}{*}{0} & \textbf{.804 (.191)} & .799 (.194) & .612 (.381) & .095 (.442) & .794 (.197) & .053 (.412) & .090 (.332) \\ 
   & \multirow{1}{*}{.25} & .759 (.154) & \textbf{.827 (.159)} & .668 (.347) & .092 (.477) & .750 (.161) & .034 (.379) & .064 (.335) \\ 
   & \multirow{1}{*}{.5} & .711 (.156) & \textbf{.839 (.140)} & .702 (.339) & .076 (.391) & .706 (.159) & .029 (.433) & .090 (.431) \\ 
   & \multirow{1}{*}{.75} & .671 (.172) & \textbf{.847 (.146)} & .731 (.340) & .049 (.448) & .672 (.172) & .075 (.371) & .056 (.340) \\ 
   \hline
\end{tabular}
\end{adjustbox}
\caption{Average $R^2$ across simulation scenarios under model misspecification, reported as the mean $R^2$ with its standard deviation in parentheses. Results are averaged over all sample sizes for each combination of response distribution (Dist) and absolute spatial dependence level ($|\rho|$). The distribution-specific spatial models (Probit-SAR for Binomial data and Poisson-SAR for Poisson data) were excluded from this summary because frequent convergence failures substantially reduced the number of valid replications in several simulation settings.}
\label{R2YYsMis}
\end{table}

Tables \ref{RMSEYYsMis} and \ref{R2YYsMis} summarize the predictive performance under model misspecification. As expected, no single method uniformly dominates across all scenarios once the assumed response distribution differs from the true data-generating mechanism. The relative performance depends on the underlying distribution and the strength of spatial dependence.

For the Binomial distribution, INLA consistently achieves the lowest RMSE and the highest $R^2$ as the spatial dependence increases, while the proposed estimator exhibits a gradual deterioration in predictive performance.
{This behavior is likely related to the bounded nature of Binomial responses. Strong positive spatial dependence tends to generate large clusters of successes or failures, producing probabilities close to zero or one. Under these conditions, the distance-based Gaussian process used in the misspecification setting may induce a dependence structure that is less accurately represented by the working covariance $(\mathbf A^\top \mathbf A)^{-1}$, which can reduce predictive efficiency. This interpretation is consistent with the discussion surrounding Corollary~\ref{cor4}, where efficiency gains depend on the adequacy of the working covariance approximation.}

In contrast, for the Gamma distribution, the proposed approach remains the best-performing method in terms of both RMSE and $R^2$ across virtually all scenarios, indicating that the estimating-equations framework is particularly robust to misspecification of the variance structure. For the Poisson distribution, GAM generally provides the best {in-sample goodness-of-fit} performance under misspecification, with the proposed estimator remaining competitive and consistently outperforming the Gaussian SAR-type alternatives.

The Gaussian spatial models (Ro-SAR, SAR, and SARAR) exhibit substantially poorer {in-sample goodness-of-fit} performance across nearly all settings. In the Poisson scenarios, these methods frequently produce extremely large prediction errors, highlighting their sensitivity to departures from the Gaussian assumptions on which they are based.

The results based on $R^2$ lead to similar conclusions. Although the proposed estimator is not uniformly optimal under misspecification, it consistently retains a substantial proportion of the explained variability and remains competitive with the more flexible approaches. Overall, these findings suggest that the proposed method exhibits a satisfactory degree of robustness to distributional misspecification, particularly considering that it is derived from a comparatively parsimonious quasi-likelihood formulation.

Overall, the simulation study demonstrates that the proposed estimator provides a favorable compromise between inferential accuracy, predictive performance, and computational stability. Under correct model specification, it consistently yields accurate parameter estimates, low prediction errors, and confidence intervals with coverage close to the nominal level. When the assumed distribution is misspecified, some loss of predictive efficiency is inevitable, and more flexible procedures such as GAM or INLA may become preferable depending on the response distribution. Nevertheless, the proposed estimator continues to provide competitive predictive performance while preserving reliable inference for the regression and spatial dependence parameters, making it a practical and robust framework for the analysis of spatially correlated non-Gaussian data.

\section{Application: 2020 U.S. presidential election}\label{sec4}
Following the approach of \cite{lacombe2018use}, this study examines the factors that influenced the 2020 U.S. presidential election results at the county level, with particular emphasis on spatial dependence. One key objective is to assess the presence of spatial interdependence in county-level voting outcomes, which may reflect the influence of unobserved spatially correlated variables, such as media exposure. The spatial logistic model includes an autoregressive parameter that enables inference regarding the existence and nature of such spatial dependence. A second objective concerns the relative magnitude of the direct and indirect (spillover) effects associated with the explanatory variables.

\begin{figure}[!ht]
\centering
\includegraphics[width=14cm]{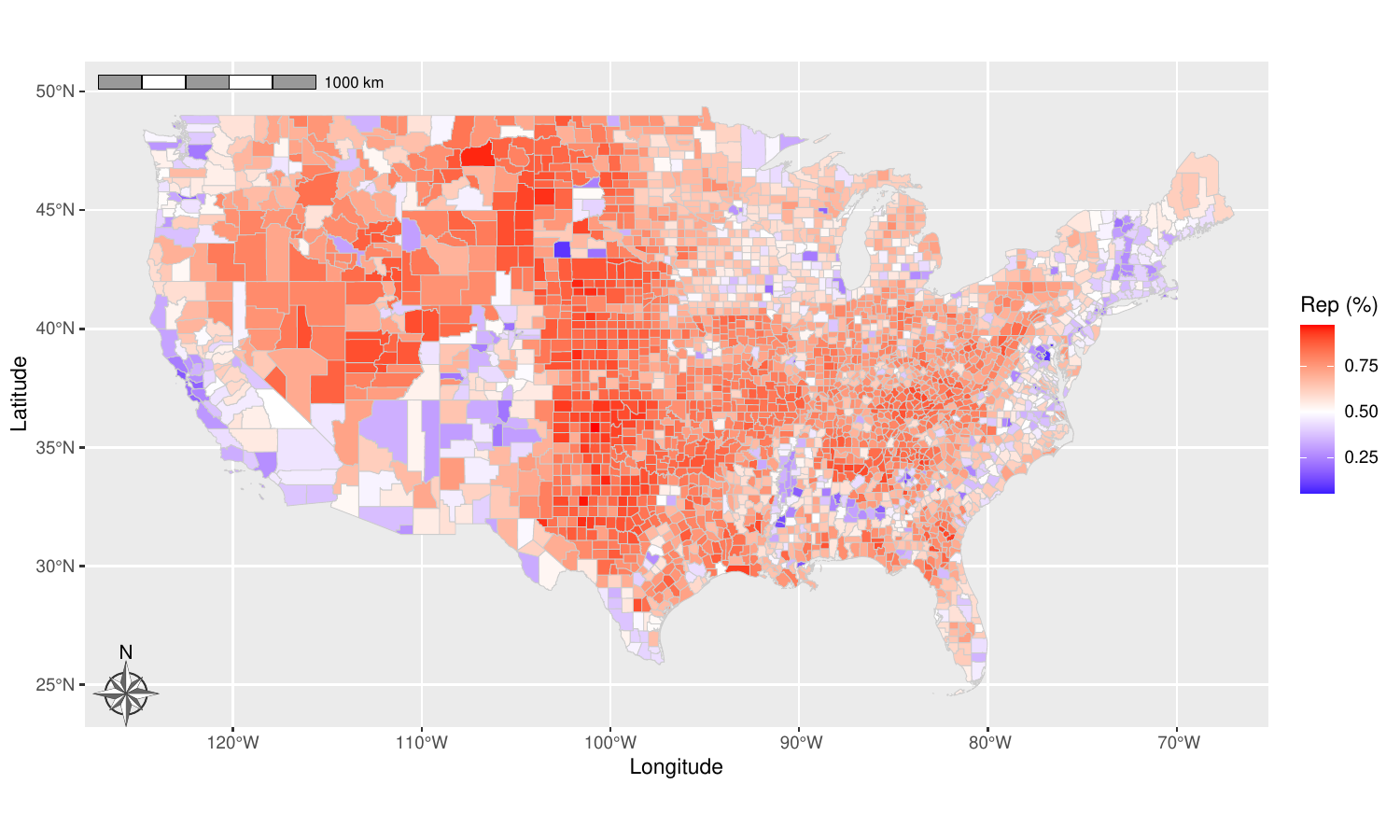}
\caption{Republican vote share by county in the 2020 U.S. presidential election. The color scale ranges from blue (100\% Democratic vote share) through white (50\% Republican -- 50\% Democratic) to red (100\% Republican vote share).}
\label{Election}
\end{figure}

There is a substantial body of literature highlighting spatial interdependence in county-level voting behavior in the United States \citep{lacombe2018use}. Interpreting the estimated effects—particularly the distinction between direct and indirect effects—is essential for understanding the mechanisms behind voting patterns, as formalized in Equation~\eqref{sk_dit}. In the present analysis, the dependent variable is the proportion of votes for the Republican Party in county~$i$, denoted by $y_i$.

The data for the dependent variable were obtained from the Massachusetts Institute of Technology (MIT) Election Lab \citep{MITElectionLab}, and include vote counts for 3155 counties. Figure~\ref{Election} displays the county-level percentage of votes cast for the Republican Party in the 2020 election.

\begin{figure}[!ht]
\centering
\includegraphics[width=14cm]{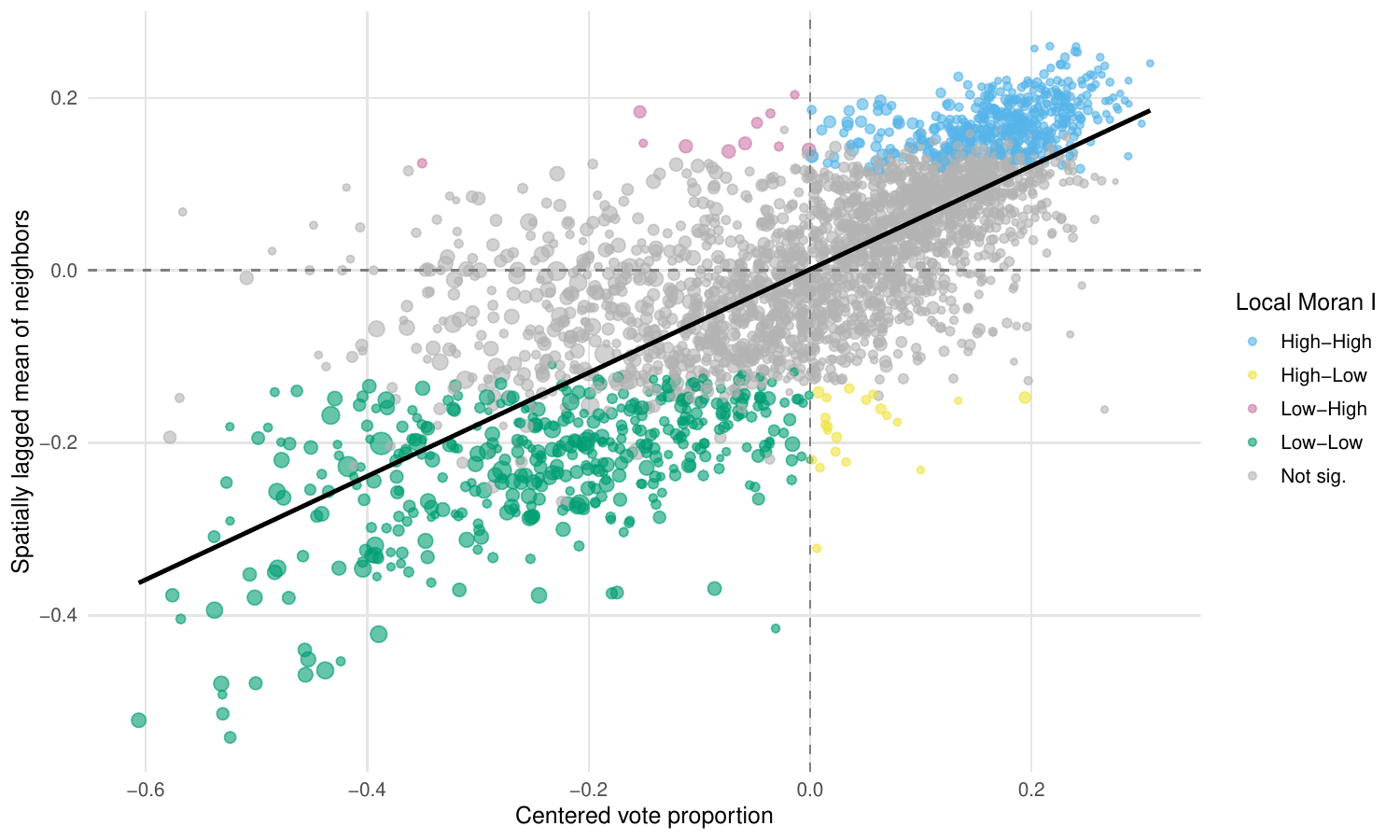}
\caption{Moran scatterplot of county-level vote proportions. Each point represents a county, with the $x$-axis showing the centered vote proportion ($p_i - \bar{p}$) and the $y$-axis showing the spatially lagged mean of neighboring counties. The size of each point is proportional to $\sqrt{\text{votes}_i}$, highlighting counties with larger voting populations. Colors indicate significant Local Moran I clusters (High-High, Low-Low, High-Low, Low-High, or not significant). The observed Moran's $I$ was $0.600 (p-value=0.001)$, indicating strong positive spatial autocorrelation.}
\label{fig:moran_scatter}
\end{figure}

Figure~\ref{fig:moran_scatter} shows the Moran scatterplot of the centered vote proportions across U.S. counties, with point sizes scaled according to the total number of votes in each county. Points are colored according to the Local Moran's I clusters, highlighting significant high-high, low-low, high-low, and low-high spatial associations. The overall positive spatial autocorrelation is confirmed by the global Moran's I statistic of 0.60, which is highly significant ($p < 0.001$). This indicates strong clustering of similar vote proportions among geographically neighboring counties, consistent with the patterns highlighted in the scatterplot.

The empirical model incorporates a set of county-level explanatory variables, based on data from the 2020 U.S. Census \citep{USCensus2020}. These include: Income: Median household income (in thousands of USD), Civilian: Proportion of the civilian population in employment or seeking employment, Unemployment: Proportion of the population unemployed, PercLess: Percentage of adults with less than a high school diploma, PercHigh: Percentage of adults with only a high school diploma, Birthrate: Number of births per 100000 population, IntMigRate: International migration rate per 100000 population, and Poverty: Proportion of the population living in poverty.

The proposed model is defined as:
\begin{equation}\label{USA}
Y_i\sim \text{Binomial}(p_i, M_i), \quad \log\left(\frac{p_i}{1-p_i}\right)=\eta_i = \rho \sum_{j=1}^n w_{ij}\eta_j + \mathbf{x}_{i}^\top \pmb{\beta},
\end{equation}
where $\mathbf{E}(Y_i)=M_i p_i=\mu_i$, $\eta_i=\log\left(\frac{p_i}{1-p_i}\right)$, and $\mbox{Var}(Y_i)=\phi_i M_ip_i(1-p_i)$,  $\mathbf{W}=\{w_{ij}\}$ is a spatial weights matrix defining county adjacency, $M_i$ is the total number of votes in county~$i$, $\mu_i$ is the expected proportion of votes for the Republican Party, and $\mathbf{x}_i$ is the vector of covariates for county~$i$.

\begin{table}[ht]
\centering
\begin{adjustbox}{width=1.1\textwidth,center}
\begin{tabular}{lccccccl}
\hline
Model &  $\hat{\rho}$ &$\phi$ & CIC & quasi-loglik & $R^2$ & $p$ & Moran's I (p-value)\\
\hline
GSCIMC 1 & 0.2517 & $1.8348\times 10^{3}$ & $8.914$ & $-5.1351\times 10^{4}$ & 0.3670&10&$0.0571(<0.001)$ \\
GSCIMC 2 & 0.2853 & $1.5050\times 10^{3}$ & $7.623$ & $-6.2268\times 10^{4}$ & 0.4406&11& $0.0457(<0.001)$\\
GSCIMC 3 & 0.3160 & $1.5245\times 10^{3}$ & $6.985$ & $-4.1513\times 10^{4}$ & 0.4670& 10& $-0.0149(0.1376)$\\
GLM  & $2.445$  & $1.6927\times 10^{3}$     & $32.967$  & $-5.5459\times 10^{4}$  & 0.4547 & 12 & $0.3152(<0.001)$\\
Ro-SAR & 0.4034  & $1.8039\times 10^{3}$  & --        & $-5.2162\times 10^{3}$   &0.6754 &10 & $0.1792(<0.001)$\\
GAM & -- & $1.2631\times 10^{3}$ &-- & $-7.3766\times 10^{4}$ & 0.5566 & 37 & $0.2389(<0.001)$\\
INLA&-- & $4.2205\times 10^{-2}$ & -- & $-2.1611\times 10^{9}$ & 0.7958 & 1561 & $-0.0761 (<0.001)$\\
\hline
\end{tabular}
\end{adjustbox}
\caption{Comparison of the fitted models. The dispersion parameter $\phi$ was computed according to Equation~\eqref{eq:phi}. The reported quasi-loglikelihood was calculated using Equation~\eqref{eq:quasilikelihood}. The CIC is reported only for the GEE-based models (GSCIMC 1--3) and for the GLM, where it is interpreted as a corresponding approximation under working independence. The coefficient of determination $R^2$ was computed between the observed proportions $Y_i/m_i$ and the fitted probabilities $\hat p_i$. The quantity $p$ denotes the effective number of estimated parameters in each model. Moran's $I$ was calculated from the deviance residuals to assess the remaining spatial autocorrelation after model fitting. The CIC is theoretically justified only for GEE models and is included for the GLM solely as a comparable approximation under the same quasi-likelihood framework.}
\label{tab:rho_cic_loglik}
\end{table}
We fitted several models to analyze the proportion of Republican votes across U.S. counties, accounting for socio-economic covariates and spatial dependence. Three GSCIMCs were estimated using a binomial family with Equation~\eqref{USA}, differing in the covariates included: Model 1 included median household income, labor force participation, unemployment rate, and educational and migration indicators; Model 2 additionally incorporated the county poverty rate; Model 3 excluded median household income. For each GSCIMC, the spatial autoregressive parameter $\rho$, dispersion $\phi$, Corrected Information Criterion (CIC), and log quasi-likelihood were estimated. 
For comparison, we also fitted a classical quasibinomial GLM with a lagged dependent variable, a robust SAR model using \texttt{sphet::spreg}, a GAM including a two-dimensional smooth term on the geographical coordinates, and a Bayesian spatial model fitted through INLA with a Besag spatial random effect.

Results are summarized in Table~\ref{tab:rho_cic_loglik}. {To facilitate comparison across competing methods, the reported quasi-loglikelihood and dispersion parameter were computed using the same definitions for all models (Equations \eqref{eq:quasilikelihood} and \eqref{eq:phi}). Consequently, these quantities are presented on a common scale, regardless of the estimation framework employed.} It is also important to note that median household income and poverty rate are conceptually related socioeconomic indicators and may exhibit substantial collinearity. Consequently, the lower CIC observed for Model 3 should not be interpreted as evidence that poverty rate is a more relevant predictor than household income. Rather, the inclusion of both variables in Model 2 may introduce redundant information, leading to a modest increase in model complexity without a corresponding improvement in fit. The similar values of $\hat{\rho}$, $R^2$, and Moran's $I$ obtained for Models 2 and 3 suggest that the two specifications capture largely the same underlying socioeconomic variation, with Model 3 providing a slightly more parsimonious representation according to the CIC criterion. 

{
Loving County (Texas) was excluded from the analysis. This county is the smallest in Texas and was excluded from the county-level census undercount analyses of \citet{castellanos20242020} because of its particular sensitivity to the Census Bureau's differential privacy adjustments. Following this precedent, we also excluded Loving County from our analysis. In addition, it exhibits several extreme demographic and socioeconomic characteristics, including the highest birth rate, exceptionally high median household income, and one of the lowest poverty rates in the sample. As a consequence, it generated unusually large Pearson and deviance residuals, exerting a disproportionate influence on residual-based diagnostics such as the estimated dispersion parameter and residual Moran's $I$. In contrast, the estimated regression coefficients and spatial dependence parameter remained highly stable. For example, the estimated spatial dependence parameter changed only from $\hat{\rho}=0.301$ to $\hat{\rho}=0.316$, while $\|\hat{\pmb{\beta}}_{\text{with}}-\hat{\pmb{\beta}}_{\text{without}}\|_2 =0.090$. Similar behavior was observed for the alternative models.
}

{The residual Moran's $I$ results show that all spatial models substantially reduce the strong spatial autocorrelation present in the original response variable. GSCIMC 3 achieves the smallest residual spatial dependence, with a Moran's $I$ of $0.0149$ that is not statistically significant, indicating that the model captures most of the spatial structure in the data. Similarly, INLA produces a negligible level of residual spatial autocorrelation. By contrast, GSCIMC Models 1 and 2 exhibit larger residual Moran's $I$ values, suggesting that the exclusion of poverty (Model 1) or the simultaneous inclusion of poverty and income (Model 2) leaves additional spatial structure unexplained. It is also worth noting that Moran's $I$ tests may detect very small departures from spatial randomness in large samples due to their high statistical power \citep{HKELEJIAN2001219, AMARAL2014773}. Overall, these results highlight a trade-off between predictive flexibility and model parsimony: although highly flexible models may achieve superior goodness of fit, GSCIMC 3 captures most of the spatial dependence with a considerably simpler specification while providing direct estimation and inference for the spatial dependence parameter.}

The comparison with alternative approaches highlights the different modeling philosophies underlying the competing methods. As expected, GAM and INLA achieve higher $R^2$ values than the GSCIMC specifications, with INLA providing the highest explanatory power. However, these improvements are obtained at the expense of substantially greater model complexity, requiring 37, and 1587 effective parameters for GAM, and INLA, respectively, compared with only 10 parameters for the GSCIMC models. {The extreme quasi-loglikelihood and dispersion values observed for INLA are likely related to overfitting. Indeed, the unadjusted $R^2$ between the observed and fitted values is close to 0.99, suggesting that the highly flexible latent spatial effect captures a large amount of sample-specific variability.}
\newline
At the same time, these improvements are accompanied by increased model complexity, as reflected by the effective degrees of freedom. In particular, the INLA  model relies on a high-dimensional latent spatial effect, resulting in a considerably larger complexity measure than the remaining models. By contrast, the GSCIMC models use a relatively small number of parameters and summarize spatial dependence through a single interpretable autoregressive coefficient. This parsimonious structure facilitates inference on the magnitude of spatial dependence and allows the computation of direct and indirect spatial effects.
\newline
Consequently, the choice of model depends on the scientific objective. When the primary goal is prediction or the reconstruction of complex spatial patterns, GAM and especially INLA provide a very flexible framework. When interest focuses on quantifying spatial dependence and interpreting spatial spillover effects within a parsimonious regression model, the GSCIMC framework offers a useful alternative.

Table~\ref{tablaREP} reports the parameter estimates, including direct, indirect, and total effects based on Equation~\eqref{sk_dit}, along with standard errors computed via the sandwich variance estimator. Coefficient estimates and standard errors for the GAM and INLA models are reported in Table~\ref{tab:coef_compare} in~\ref{apen00}. The three approaches identify similar substantive effects for the main covariates, particularly unemployment rate, educational attainment, migration, and poverty. Although the direction and statistical significance of the effects are largely consistent across specifications, some differences in magnitude are observed, especially for INLA, where the latent spatial component absorbs a larger share of the local spatial variation. Consequently, coefficient magnitudes should be interpreted with caution, as the models differ in both their spatial structure and estimation framework.
\begin{table}[ht]
\centering
\begin{adjustbox}{width=1.1\textwidth,center}
\begin{tabular}{rrrrrrrr}
  \hline
 & Estim & Robust SE & $z$ & $p(|Z|>z)$ & Direct & Indirect & Total \\ 
  \hline
(Intercept) & 2.1324 & 0.2220 & 9.6063 & $<$0.001 & 2.1748 & 0.9397 & 3.1145 \\ 
  Civ\_lab\_force & -4.6584 & 0.3480 & -13.3875 & $<$0.001  & -4.7510 & -2.0529 & -6.8039 \\ 
  Unemp\_rate & -0.0987 & 0.0059 & -16.6749 & $<$0.001  & -0.1006 & -0.0435 & -0.1441 \\ 
  PorcLess & 0.0198 & 0.0035 & 5.6972 & $<$0.001  & 0.0202 & 0.0087 & 0.0289 \\ 
  PorcHigh & 0.0473 & 0.0019 & 25.2041 &$<$0.001  & 0.0483 & 0.0209 & 0.0691 \\ 
  BirthRate & 0.0004 & 0.0003 & 1.2850 & 0.1988 & 0.0004 & 0.0002 & 0.0006 \\ 
  PorcMigInt & -0.0077 & 0.0019 & -3.9771 & $<$0.001  & -0.0079 & -0.0034 & -0.0113 \\ 
  Poverty & -5.8740 & 0.4456 & -13.1816 & $<$0.001 & -5.9908 & -2.5885 & -8.5794 \\ 
  \hline
\end{tabular}
\end{adjustbox}
\caption{Parameter estimates, standard errors, and spatial effects obtained from the selected GSCIMC model for the county-level Republican vote share in the 2020 U.S. presidential election. The estimated spatial dependence parameter was {$\hat{\rho}=0.3160$ (naive SE $=0.0002$; robust SE $=0.0164$)}, indicating a moderate positive association among neighboring counties. {The robust standard error was computed using Equation~\eqref{eq:varrhosand}, whereas the naive standard error was obtained from Equation~\eqref{eq:varrho1}.}}
\label{tablaREP}
\end{table}

The model formulation in Equation~\eqref{USA} gives greater weight to counties with larger populations, resulting in a model that differs from that proposed by \cite{lacombe2018use} for the 2004 election. In their model, the dependent variable is defined as $y^*_i = \mathbf{1}(y_i \geq 0.5)$, which, as shown in simulation studies, tends to underestimate spatial dependence in binomial models when the number of trials per unit (i.e., total votes) is not constant.

The estimate for the poverty rate presents the strongest direct effect on the Republican vote share, with a coefficient of $-5.9908$. This suggests that one percentage point increase in the poverty rate reduces the odds of voting Republican by approximately 5.{81}\%, as indicated by $\exp(0.01 \times -5.9908) = 0.9418$.  Although the indirect effect ($-2.585$)  slightly increases the total impact, it remains large and negative at $-8.5794$.

This implies that counties with higher Poverty rates tend to vote less for the Republican Party, which may reflect patterns of economic polarization in political behavior, as discussed \cite{kulachai2023factors}. The variables PercLess and PercHigh, capturing the proportion of adults with lower or High levels of educational attainment, show positive associations with Republican vote share. Their total effects are $0.028$ and $0.069$, respectively, consistent with theories linking intermediate educational attainment with populist voting tendencies \citep{bor2017diverging}.   Both variables also exhibit slightly positive indirect effects ($0.009$ and $0.021$), suggesting related patterns in neighboring counties. Other demographic factors, such as the birthrate ($p-value = 0.198$), appear to have weak effects on voting outcomes. In contrast, the international migration rate shows a negative  and statistically significant effect ($p-value < 0.001$), although the estimated magnitude is very small. A direct effect of $-0.008$, indicating that counties with higher migration rates tend to exhibit slightly lower Republican vote shares, and an indirect effect of $-0.003$.
Finally, the unemployment rate shows a statistically significant ($p-value<0.001$) negative effect (direct = $-0.098$, indirect = $-0.043$), indicating that counties with higher unemployment were less supportive of the Republican candidate. This finding contrasts with certain political narratives emphasizing Republican support among economically distressed populations. In this way, the civilian labor force variable is negative (direct=$-4.751$, indirect=$-2.053$) and statistically significant ($p-value<0.001$), indicating that counties with larger labor force participation tend to exhibit lower Republican vote shares. One possible explanation is that counties with larger labor forces are typically more urbanized and economically diversified, characteristics commonly associated with more Democratic voting patterns. In this sense, the variable may be capturing structural differences between large metropolitan counties and smaller rural areas. Moreover, once unemployment is controlled for, the labor force variable may reflect broader labor market dynamics rather than economic distress itself.
Finally, when Model 1 is considered—where median household income is included but the poverty rate is excluded—the coefficient for median income becomes positive and significant ($0.00559 (p-value =0.008)$). Since income and poverty capture closely related dimensions of local economic conditions, these results are consistent: higher income levels and lower poverty rates are both associated with higher Republican vote shares.
\begin{remark}
 To improve the numerical stability of the estimation procedure, the optimization algorithm was initialized using estimates obtained from a robust spatial model fitted to a family-specific transformed response. Furthermore, numerical stabilization procedures were incorporated to account for the substantial heterogeneity in county vote totals and the highly variable binomial denominators, preventing the appearance of extremely small quantities during intermediate computations. These refinements were motivated by the empirical application, where a small number of counties contributed exceptionally large weights because of their much larger vote totals. In this setting, numerical instabilities could occasionally lead the optimization routine to converge to implausible solutions, including estimates of the spatial dependence parameter that were inconsistent with the strong positive spatial autocorrelation observed in the data. The computational refinements substantially improved the robustness and reproducibility of the estimation procedure, yielding stable positive estimates of the spatial dependence parameter that are consistent with the observed spatial clustering.   
\end{remark}

\section{Conclusions}\label{sec5}

The GSCIMC model presented in this work provides a novel and flexible framework, grounded in GEE, for incorporating spatial dependence directly into the linear predictor of response variables from the exponential family. Rather than extending the classical SAR model, GSCIMC offers a working alternative that recovers asymptotically the SAR estimator as a special case under Gaussian responses (Corollary \ref{cor2}) while extending naturally to non-Gaussian outcomes.

The proposed methodology is grounded in log quasi-likelihood theory and GEE, ensuring both consistency and asymptotic efficiency of the estimators. Theoretical results, including consistency, asymptotic normality, and valid sandwich variance formulas, are formally established and supported by simulation evidence. These asymptotic properties enable valid inference for both mean and spatial parameters, including the autoregressive coefficient $\rho$, across a variety of distributions and sample sizes. Simulation studies across Poisson, Gamma, and Binomial responses demonstrate the practical benefits of the proposed approach. The GSCIMC consistently achieves full convergence across all scenarios, in contrast with existing methods—particularly the Pois-SAR and Probit-SAR models—which suffer from severe convergence issues. The RMSE and coverage metrics confirm the robustness of the proposed estimator even under strong spatial dependence ($\rho \approx \pm 0.75$) or small sample sizes.

Overall, the comparison with GAM and INLA highlights the different objectives of these approaches. While the more flexible models may provide improved predictive performance in some settings, particularly under model misspecification, the proposed GSCIMC achieves competitive performance with a substantially more parsimonious formulation while enabling direct statistical inference for the spatial dependence parameter.

For Poisson-distributed outcomes, the proposed model clearly outperforms traditional GLMs and homoscedastic SAR models by more accurately estimating $\rho$, yielding lower bias and variance. Its performance remains stable even in complex scenarios with strong autocorrelation, outperforming the Pois-SAR model across the board. In the Gamma setting, the GSCIMC exhibits low bias and high coverage, matching or exceeding the performance of Ro-SARAR alternatives, particularly in small or moderate samples. Under the Binomial distribution, the method avoids the limitations of Probit-SAR and GLM, demonstrating consistent estimation and reliable inference, with coverage rates near the nominal level.

The implementation is provided through the \texttt{GSCIMC} function in the \texttt{spatemR} package for \texttt{R}, allowing practitioners to apply the model in applied spatial problems easily. The routine has proven to be numerically stable and efficient, adapting well to various spatial structures, sample sizes, and response types. An empirical application to spatial modeling of Republican vote shares further highlights the interpretability and flexibility of the method. The proposed model effectively accounts for spatial dependencies in the data and enables direct interpretation of the effects of covariates on the mean structure, analogous to results obtained from GLMs, while correcting for spatial bias. This case study underscores the utility of the GSCIMC in real-world policy or political science settings.

In summary, the GSCIMC provides a general, consistent, and computationally efficient framework for analyzing Gaussian and non-Gaussian spatial data. Its solid theoretical foundations, robust empirical performance, and flexible implementation make it a valuable contribution to the spatial statistics literature and a powerful tool for applied researchers.

Future research will extend the framework to incorporate nonlinear covariate effects using additive components, following the GAM philosophy. Further developments may also consider multivariate spatial responses, spatio-temporal extensions, or Bayesian formulations of the proposed methodology.

\phantomsection\label{supplementary-material}
\bigskip

\begin{center}

{\large\bf SUPPLEMENTARY MATERIAL}

\end{center}

\begin{description}
\item[Supplementary File 1:]
R scripts with the simulation routines used in the paper, including data generation, model fitting, and result visualization. All methods can be reproduced using the \texttt{GSCIMC} function available in the open-source \texttt{spatemR} R package, which can be accessed via CRAN at \url{https://cran.r-project.org/package=spatemR}. The full source code is also available on GitHub at \url{https://github.com/Cruzalirio/spatemR/blob/main/R/GSCIMC.R}.

\item[Voting data set:]
Data set used in the empirical analysis of voting patterns presented in Section~\ref{sec4}. It contains spatially referenced voting results with explanatory covariates for each location. (.csv file)

\end{description}

\appendix
\section{Complementary Tables}\label{apen00}

\begin{table}[ht]
\footnotesize
\centering
\begin{adjustbox}{width=1.2\textwidth,center}
\begin{tabular}{llrlllllll}
  \hline
Dist & $|\rho|$ & n & Proposed & Probit-SAR & Pois-SAR  & Ro-SAR & GLM & SAR & SARAR \\ 
  \hline
\multirow{16}{*}{Binomial} & \multirow{4}{*}{0} &  49 & \textbf{.007 (.005)} & .608 (.184) & --& .013 (.014) & .066 (.071)& .013 (.012) & .023 (.074) \\ 
   &  &  81 & \textbf{.005 (.004)} & .638 (.138)& -- & .010 (.009) & .048 (.047) & .010 (.009) & .027 (.109) \\ 
   &  & 144 & \textbf{.004 (.003)} & .657 (.062) & --& .007 (.006) & .034 (.031) & .007 (.006) & .035 (.133) \\ 
   &  & 400 & \textbf{.002 (.002)} & .657 (.024)& -- & .004 (.004) & .023 (.024) & .004 (.004) & .073 (.202) \\ 
   & \multirow{4}{*}{.25} &  49 & \textbf{.007 (.005)} & .551 (.248)& -- & .012 (.012) & 1.019 (.100) & .012 (.012) & .020 (.077) \\ 
   &  &  81 & \textbf{.005 (.004)} & .588 (.223) & --& .009 (.008) & 1.017 (.086) & .010 (.008) & .029 (.119) \\ 
   &  & 144 & \textbf{.004 (.003)} & .613 (.207)& -- & .007 (.006) & 1.021 (.064) & .008 (.007) & .034 (.134) \\ 
   &  & 400 & \textbf{.002 (.002)} & .640 (.203)& -- & .004 (.003) & 1.020 (.041) & .005 (.004) & .089 (.243) \\ 
   & \multirow{4}{*}{.5} &  49 & \textbf{.009 (.018)} & .434 (.336)& -- & \textbf{.009 (.009)} & 2.165 (.206) & .011 (.010) & .023 (.119) \\  
   &  &  81 & \textbf{.006 (.010)} & .456 (.338)& -- & .007 (.007) & 2.164 (.176) & .009 (.009) & .036 (.159) \\ 
   &  & 144 & \textbf{.004 (.004)} & .452 (.313)& -- & .005 (.005) & 2.166 (.139) & .007 (.006) & .049 (.189) \\ 
   &  & 400 & \textbf{.002 (.002)} & .491 (.328)& -- & .003 (.003) & 2.164 (.108) & .005 (.004) & .132 (.323) \\ 
   & \multirow{4}{*}{.75} &  49 & .055 (.102) & .282 (.347)& -- & \textbf{.007 (.008)} & 4.067 (1.066) & .011 (.013) & .076 (.278) \\ 
   &  &  81 & .039 (.080) & .252 (.335)& -- & \textbf{.006 (.005)} & 4.015 (.809) & .008 (.009) & .092 (.297) \\ 
   &  & 144 & .023 (.050) & .231 (.308)& -- & \textbf{.004 (.003)} & 3.951 (.616) & .007 (.007) & .181 (.416) \\ 
   &  & 400 & \textbf{.001 (.006)} & .428 (.195)& -- & .002 (.002) & 3.425 (.164) & .004 (.003) & .199 (.431) \\ 
  \hline
  \multirow{16}{*}{Gamma} & \multirow{4}{*}{0} &  49 & .203 (.159) & -- & --& .277 (.222) & \textbf{.060 (.062)} & .132 (.103) & .217 (.153) \\ 
   &  &  81 & .152 (.113) & -- & --& .198 (.149) & \textbf{.040 (.039)} & .098 (.076) & .169 (.121) \\ 
   &  & 144 & .115 (.091) & -- & --& .143 (.116) & \textbf{.024 (.024)} & .074 (.057) & .131 (.098) \\ 
   &  & 400 & .066 (.051) & --& -- & .088 (.069) & \textbf{.013 (.010)} & .047 (.036) & .084 (.066) \\ 
   & \multirow{4}{*}{.25} &  49 & \textbf{.195 (.150)} & -- & --& .266 (.236) & .223 (.073) & .207 (.141) & \textbf{.195 (.159)} \\ 
   &  &  81 & \textbf{.137 (.109)} & --& -- & .191 (.154) & .221 (.053) & .185 (.113) & .150 (.126) \\ 
   &  & 144 & \textbf{.110 (.084)} & -- & --& .142 (.107) & .226 (.034) & .181 (.092) & .119 (.096) \\ 
   &  & 400 & \textbf{.063 (.047)} & -- & --& .080 (.062) & .229 (.018) & .178 (.059) & .073 (.057) \\ 
   & \multirow{4}{*}{.5} &  49 & \textbf{.143 (.132)} & --& -- & .196 (.182) & .437 (.070) & .318 (.152) & .168 (.159) \\ 
   &  &  81 & \textbf{.109 (.095)} & -- & --& .139 (.121) & .447 (.053) & .323 (.123) & .141 (.121) \\ 
   &  & 144 & \textbf{.073 (.060)} & -- & --& .099 (.082) & .454 (.037) & .321 (.092) & .112 (.085) \\ 
   &  & 400 & \textbf{.045 (.035)} & --& -- & .061 (.048) & .462 (.022) & .321 (.058) & .085 (.056) \\ 
   & \multirow{4}{*}{.75} &  49 & \textbf{.086 (.102)} & --& -- & .111 (.106) & .683 (.076) & .378 (.156) & .159 (.132) \\ 
   &  &  81 & \textbf{.067 (.072)} & --& -- & .078 (.078) & .695 (.058) & .370 (.124) & .138 (.094) \\ 
   &  & 144 & \textbf{.045 (.040)} & --& -- & .056 (.045) & .706 (.044) & .365 (.096) & .118 (.062) \\ 
   &  & 400 & \textbf{.027 (.026)} & --& -- & .033 (.025) & .719 (.027) & .359 (.056) & .108 (.038) \\ 
   \hline
  \multirow{16}{*}{Poisson} & \multirow{4}{*}{0} &  49 & .130 (.099) & --& \textbf{.039 (.031)} & .164 (.124) & .042 (.041) & .105 (.080) & .143 (.107) \\ 
   &  &  81 & .092 (.069) & --& \textbf{.028 (.021)} & .125 (.094) & .029 (.025) & .085 (.063) & .117 (.090) \\ 
   &  & 144 & .070 (.057) & --& \textbf{.020 (.016)} & .096 (.071) & .021 (.017) & .064 (.048) & .094 (.067) \\ 
   &  & 400 & .036 (.028) & --& .011 (.008) & .051 (.041) & \textbf{.010 (.007)} & .037 (.028) & .050 (.041) \\ 
   & \multirow{4}{*}{.25} &  49 & \textbf{.118 (.095)}& -- & .181 (.042) & .148 (.119) & .196 (.059) & .152 (.108) & .132 (.108) \\ 
   &  &  81 & \textbf{.085 (.065)} & --& .182 (.029) & .109 (.086) & .200 (.044) & .129 (.088) & .099 (.081) \\ 
   &  & 144 & \textbf{.060 (.047)}& -- & .185 (.021) & .085 (.065) & .204 (.030) & .125 (.073) & .079 (.061) \\ 
   &  & 400 & \textbf{.036 (.027)} & --& .186 (.012) & .049 (.038) & .209 (.019) & .121 (.048) & .046 (.036) \\ 
   & \multirow{4}{*}{.5} &  49 & \textbf{.088 (.076)} & --& .370 (.036) & .116 (.102) & .396 (.088) & .220 (.130) & .112 (.105) \\ 
   &  &  81 & \textbf{.061 (.050)} & --& .373 (.024) & .080 (.067) & .406 (.070) & .208 (.100) & .078 (.068) \\ 
   &  & 144 & \textbf{.046 (.036)} & --& .374 (.016) & .061 (.046) & .412 (.060) & .205 (.082) & .063 (.051) \\ 
   &  & 400 & \textbf{.024 (.019)}& -- & .375 (.010) & .035 (.027) & .420 (.051) & .207 (.060) & .045 (.034) \\ 
   & \multirow{4}{*}{.75} &  49 & \textbf{.050 (.046)}& -- & .561 (.033) & .061 (.055) & .594 (.166) & .230 (.139) & .076 (.073) \\ 
   &  &  81 & \textbf{.034 (.033)} & --& .561 (.022) & .042 (.036) & .594 (.161) & .217 (.129) & .057 (.051) \\ 
   &  & 144 & \textbf{.022 (.020)}& -- & .561 (.013) & .030 (.025) & .602 (.151) & .214 (.126) & .051 (.040) \\ 
   &  & 400 & \textbf{.012 (.011)} & --& .562 (.005) & .019 (.015) & .607 (.141) & .215 (.123) & .046 (.030) \\ 
   \hline
\end{tabular}
\end{adjustbox}
\caption{$RMSE(\hat{\rho})$ of the spatial parameter $\hat{\rho}$ across simulation scenarios. Rows are indexed by the distribution (Dist), the absolute value of the true spatial parameter ($|\rho|$), and the sample size ($n$), while columns correspond to the different models. Each cell reports the mean RMSE across replications, with the corresponding standard deviation in parentheses. Leading zeros in decimal values have been omitted for compactness.}
\label{RMSERhoAA}
\end{table}

\begin{table}[ht]
\footnotesize
\centering
\begin{adjustbox}{width=1.2\textwidth,center}
\begin{tabular}{llrlllllllll}
  \hline
Dist &$|\rho|$ & n & Proposed & Probit-SAR &Pois-SAR  & Ro-SAR & GLM & SAR & SARAR & GAM & INLA \\ 
  \hline
\multirow{16}{*}{Binomial} & \multirow{4}{*}{0} &  49 & .005 (.004) & 1.078 (.712) & --& .010 (.010) & .040 (.042) & .010 (.010) & .012 (.017) & \textbf{.004 (.004)} & \textbf{.004 (.003)} \\ 
   &  &  81 & .004 (.003) & 1.507 (1.133)& -- & .008 (.007) & .029 (.028) & .008 (.007) & .011 (.017) & \textbf{.003 (.002)} & \textbf{.003 (.002)} \\ 
   &  & 144 & .003 (.002) & 1.645 (NA)& -- & .006 (.005) & .021 (.019) & .006 (.005) & .011 (.022) & \textbf{.002 (.002)} & \textbf{.002 (.002)} \\ 
   &  & 400 & .002 (.001) & --& -- & .004 (.003) & .014 (.015) & .004 (.003) & .017 (.033) & \textbf{.001 (.001)} & \textbf{.001 (.001)} \\ 
   & \multirow{4}{*}{.25} &  49 & \textbf{.006 (.005)} & .864 (.663)& -- & .010 (.010) & .633 (.060) & .011 (.011) & .012 (.016) & .135 (.068) & .134 (.063) \\ 
   &  &  81 & \textbf{.004 (.003)} & 1.248 (.740)& -- & .009 (.007) & .632 (.052) & .009 (.007) & .013 (.022) & .132 (.061) & .132 (.059) \\ 
   &  & 144 & \textbf{.003 (.002)} & 1.660 (NA)& -- & .006 (.006) & .634 (.038) & .007 (.006) & .012 (.023) & .133 (.041) & .133 (.042) \\ 
   &  & 400 & \textbf{.002 (.001)} & --& -- & .004 (.003) & .634 (.024) & .004 (.003) & .019 (.038) & .131 (.036) & .131 (.038) \\ 
   & \multirow{4}{*}{.5} &  49 & .013 (.027) & 1.542 (1.105)& -- & \textbf{.011 (.011)} & 1.338 (.127) & .012 (.012) & .014 (.026) & .344 (.214) & .340 (.207) \\ 
   &  &  81 & \textbf{.008 (.014)} & .752 (.398) & --& .009 (.008) & 1.335 (.107) & .010 (.009) & .015 (.032) & .332 (.203) & .332 (.200) \\ 
   &  & 144 & \textbf{.005 (.005)} & 1.296 (.468)& -- & .007 (.006) & 1.335 (.083) & .008 (.007) & .016 (.039) & .334 (.184) & .334 (.191) \\ 
   &  & 400 & \textbf{.003 (.003)} & --& -- & .004 (.003) & 1.335 (.062) & .005 (.004) & .032 (.061) & .330 (.159) & .332 (.170) \\ 
   & \multirow{4}{*}{.75} &  49 & .127 (.257) & 1.405 (.663)& -- & \textbf{.018 (.019)} & 2.523 (.819) & .024 (.030) & .036 (.077) & .840 (.710) & .817 (.715) \\ 
   &  &  81 & .090 (.201) & 1.148 (.519)& -- & \textbf{.014 (.012)} & 2.467 (.586) & .018 (.020) & .044 (.094) & .859 (.673) & .852 (.681) \\ 
   &  & 144 & .052 (.117) & 1.135 (.322)& -- & \textbf{.010 (.008)} & 2.417 (.424) & .014 (.014) & .067 (.117)& .875 (.676) & .875 (.699) \\ 
   &  & 400 & \textbf{.003 (.013)} & .734 (.329) & --& .006 (.005) & 2.085 (.098) & .006 (.006) & .017 (.020) & .836 (.602) & .856 (.649) \\ 
   \hline
  \multirow{16}{*}{Gamma} & \multirow{4}{*}{0} &  49 & .160 (.125) & --& -- & .566 (.211) & .195 (.166) & .564 (.196) & .563 (.200) & .126 (.088) & \textbf{.107 (.075)} \\ 
   &  &  81 & .123 (.095) & -- & --& .571 (.158) & .150 (.112) & .573 (.150) & .571 (.156) & .093 (.068) & \textbf{.089 (.062)} \\ 
   &  & 144 & .092 (.068) & -- & --& .573 (.111) & .098 (.076) & .573 (.109) & .573 (.111) & .077 (.054) & \textbf{.070 (.049)} \\ 
   &  & 400 & .050 (.038) & --& -- & .577 (.064) & .055 (.042) & .578 (.063) & .577 (.063) & \textbf{.039 (.032)} & .041 (.031) \\ 
   & \multirow{4}{*}{.25} &  49 & .165 (.129) & --& -- & .558 (.254) & .203 (.157) & .544 (.238) & .555 (.241) & .162 (.129) & \textbf{.157 (.109)} \\ 
   &  &  81 & \textbf{.127 (.097)} & --& -- & .581 (.236) & .160 (.121) & .559 (.221) & .576 (.231) & .147 (.100) & .156 (.108) \\ 
   &  & 144 & \textbf{.091 (.067)} & --& -- & .572 (.177) & .114 (.090) & .550 (.172) & .568 (.176) & .147 (.088) & .149 (.090) \\ 
   &  & 400 & \textbf{.052 (.040)} & --& -- & .574 (.161) & .088 (.063) & .553 (.154) & .572 (.160) & .134 (.065) & .134 (.070) \\ 
   & \multirow{4}{*}{.5} &  49 & \textbf{.189 (.157)} & --& -- & .575 (.362) & .281 (.225) & .500 (.353) & .544 (.359) & .322 (.208) & .346 (.249) \\ 
   &  &  81 & \textbf{.132 (.109)} & -- & --& .568 (.330) & .260 (.193) & .480 (.336) & .538 (.332) & .331 (.200) & .351 (.242) \\ 
   &  & 144 & \textbf{.096 (.076)} & --& -- & .564 (.312) & .247 (.185) & .468 (.331) & .537 (.316) & .316 (.219) & .328 (.247) \\ 
   &  & 400 & \textbf{.057 (.043)} & --& -- & .574 (.298) & .226 (.176) & .472 (.322) & .550 (.303) & .322 (.213) & .324 (.227) \\ 
   & \multirow{4}{*}{.75} &  49 & \textbf{.296 (.287)} & --& -- & .591 (.461) & .826 (.636) & .660 (.375) & .556 (.457) & .777 (.695) & .834 (.743) \\ 
   &  &  81 & \textbf{.269 (1.015)} & --& -- & .588 (.469) & .841 (.595) & .668 (.338) & .548 (.467) & .799 (.688) & .818 (.719) \\ 
   &  & 144 & \textbf{.161 (.171)} & -- & --& .583 (.463) & .844 (.601) & .660 (.312) & .535 (.474) & .819 (.736) & .825 (.736) \\ 
   &  & 400 & \textbf{.076 (.078)} & -- & --& .572 (.440) & .822 (.615) & .637 (.281) & .511 (.469) & .822 (.756) & .767 (.709) \\ 
   \hline
  \multirow{16}{*}{Poisson} & \multirow{4}{*}{0} &  49 & .128 (.098) & --& .122 (.093) & .503 (.217) & .162 (.121) & .525 (.152) & .507 (.198) & .104 (.085) & \textbf{.100 (.080)} \\ 
   &  &  81 & .093 (.070)& -- & .095 (.074) & .512 (.170) & .112 (.091) & .520 (.122) & .512 (.163) & .084 (.063) & \textbf{.081 (.060)} \\ 
   &  & 144 & .072 (.056) & --& .071 (.051) & .499 (.127) & .086 (.068) & .510 (.092) & .499 (.124) & .056 (.044) & \textbf{.055 (.042)} \\ 
   &  & 400 & .038 (.030)& -- & .040 (.029) & .509 (.071) & .044 (.035) & .512 (.054) & .509 (.071) & \textbf{.033 (.024)} & \textbf{.033 (.024)} \\ 
   & \multirow{4}{*}{.25} &  49 & .132 (.107) & --& \textbf{.125 (.098)} & .530 (.260) & .169 (.133) & .533 (.167) & .526 (.232) & .167 (.122) & .161 (.116) \\ 
   &  &  81 & \textbf{.091 (.073)}& --& \textbf{.091 (.072)} & .517 (.232) & .127 (.098) & .528 (.129) & .517 (.210) & .138 (.086) & .137 (.084) \\ 
   &  & 144 & .069 (.054)& -- & \textbf{.068 (.055)} & .523 (.211) & .097 (.072) & .525 (.105) & .522 (.195) & .140 (.080) & .142 (.081) \\ 
   &  & 400 & \textbf{.041 (.034)} & --& .043 (.034) & .525 (.192) & .073 (.049) & .525 (.077) & .525 (.181) & .135 (.061) & .135 (.062) \\ 
   & \multirow{4}{*}{.5} &  49 & \textbf{.143 (.123)} & --& .154 (.161) & .564 (.387) & .271 (.184) & .565 (.215) & .560 (.341) & .424 (1.097) & .346 (.225) \\ 
   &  &  81 & \textbf{.103 (.089)} & --& .111 (.088) & .557 (.382) & .277 (.167) & .557 (.189) & .555 (.335) & .331 (.230) & .323 (.228) \\ 
   &  & 144 & \textbf{.075 (.067)}& -- & .083 (.068) & .553 (.381) & .271 (.134) & .553 (.173) & .552 (.338) & .331 (.219) & .332 (.211) \\ 
   &  & 400 & \textbf{.044 (.038)}& -- & .048 (.039) & .553 (.369) & .277 (.105) & .548 (.149) & .551 (.329) & .324 (.215) & .324 (.188) \\ 
   & \multirow{4}{*}{.75} &  49 & \textbf{.162 (.155)} & --& .315 (.262) & .715 (.629) & .964 (.476) & .649 (.384) & .690 (.574) & 1.558 (5.965) & .838 (.724) \\ 
   &  &  81 & \textbf{.108 (.103)} & --& .236 (.189) & .693 (.624) & 1.035 (.423) & .629 (.370) & .669 (.571) & .834 (.720) & .840 (.704) \\ 
   &  & 144 & \textbf{.074 (.078)}& -- & .179 (.128) & .678 (.618) & 1.072 (.421) & .612 (.359) & .658 (.565) & .840 (.773) & .851 (.731) \\ 
   &  & 400 & \textbf{.043 (.045)}& -- & .119 (.076) & .668 (.613) & 1.110 (.396) & .601 (.344) & .656 (.558) & .832 (.772) & .831 (.688) \\ 
   \hline
\end{tabular}
\end{adjustbox}
\caption{$RMSE(\hat{\beta}_0)$ of the parameter $\beta_0$ across simulation scenarios. Rows are indexed by the distribution (Dist), the absolute value of the true spatial parameter ($|\rho|$), and the sample size ($n$), while columns correspond to the different models. Each cell reports the mean RMSE across replications, with the corresponding standard deviation in parentheses. Leading zeros in decimal values have been omitted for compactness.}
\label{RMSEBetaAA}
\end{table}

\begin{table}[ht]
\footnotesize
\centering
\begin{adjustbox}{width=1.2\textwidth,center}
\begin{tabular}{llrlllllllll}
  \hline
Dist &$|\rho|$ & n & Proposed & Probit-SAR & Pois-SAR  & Ro-SAR & GLM & SAR & SARAR & GAM & INLA \\ 
  \hline
\multirow{16}{*}{Binomial} & \multirow{4}{*}{0} &  49 & \textbf{.004 (.003)} & 1.240 (.756) & --& .008 (.008) & .008 (.010) & .008 (.008) & .017 (.058) & \textbf{.004 (.003)} & \textbf{.004 (.003)} \\ 
   &  &  81 & \textbf{.003 (.002)} & 1.375 (.687)& -- & .006 (.005) & .006 (.006) & .006 (.005) & .019 (.070) & \textbf{.003 (.002)} & \textbf{.003 (.002)} \\ 
   &  & 144 & \textbf{.002 (.002)} & 2.030 (NA) & --& .005 (.004) & .005 (.004) & .005 (.004) & .026 (.091) & \textbf{.002 (.002)} & \textbf{.002 (.002)} \\ 
   &  & 400 & \textbf{.001 (.001)} & -- & -- & .003 (.002) & .003 (.003) & .003 (.002) & .060 (.141) & \textbf{.001 (.001)} & \textbf{.001 (.001)} \\ 
   & \multirow{4}{*}{.25} &  49 & \textbf{.004 (.003)} & 1.178 (.658)& -- & .008 (.008) & .009 (.008) & .008 (.008) & .014 (.048) & .064 (.032) & .035 (.025) \\ 
   &  &  81 & \textbf{.003 (.003)} & 1.525 (.753)& -- & .006 (.005) & .007 (.006) & .006 (.006) & .022 (.078) & .042 (.020) & .032 (.019) \\ 
   &  & 144 & \textbf{.002 (.002)} & 1.385 (NA) & --& .005 (.005) & .005 (.005) & .005 (.005) & .027 (.092) & .025 (.014) & .032 (.016) \\ 
   &  & 400 & \textbf{.001 (.001)} & -- & -- & .003 (.002) & .003 (.003) & .003 (.002) & .065 (.145) & .011 (.008) & .032 (.012) \\ 
   & \multirow{4}{*}{.5} &  49 & \textbf{.005 (.006)} & 1.293 (.671)& -- & .009 (.008) & .012 (.010) & .009 (.009) & .016 (.056) & .152 (.101) & .086 (.069) \\ 
   &  &  81 & \textbf{.003 (.003)} & .911 (.439)& -- & .007 (.006) & .010 (.008) & .007 (.006) & .025 (.084) & .097 (.057) & .081 (.063) \\ 
   &  & 144 & \textbf{.003 (.002)} & .759 (.325)& -- & .005 (.004) & .007 (.006) & .005 (.005) & .035 (.106) & .060 (.043) & .081 (.054) \\ 
   &  & 400 & \textbf{.001 (.001)} & -- & -- & .003 (.002) & .004 (.004) & .003 (.003) & .089 (.165) & .030 (.024) & .077 (.046) \\ 
   & \multirow{4}{*}{.75} &  49 & .016 (.033) & .556 (.465) & --& \textbf{.011 (.011)} & .028 (.028) & .013 (.013) & .046 (.116) & .281 (.240) & .180 (.176) \\ 
   &  &  81 & .010 (.019) & .340 (.243) & --& \textbf{.009 (.007)} & .022 (.019) & .011 (.010) & .060 (.138) & .168 (.141) & .171 (.167) \\ 
   &  & 144 & .007 (.011) & .345 (.207) & --& \textbf{.006 (.005)} & .016 (.015) & .008 (.007) & .103 (.177) & .107 (.093) & .164 (.147) \\ 
   &  & 400 & \textbf{.002 (.002)} & .192 (.138)& -- & .003 (.003) & .006 (.005) & .004 (.003) & .109 (.174) & .059 (.041) & .146 (.123) \\ 
   \hline
  \multirow{16}{*}{Gamma} & \multirow{4}{*}{0} &  49 & .123 (.101) & -- & -- & .158 (.123) & .123 (.101) & .156 (.120) & .157 (.122) & \textbf{.110 (.091)} & .115 (.090) \\ 
   &  &  81 & .089 (.072) & -- & -- & .113 (.086) & .090 (.072) & .113 (.086) & .113 (.086) & .090 (.068) & \textbf{.086 (.065)} \\ 
   &  & 144 & \textbf{.066 (.053)} & -- & -- & .085 (.066) & \textbf{.066 (.054)} & .084 (.066) & .085 (.066) & .081 (.056) & .078 (.054) \\ 
   &  & 400 & \textbf{.041 (.030)} & -- & -- & .051 (.037) & \textbf{.041 (.031)} & .052 (.037) & .051 (.037) & .044 (.033) & .043 (.033) \\ 
   & \multirow{4}{*}{.25} &  49 & .130 (.101) & -- & -- & .154 (.118) & .129 (.100) & .151 (.119) & .155 (.119) & .127 (.097) & \textbf{.121 (.090)} \\ 
   &  &  81 & .091 (.069) & -- & -- & .117 (.090) & \textbf{.090 (.071)} & .115 (.089) & .116 (.089) & .098 (.080) & .093 (.074) \\ 
   &  & 144 & \textbf{.067 (.055)} & -- & -- & .085 (.065) & \textbf{.067 (.056)} & .086 (.066) & .085 (.065) & .073 (.055) & .074 (.055) \\ 
   &  & 400 & \textbf{.041 (.030)} & -- & -- & .053 (.040) & \textbf{.041 (.030)} & .053 (.040) & .053 (.040) & \textbf{.041 (.033)} & \textbf{.041 (.032)} \\ 
   & \multirow{4}{*}{.5} &  49 & \textbf{.141 (.106)} & -- & -- & .155 (.116) & .142 (.109) & .162 (.122) & .152 (.118) & .155 (.122) & .146 (.115) \\ 
   &  &  81 & \textbf{.099 (.075)} & -- & -- & .112 (.088) & .104 (.080) & .118 (.093) & .112 (.087) & .112 (.081) & .113 (.082) \\ 
   &  & 144 & \textbf{.074 (.058)} & -- & -- & .086 (.068) & .078 (.063) & .094 (.072) & .087 (.067) & .081 (.062) & .081 (.061) \\ 
   &  & 400 & \textbf{.042 (.031)} & -- & -- & .048 (.036) & .049 (.036) & .055 (.041) & .049 (.037) & .058 (.044) & .058 (.044) \\ 
   & \multirow{4}{*}{.75} &  49 & .143 (.115) & -- & -- & \textbf{.141 (.113)} & .207 (.778) & .167 (.127) & .148 (.119) & .208 (.155) & .204 (.161) \\ 
   &  &  81 & .152 (1.173) & -- & -- & \textbf{.109 (.087)} & .143 (.114) & .133 (.106) & .113 (.091) & .148 (.126) & .157 (.132) \\ 
   &  & 144 & \textbf{.080 (.064)} & -- & -- & \textbf{.080 (.059)} & .118 (.093) & .102 (.078) & .085 (.064) & .124 (.100) & .125 (.102) \\ 
   &  & 400 & .048 (.035) & -- & -- & \textbf{.046 (.035)} & .092 (.066) & .071 (.053) & .056 (.044) & .091 (.065) & .090 (.066) \\ 
   \hline
  \multirow{16}{*}{Poisson} & \multirow{4}{*}{0} &  49 & .080 (.067) & --& .078 (.060) & .204 (.075) & .079 (.066) & .201 (.073) & .203 (.075) & .075 (.054) & \textbf{.071 (.049)} \\ 
   &  &  81 & .058 (.043)& -- & .056 (.044) & .197 (.058) & .058 (.044) & .197 (.058) & .197 (.058) & .053 (.046) & \textbf{.050 (.040)} \\ 
   &  & 144 & \textbf{.040 (.030)} & --& \textbf{.040 (.028)} & .199 (.040) & \textbf{.040 (.030)} & .198 (.040) & .199 (.040) & .046 (.032) & \textbf{.040 (.029)} \\ 
   &  & 400 & \textbf{.021 (.017)}& -- & .022 (.016) & .195 (.026) & .022 (.017) & .195 (.026) & .195 (.026) & .022 (.017) & .022 (.017) \\ 
   & \multirow{4}{*}{.25} &  49 & .080 (.064)& -- & .076 (.059) & .196 (.076) & .083 (.064) & .194 (.075) & .195 (.075) & .080 (.071) & \textbf{.073 (.060)} \\ 
   &  &  81 & .058 (.047) & --& \textbf{.054 (.044)} & .198 (.060) & .059 (.047) & .195 (.061) & .197 (.060) & .058 (.044) & .055 (.042) \\ 
   &  & 144 & .042 (.033)& -- & \textbf{.041 (.033)} & .194 (.045) & .043 (.034) & .191 (.046) & .193 (.045) & .046 (.035) & .044 (.035) \\ 
   &  & 400 & .025 (.019)& -- & \textbf{.023 (.017)} & .195 (.029) & .026 (.019) & .193 (.029) & .195 (.029) & .025 (.017) & .025 (.017) \\ 
   & \multirow{4}{*}{.5} &  49 & .076 (.064)& -- & \textbf{.072 (.062)} & .185 (.080) & .089 (.071) & .177 (.082) & .179 (.079) & .142 (.641) & .088 (.069) \\ 
   &  &  81 & .056 (.046) & --& \textbf{.055 (.042)} & .185 (.066) & .066 (.052) & .176 (.068) & .180 (.065) & .079 (.065) & .072 (.062) \\ 
   &  & 144 & .040 (.032) & --& \textbf{.038 (.031)} & .186 (.055) & .049 (.040) & .176 (.057) & .181 (.054) & .058 (.044) & .053 (.042) \\ 
   &  & 400 & \textbf{.021 (.017)}& -- & .023 (.017) & .183 (.041) & .035 (.027) & .174 (.040) & .179 (.040) & .043 (.031) & .041 (.032) \\ 
   & \multirow{4}{*}{.75} &  49 & \textbf{.067 (.063)} & --& .103 (.082) & .161 (.095) & .130 (.116) & .143 (.093) & .149 (.090) & .397 (1.562) & .136 (.131) \\ 
   &  &  81 & \textbf{.044 (.049)} & --& .077 (.062) & .156 (.083) & .111 (.094) & .135 (.082) & .143 (.079) & .151 (.131) & .135 (.121) \\ 
   &  & 144 & \textbf{.031 (.030)} & --& .051 (.043) & .157 (.078) & .096 (.086) & .137 (.076) & .146 (.073) & .105 (.091) & .108 (.107) \\ 
   &  & 400 & \textbf{.017 (.016)}& -- & .025 (.022) & .156 (.072) & .077 (.065) & .136 (.066) & .145 (.068) & .086 (.061) & .090 (.081) \\ 
   \hline
\end{tabular}
\end{adjustbox}
\caption{$RMSE(\hat{\beta}_1)$ of the parameter $\beta_1$ across simulation scenarios. Rows are indexed by the distribution (Dist), the absolute value of the true spatial parameter ($|\rho|$), and the sample size ($n$), while columns correspond to the different models. Each cell reports the mean RMSE across replications, with the corresponding standard deviation in parentheses. Leading zeros in decimal values have been omitted for compactness.}
\label{RMSEBeta1AA}
\end{table}

\begin{table}[ht]
\footnotesize
\centering
\begin{adjustbox}{width=1.2\textwidth,center}
\begin{tabular}{llrlllllllll}
  \hline
Dist &$|\rho|$ & n & Proposed & Probit-SAR & Pois-SAR & Ro-SAR & GLM & SAR & SARAR & GAM & INLA \\ 
  \hline
\multirow{16}{*}{Binomial} & \multirow{4}{*}{0} &  49 & \textbf{.004 (.003)} & 1.240 (.756)& -- & .008 (.008) & .008 (.010) & .008 (.008) & .017 (.058) & \textbf{.004 (.003)} & \textbf{.004 (.003)} \\ 
   &  &  81 & \textbf{.003 (.002)} & 1.375 (.687)& -- & .006 (.005) & .006 (.006) & .006 (.005) & .019 (.070) & \textbf{.003 (.002)} & \textbf{.003 (.002)} \\ 
   &  & 144 & \textbf{.002 (.002)} & 2.030 (NA) & --& .005 (.004) & .005 (.004) & .005 (.004) & .026 (.091) & \textbf{.002 (.002)} & \textbf{.002 (.002)} \\ 
   &  & 400 & \textbf{.001 (.001)} & -- & -- & .003 (.002) & .003 (.003) & .003 (.002) & .060 (.141) & \textbf{.001 (.001)} & \textbf{.001 (.001)} \\ 
   & \multirow{4}{*}{.25} &  49 & \textbf{.004 (.003)} & 1.178 (.658)& -- & .008 (.008) & .009 (.008) & .008 (.008) & .014 (.048) & .064 (.032) & .035 (.025) \\ 
   &  &  81 & \textbf{.003 (.003)} & 1.525 (.753)& -- & .006 (.005) & .007 (.006) & .006 (.006) & .022 (.078) & .042 (.020) & .032 (.019) \\ 
   &  & 144 & \textbf{.002 (.002)} & 1.385 (NA)& -- & .005 (.005) & .005 (.005) & .005 (.005) & .027 (.092) & .025 (.014) & .032 (.016) \\ 
   &  & 400 & \textbf{.001 (.001)} & -- & -- & .003 (.002) & .003 (.003) & .003 (.002) & .065 (.145) & .011 (.008) & .032 (.012) \\ 
   & \multirow{4}{*}{.5} &  49 & \textbf{.005 (.006)} & 1.293 (.671) & --& .009 (.008) & .012 (.010) & .009 (.009) & .016 (.056) & .152 (.101) & .086 (.069) \\ 
   &  &  81 & \textbf{.003 (.003)} & .911 (.439)& --  & .007 (.006)& .010 (.008) & .007 (.006) & .025 (.084) & .097 (.057) & .081 (.063) \\ 
   &  & 144 & \textbf{.003 (.002)} & .759 (.325) & --& .005 (.004) & .007 (.006) & .005 (.005) & .035 (.106) & .060 (.043) & .081 (.054) \\ 
   &  & 400 & \textbf{.001 (.001)} & -- & -- & .003 (.002) & .004 (.004) & .003 (.003) & .089 (.165) & .030 (.024) & .077 (.046) \\ 
   & \multirow{4}{*}{.75} &  49 & .016 (.033) & .556 (.465) & --& \textbf{.011 (.011)} & .028 (.028) & .013 (.013) & .046 (.116) & .281 (.240) & .180 (.176) \\ 
   &  &  81 & .010 (.019) & .340 (.243)& -- & \textbf{.009 (.007)} & .022 (.019) & .011 (.010) & .060 (.138) & .168 (.141) & .171 (.167) \\ 
   &  & 144 & .007 (.011) & .345 (.207)& -- & \textbf{.006 (.005)} & .016 (.015) & .008 (.007) & .103 (.177) & .107 (.093) & .164 (.147) \\ 
   &  & 400 & \textbf{.002 (.002)} & .192 (.138)& -- & .003 (.003) & .006 (.005) & .004 (.003) & .109 (.174) & .059 (.041) & .146 (.123) \\ 
   \hline
  \multirow{16}{*}{Gamma} & \multirow{4}{*}{0} &  49 & .123 (.101) & -- & -- & .158 (.123) & .123 (.101) & .156 (.120) & .157 (.122) & \textbf{.110 (.091)} & .115 (.090) \\ 
   &  &  81 & .089 (.072) & -- & -- & .113 (.086) & .090 (.072) & .113 (.086) & .113 (.086) & .090 (.068) & \textbf{.086 (.065)} \\ 
   &  & 144 & \textbf{.066 (.053)} & -- & -- & .085 (.066) & \textbf{.066 (.054)} & .084 (.066) & .085 (.066) & .081 (.056) & .078 (.054) \\ 
   &  & 400 & \textbf{.041 (.030)} & -- & -- & .051 (.037) & \textbf{.041 (.031)} & .052 (.037) & .051 (.037) & .044 (.033) & .043 (.033) \\ 
   & \multirow{4}{*}{.25} &  49 & .130 (.101) & -- & -- & .154 (.118) & .129 (.100) & .151 (.119) & .155 (.119) & .127 (.097) & \textbf{.121 (.090)} \\ 
   &  &  81 & .091 (.069) & -- & -- & .117 (.090) & \textbf{.090 (.071)} & .115 (.089) & .116 (.089) & .098 (.080) & .093 (.074) \\ 
   &  & 144 & \textbf{.067 (.055)} & -- & -- & .085 (.065) & \textbf{.067 (.056)} & .086 (.066) & .085 (.065) & .073 (.055) & .074 (.055) \\ 
   &  & 400 & \textbf{.041 (.030)} & -- & -- & .053 (.040) & \textbf{.041 (.030)} & .053 (.040) & .053 (.040) & \textbf{.041 (.033)} & \textbf{.041 (.032)} \\ 
   & \multirow{4}{*}{.5} &  49 & \textbf{.141 (.106)} & -- & -- & .155 (.116) & .142 (.109) & .162 (.122) & .152 (.118) & .155 (.122) & .146 (.115) \\ 
   &  &  81 & \textbf{.099 (.075)} & -- & -- & .112 (.088) & .104 (.080) & .118 (.093) & .112 (.087) & .112 (.081) & .113 (.082) \\ 
   &  & 144 & \textbf{.074 (.058)} & -- & -- & .086 (.068) & .078 (.063) & .094 (.072) & .087 (.067) & .081 (.062) & .081 (.061) \\ 
   &  & 400 & \textbf{.042 (.031)} & -- & -- & .048 (.036) & .049 (.036) & .055 (.041) & .049 (.037) & .058 (.044) & .058 (.044) \\ 
   & \multirow{4}{*}{.75} &  49 & .143 (.115) & -- & -- & \textbf{.141 (.113)} & .207 (.778) & .167 (.127) & .148 (.119) & .208 (.155) & .204 (.161) \\ 
   &  &  81 & .152 (1.173) & -- & -- & \textbf{.109 (.087)} & .143 (.114) & .133 (.106) & .113 (.091) & .148 (.126) & .157 (.132) \\ 
   &  & 144 & \textbf{.080 (.064)} & -- & -- & \textbf{.080 (.059)} & .118 (.093) & .102 (.078) & .085 (.064) & .124 (.100) & .125 (.102) \\ 
   &  & 400 & .048 (.035) & -- & -- & \textbf{.046 (.035)} & .092 (.066) & .071 (.053) & .056 (.044) & .091 (.065) & .090 (.066) \\ 
   \hline
  \multirow{16}{*}{Poisson} & \multirow{4}{*}{0} &  49 & .080 (.067)& --  & .078 (.060) & .204 (.075) & .079 (.066) & .201 (.073) & .203 (.075) & .075 (.054) & \textbf{.071 (.049)} \\ 
   &  &  81 & .058 (.043) & -- & .056 (.044) & .197 (.058) & .058 (.044) & .197 (.058) & .197 (.058) & .053 (.046) & \textbf{.050 (.040)} \\ 
   &  & 144 & \textbf{.040 (.030)}& --  & \textbf{.040 (.028)} & .199 (.040) & \textbf{.040 (.030)} & .198 (.040) & .199 (.040) & .046 (.032) & \textbf{.040 (.029)} \\ 
   &  & 400 & \textbf{.021 (.017)}& --  & .022 (.016) & .195 (.026) & .022 (.017) & .195 (.026) & .195 (.026) & .022 (.017) & .022 (.017) \\ 
   & \multirow{4}{*}{.25} &  49 & .080 (.064) & -- & .076 (.059) & .196 (.076) & .083 (.064) & .194 (.075) & .195 (.075) & .080 (.071) & \textbf{.073 (.060)} \\ 
   &  &  81 & .058 (.047)& --  & \textbf{.054 (.044)} & .198 (.060) & .059 (.047) & .195 (.061) & .197 (.060) & .058 (.044) & .055 (.042) \\ 
   &  & 144 & .042 (.033)& --  & \textbf{.041 (.033)} & .194 (.045) & .043 (.034) & .191 (.046) & .193 (.045) & .046 (.035) & .044 (.035) \\ 
   &  & 400 & .025 (.019) & -- & \textbf{.023 (.017)} & .195 (.029) & .026 (.019) & .193 (.029) & .195 (.029) & .025 (.017) & .025 (.017) \\ 
   & \multirow{4}{*}{.5} &  49 & .076 (.064) & -- & \textbf{.072 (.062)} & .185 (.080) & .089 (.071) & .177 (.082) & .179 (.079) & .142 (.641) & .088 (.069) \\ 
   &  &  81 & .056 (.046) & -- & \textbf{.055 (.042)} & .185 (.066) & .066 (.052) & .176 (.068) & .180 (.065) & .079 (.065) & .072 (.062) \\ 
   &  & 144 & .040 (.032) & -- & \textbf{.038 (.031)} & .186 (.055) & .049 (.040) & .176 (.057) & .181 (.054) & .058 (.044) & .053 (.042) \\ 
   &  & 400 & \textbf{.021 (.017)} & -- & .023 (.017) & .183 (.041) & .035 (.027) & .174 (.040) & .179 (.040) & .043 (.031) & .041 (.032) \\ 
   & \multirow{4}{*}{.75} &  49 & \textbf{.067 (.063)} & -- & .103 (.082) & .161 (.095) & .130 (.116) & .143 (.093) & .149 (.090) & .397 (1.562) & .136 (.131) \\ 
   &  &  81 & \textbf{.044 (.049)} & -- & .077 (.062) & .156 (.083) & .111 (.094) & .135 (.082) & .143 (.079) & .151 (.131) & .135 (.121) \\ 
   &  & 144 & \textbf{.031 (.030)} & -- & .051 (.043) & .157 (.078) & .096 (.086) & .137 (.076) & .146 (.073) & .105 (.091) & .108 (.107) \\ 
   &  & 400 & \textbf{.017 (.016)} & -- & .025 (.022) & .156 (.072) & .077 (.065) & .136 (.066) & .145 (.068) & .086 (.061) & .090 (.081) \\ 
   \hline
\end{tabular}
\end{adjustbox}
\caption{$RMSE(\hat{\beta}_2)$ of the parameter $\beta_1$ across simulation scenarios. Rows are indexed by the distribution (Dist), the absolute value of the true spatial parameter ($|\rho|$), and the sample size ($n$), while columns correspond to the different models. Each cell reports the mean RMSE across replications, with the corresponding standard deviation in parentheses. Leading zeros in decimal values have been omitted for compactness.}
\label{RMSEBeta2AA}
\end{table}

\begin{table}[ht] 
\centering 
\footnotesize 
\begin{adjustbox}{width=1.2\textwidth,center}
\begin{tabular}{lccc}
\hline
Variable & GSCIMC & GAM & INLA\\ 
\hline (Intercept) & $2.132\,(2.22\times10^{-1})^{***}$ & $1.984\,(1.34\times10^{-1})^{***}$ & $1.018\,(1.07\times10^{-1})^{***}$ \\  
Civilian labor force & $-4.658\,(3.48\times10^{-1})^{***}$ & $-4.060\,(2.02\times10^{-1})^{***}$ & $-7.15\times10^{-1}\,(1.46\times10^{-1})^{***}$ \\  
Unemployment rate & $-9.87\times10^{-2}\,(5.90\times10^{-3})^{***}$ & $-7.84\times10^{-2}\,(4.41\times10^{-3})^{***}$ & $-1.07\times10^{-1}\,(5.57\times10^{-3})^{***}$ \\ 
PorcLess & $1.98\times10^{-2}\,(3.50\times10^{-3})^{***}$ & $1.67\times10^{-3}\,(2.45\times10^{-3})$ & $1.45\times10^{-2}\,(2.01\times10^{-3})^{***}$ \\  
PorcHigh & $4.73\times10^{-2}\,(1.90\times10^{-3})^{***}$ & $6.08\times10^{-2}\,(1.30\times10^{-3})^{***}$ & $4.34\times10^{-2}\,(1.33\times10^{-3})^{***}$ \\ 
Birth Rate & $4.00\times10^{-4}\,(3.00\times10^{-4})$ & $-1.02\times10^{-4}\,(1.81\times10^{-4})$ & $-2.44\times10^{-4}\,(1.03\times10^{-4})^{*}$ \\  
PorcMigInt & $-7.70\times10^{-3}\,(1.90\times10^{-3})^{***}$ & $-8.86\times10^{-3}\,(1.33\times10^{-3})^{***}$ & $-7.94\times10^{-3}\,(1.58\times10^{-3})^{***}$ \\  
Poverty & $-5.874\,(4.46\times10^{-1})^{***}$ & $-7.053\,(2.49\times10^{-1})^{***}$ & $-5.565\,(2.40\times10^{-1})^{***}$ \\
\hline 
\end{tabular}
\end{adjustbox}
\caption{Comparison of regression coefficient estimates across the GSCIMC, GAM, and INLA models. Entries are reported as estimate (standard error). For INLA, posterior means and posterior standard deviations are shown. $^{***}p<0.001$, $^{**}p<0.01$, $^{*}p<0.05$.}
\label{tab:coef_compare} 
\end{table}
\noindent

\clearpage
\section{Complementary Figures}\label{apen0}

\begin{figure}[!ht]
\centering
\includegraphics[width=14cm]{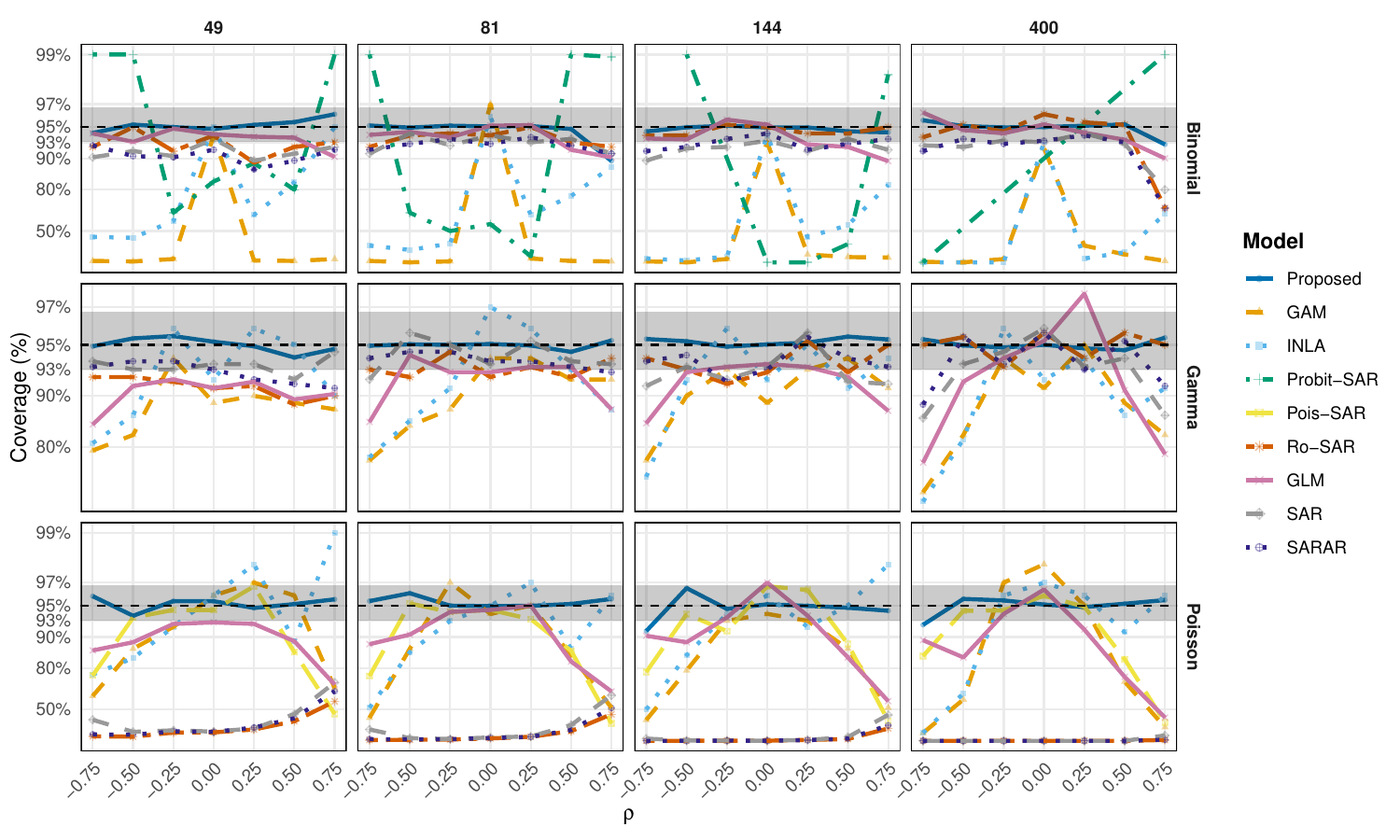}
\caption{Coverage rates of 95\% confidence intervals for the estimated parameter $\hat{\beta}_1$ (y-axis), across true $\rho$ values (x-axis), by model, distribution (rows), and sample size (columns). The dashed horizontal line indicates the nominal 95\% level. Shaded bands represent the approximate confidence interval for a binomial proportion with $0.95$ and $500$ replications. Additionally, the y-axis is transformed using a log-scale of the complement, $-\log(1 - y)$, to enhance the visualization of deviations from the nominal level.}
\label{coverbeta1}
\end{figure}

\begin{figure}[!ht]
\centering
\includegraphics[width=12cm]{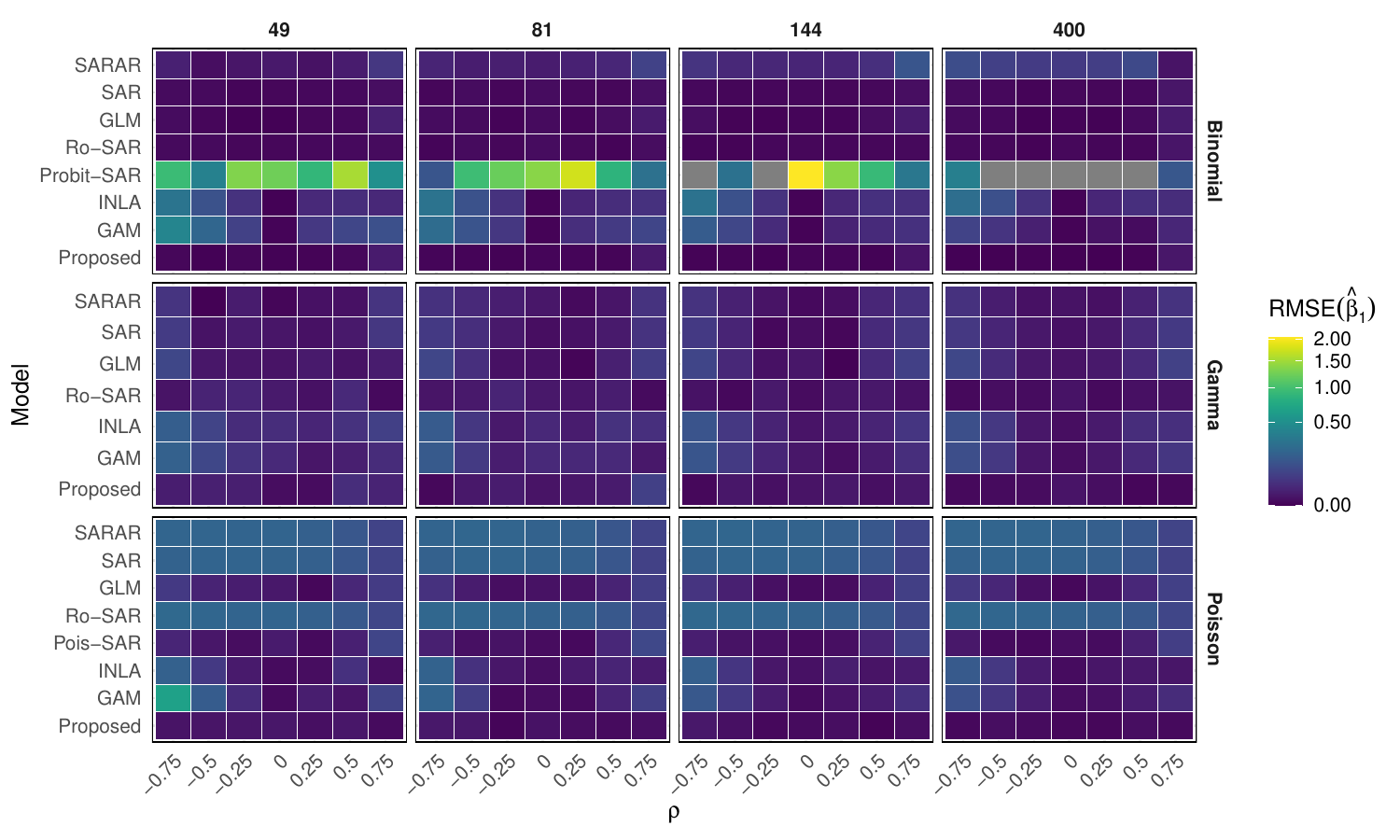}
\caption{Heatmap of RMSE (log scale) for the estimated parameter $\hat{\beta}_1$ at different true $\rho$ values (x-axis), by model (y-axis), distribution (rows), and sample size (columns). }
\label{rmsebeta1}
\end{figure}
\begin{figure}[!ht]
\centering
\includegraphics[width=12cm]{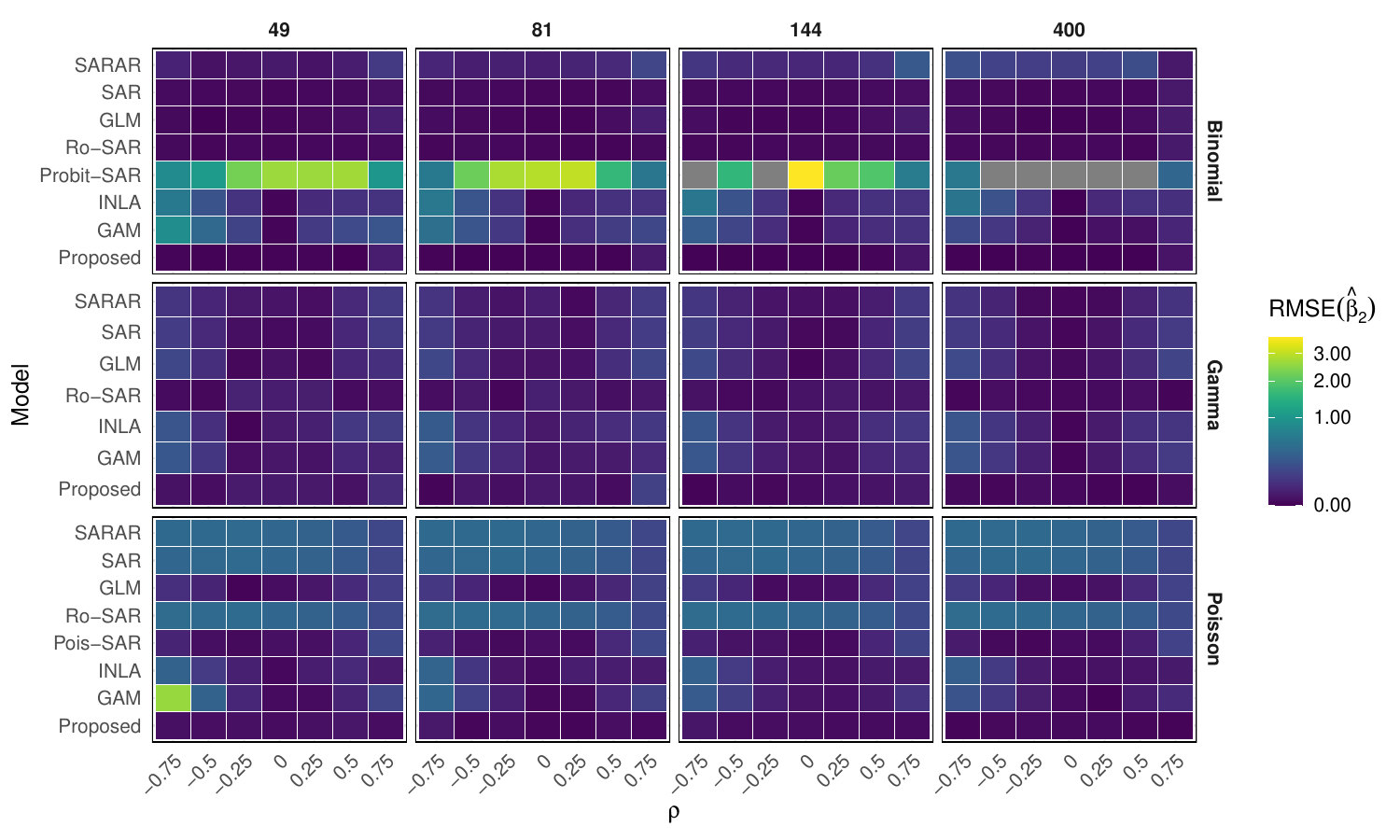}
\caption{Heatmap of RMSE (log scale) for the estimated parameter $\hat{\beta}_2$ at different true $\rho$ values (x-axis), by model (y-axis), distribution (rows), and sample size (columns). }
\label{rmsebeta2}
\end{figure}

\begin{figure}[!ht]
\centering
\includegraphics[width=14cm]{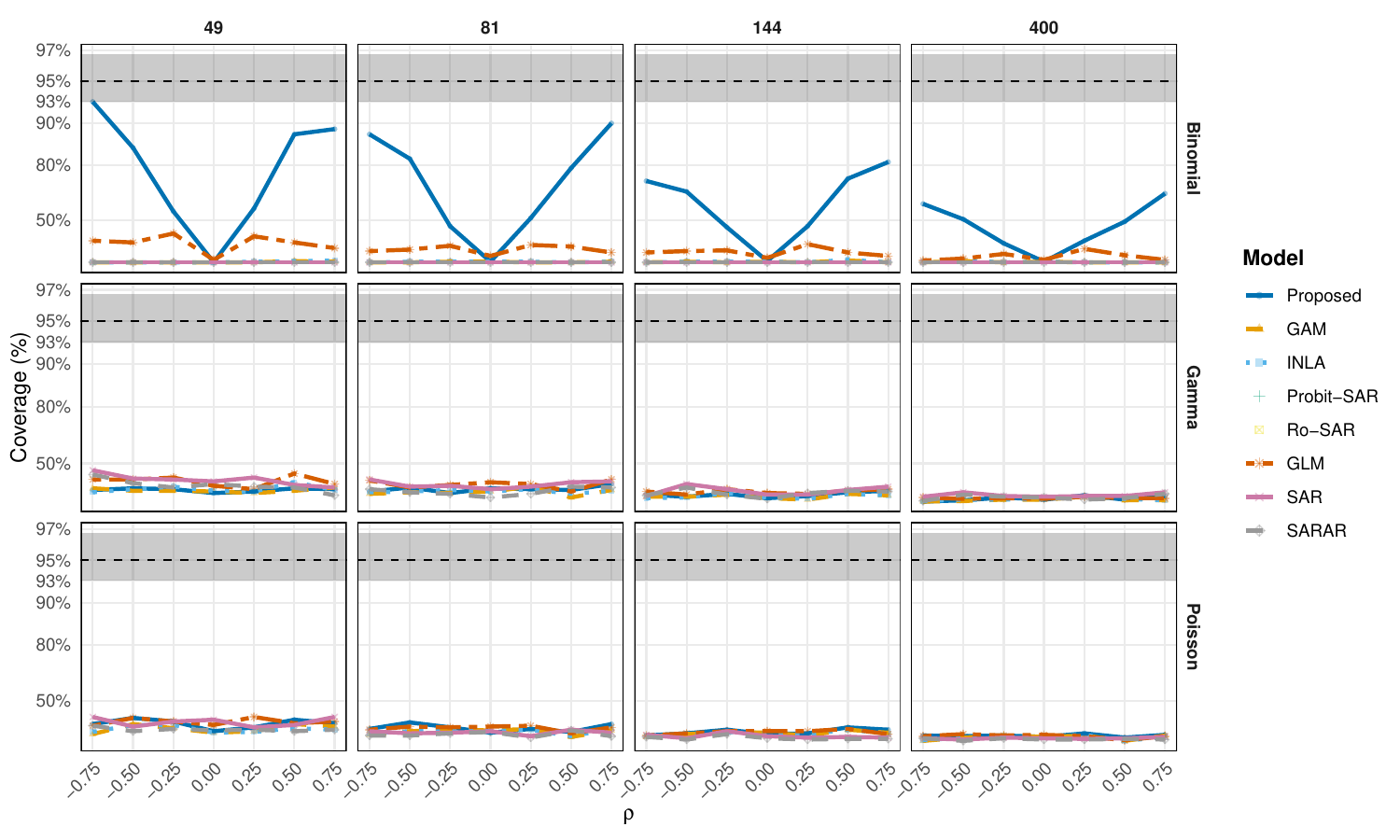}
\caption{Coverage rates of 95\% confidence intervals for the estimated parameter $\hat{\beta}_0$ (y-axis) under model misspecification (Equation \eqref{eq:miss}), across true $\rho$ values (x-axis), by model, distribution (rows), and sample size (columns). The dashed horizontal line indicates the nominal 95\% level. Shaded bands represent the approximate confidence interval for a binomial proportion with $0.95$ and $500$ replications. Additionally, the y-axis is transformed using a log-scale of the complement, $-\log(1 - y)$, to enhance the visualization of deviations from the nominal level.}
\label{coverbeta02}
\end{figure}

\begin{figure}[!ht]
\centering
\includegraphics[width=14cm]{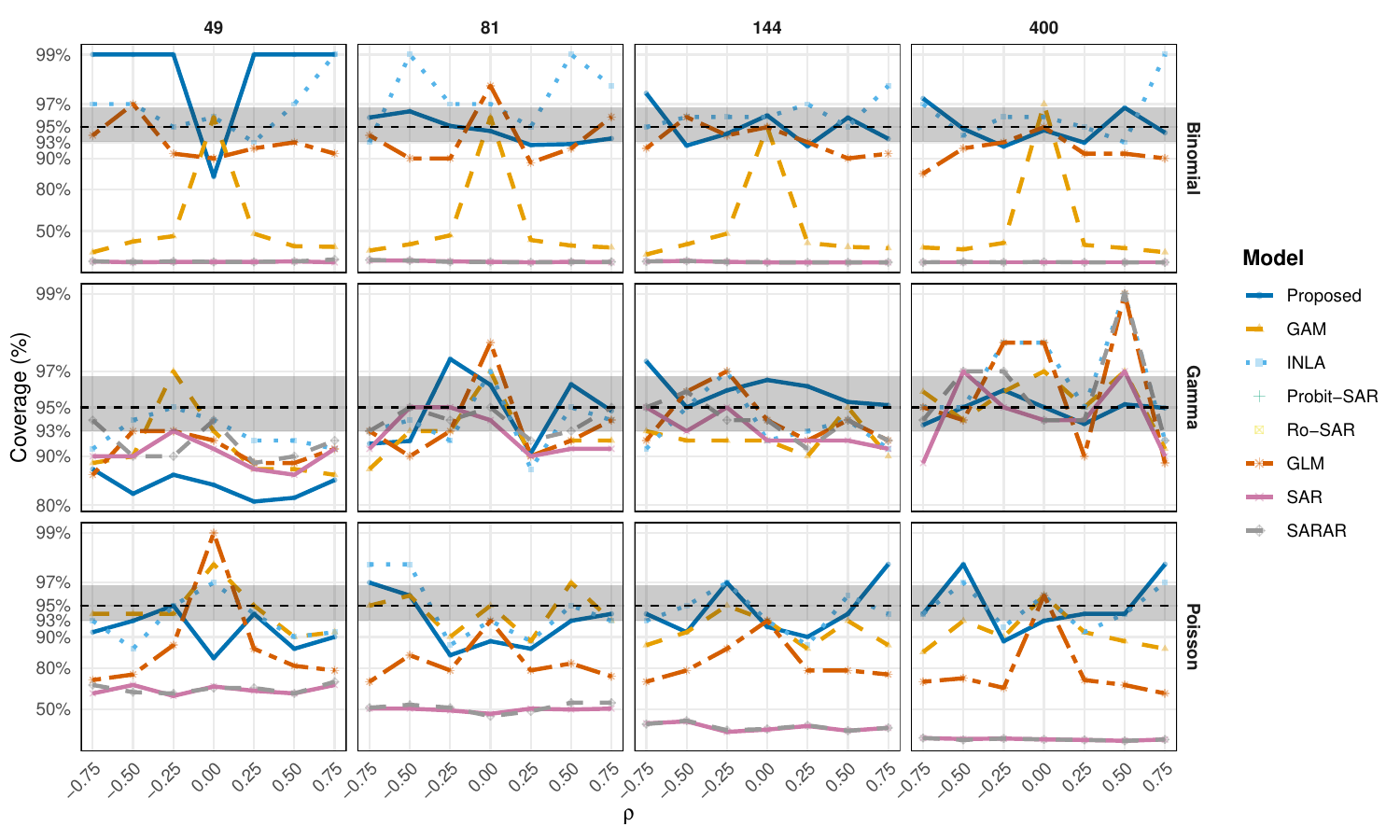}
\caption{Coverage rates of 95\% confidence intervals for the estimated parameter $\hat{\beta}_1$ (y-axis) under model misspecification (Equation \eqref{eq:miss}), across true $\rho$ values (x-axis), by model, distribution (rows), and sample size (columns). The dashed horizontal line indicates the nominal 95\% level. Shaded bands represent the approximate confidence interval for a binomial proportion with $0.95$ and $500$ replications. Additionally, the y-axis is transformed using a log-scale of the complement, $-\log(1 - y)$, to enhance the visualization of deviations from the nominal level.}
\label{coverbeta01}
\end{figure}

\clearpage
\section{Proofs}
The proofs of Theorem and Corollaries rely on the asymptotic theory of GEE developed by \cite{liang1986longitudinal} and the robust covariance results of \cite{pan2001robust}. For completeness, we restate these results below using the notation adopted throughout this paper and combine them with Theorem~\ref{te1}.
\begin{theorem}[Liang and Zeger, 1986]\label{thm:lz1}
Consider the estimating equation
\[
\mathbf{U}(\pmb{\beta})
=
\tilde{\mathbf X}^{\top}
\mathbf{V}(\pmb{\mu})^{-1/2}
\mathbf{R}^{-1}
\mathbf{V}(\pmb{\mu})^{-1/2}
(\mathbf Y-\pmb{\mu}),
\]
where $\mathbf{R}$ is a working correlation matrix. Under Assumptions
\ref{as:gmu}--\ref{as:lee}, if the marginal mean model is correctly specified,
then the solution $\hat{\pmb{\beta}}$ of
\[
\mathbf U(\pmb{\beta})=\mathbf 0
\]
is consistent and asymptotically normal, satisfying
\[
\sqrt n
\left(
\hat{\pmb{\beta}}-\pmb{\beta}
\right)
\xrightarrow{d}
N\!\left(
0,
\mathbf J^{-1}\mathbf B\mathbf J^{-1}
\right).
\]
Furthermore, among estimating equations sharing the same marginal mean
specification, working covariance structures that better approximate the true
dependence pattern lead to asymptotic covariance matrices that are no larger
than those obtained under a working independence specification.
\end{theorem}
\begin{proof}
    See Theorem 1 of \citet[pp.~15]{liang1986longitudinal}
\end{proof}
\begin{theorem}[Adapted from Theorem 2 of Liang and Zeger (1986)]
\label{thm:lz2}

Under the assumptions of Theorem~\ref{thm:lz1},

\[
\sqrt n
\left(
\hat{\pmb\beta}-\pmb\beta
\right)
\xrightarrow{d}
N\Big(
0,
\mathbf J^{-1}\mathbf B\mathbf J^{-1}
\Big).
\]

Furthermore, among estimating equations sharing the same marginal mean specification, working covariance structures that more closely approximate the true dependence structure yield asymptotic covariance matrices that are no larger than those obtained under a working independence specification.
\end{theorem}
\begin{proof}
  See Theorem 2 of \citet[pp.~15]{liang1986longitudinal}   
\end{proof}
\begin{theorem}[Pan, 2001]\label{thm:Pan}
Under the conditions of Theorem~\ref{thm:lz1}, the sandwich estimator
\[
\widehat{\mathrm{Var}}
\left(
\hat{\pmb{\beta}}
\right)
=
\hat{\mathbf J}^{-1}
\hat{\mathbf B}
\hat{\mathbf J}^{-1}
\]
is a consistent estimator of the asymptotic covariance matrix of
$\hat{\pmb{\beta}}$, even when the working covariance structure is misspecified.
Consequently, valid large-sample inference may be obtained without requiring
the covariance matrix in Equation~\eqref{eq:covariance} to coincide with the
true covariance of the data-generating process.
\end{theorem}
\begin{proof}
  See  Theorem 2 of \citet[pp.~903]{pan2001robust}
\end{proof}

\begin{theorem}[Adapted from Nie (2006)]
\label{thm:nie}

Suppose the objective function
$\ell(\rho,\pmb{\beta})$
is twice continuously differentiable, identifiable, and satisfies the usual
regularity conditions. Then the maximizer

\[
\hat{\rho}
=
\arg\max_{\rho}
\ell(\rho,\hat{\pmb{\beta}})
\]

is a consistent estimator of $\rho$.

\end{theorem}
\begin{proof}
 See Theorem 2.1 of \citet[pp.~127]{nie2006strong}
\end{proof}
\begin{theorem}[Adapted from Yin et al. (2006)]
\label{thm:yin}

Consider the quasi-likelihood estimating equation

\[
\mathbf{U}(\pmb{\theta})
=
\sum_{i=1}^{n}
\frac{\partial \ell_i(\pmb{\theta})}
{\partial \pmb{\theta}}
=
\mathbf 0,
\]

where $\pmb{\theta}$ denotes a finite-dimensional parameter vector.
Suppose that:

\begin{enumerate}
\item the quasi-likelihood is sufficiently smooth;
\item the mean specification is correctly specified;
\item the information matrix is positive definite;
\item the covariates and moments satisfy the regularity conditions required for generalized linear models.
\end{enumerate}

Then there exists an estimator $\hat{\pmb{\theta}}$
satisfying

\[
\mathbf U(\hat{\pmb{\theta}})=0,
\]

such that

\[
\hat{\pmb{\theta}}
\xrightarrow[]{p}
\pmb{\theta},
\]

and

\[
\sqrt n
\left(
\hat{\pmb{\theta}}
-
\pmb{\theta}
\right)
\xrightarrow[]{d}
N
\left(
0,
\mathbf I(\pmb{\theta})^{-1}
\right),
\]

where $\mathbf I(\pmb{\theta})$ denotes the asymptotic information matrix.
In particular,
Let

\[
\hat{\rho}
=
\arg\max_{\rho}
\ell(\rho,\hat{\pmb{\beta}},
\mathbf Y_n,\mathbf X),
\]

where $\ell(\cdot)$ is the quasi-likelihood defined in
Equation~\eqref{eq:quasilikelihood}.

Under Assumptions
\ref{as:gmu}--\ref{as:lee},
there exists a solution $\hat{\rho}$ such that

\[
\hat{\rho}
\xrightarrow[]{p}
\rho,
\]

and

\[
\sqrt n
(\hat{\rho}-\rho)
\xrightarrow[]{d}
N
\left(
0,
d_n^{-1}
\right),
\]

where

\[
d_n
=
-\sum_{i=1}^{n}
\left(
\frac{\partial \mu_i}{\partial \rho}
\right)^2
E
\left[
\frac{\partial^2 \ell_i}
{\partial \mu_i^2}
\right].
\]

\end{theorem}
\begin{proof}
    See \cite{yin2006asymptotic}
\end{proof}
\begin{theorem}[Multivariate Delta Method]
\label{thm:delta}

Let

\[
\sqrt n
(\hat{\pmb{\theta}}-\pmb{\theta})
\xrightarrow{d}
\mathcal{N}(0,\pmb{\Sigma}).
\]

If $h(\cdot)$ is continuously differentiable, then

\[
\sqrt n
\big(
h(\hat{\pmb{\theta}})-h(\pmb{\theta})
\big)
\xrightarrow{d}
\mathcal{N}
\left(
0,
\mathbf{D}h(\pmb{\theta})\pmb{\Sigma} \mathbf{D}h(\pmb{\theta})^\top
\right).
\]
\end{theorem}
\begin{proof}
     See Theorem 3.9.4 of \citet[pp.~374]{van1996weak}
\end{proof}
\begin{theorem}[Mertens]
\label{thm:mertens}
Let
\[
\sum_{j=0}^{\infty} \mathbf{B}_j
\qquad\text{and}\qquad
\sum_{j=0}^{\infty} \mathbf{C}_j
\]
be two absolutely convergent matrix series. Then their Cauchy product
\[
\sum_{j=0}^{\infty}
\left(
\sum_{k=0}^{j}
\mathbf{B}_k \mathbf{C}_{j-k}
\right)
\]
is also absolutely convergent and converges to
\[
\left(\sum_{j=0}^{\infty}\mathbf{B}_j\right)
\left(\sum_{j=0}^{\infty}\mathbf{C}_j\right).
\]
\end{theorem}
\begin{proof}
    See Theorem 8.46 of \cite[p.~204]{apostol_analysis}
\end{proof}

\begin{theorem}[Adapted from Theorem 5.8 of Casella and Berger (2002)]
\label{thm:casella}

Let $Y$ belong to a regular exponential family with density or mass function

\[
f(y;\theta)
=
\exp\{y\theta-a(\theta)+b(y)\},
\]

where the natural parameter space is an open subset of
$\mathbb{R}$.

Then:

\begin{enumerate}
\item The log-likelihood function is continuously differentiable with respect to $\theta$.

\item The first and second derivatives of the log-likelihood exist and are continuous.

\item The expectation and variance satisfy

\[
E(Y)=a'(\theta),
\qquad
Var(Y)=a''(\theta).
\]

\item If $a(\theta)$ is sufficiently smooth, all higher-order derivatives required for asymptotic likelihood theory exist.

\item The Fisher information

\[
I(\theta)
=
E\!\left[
-\frac{\partial^2}{\partial\theta^2}
\log f(Y;\theta)
\right]
=
a''(\theta)
\]

is finite and positive.
\end{enumerate}

\end{theorem}
\begin{proof}
    See Theorem 5.8 of \cite[pp.~27]{casella2002point}
\end{proof}
\subsection{Proof of theorem \ref{te1}}\label{proofte1}
\begin{proof}
Let $\rho \in (-1,1)$ and assume that $\mathbf{W}$ is row-normalized, that is, $\mathbf{W}\mathbf{1}_n = \mathbf{1}_n$. Under Assumption~\ref{as::w1}, the series:
\begin{equation*}
 \mathbf{A}^{-1}=\sum _{j=0}^{\infty} \rho^j\mathbf{W}^j
\end{equation*}
is absolutely convergent. 
Then, consider the series
\[
\sum_{j=0}^{\infty} \rho^j \mathbf{W}^{j+1}= \mathbf{W} \sum_{j=0}^{\infty} \rho^j \mathbf{W}^j=\mathbf{W}\mathbf{A}^{-1}.
\]
and, using a infinite norm $\|\cdot\|_{\infty}$ and Cauchy-Schwarz inequality, we have
\[
\sum_{j=0}^{\infty} \|\rho^j \mathbf{W}^{j+1}\|_{\infty}
\leq \|\mathbf{W}\|_{\infty} \sum_{j=0}^{\infty} \|\rho^j \mathbf{W}^j\|_\infty = 1 \sum_{j=0}^\infty |\rho^j| = \frac{1}{1-|\rho|}< \infty,
\]
which shows that the $\mathbf{WA}^{-1}$ is also absolutely convergent.
By Mertens' theorem, which guarantees the convergence of the product of convergent series, (See Theorem 8.46 of \cite[p.~204]{apostol_analysis}) the following series are absolutely convergent:
\begin{align*}
 \mathbf{WA}^{-1}\times \cdots \times \mathbf{WA}^{-1}&=\mathbf{W}\left(\sum _{j=0}^{\infty} \rho ^j\mathbf{W}^j\right)\times \cdots \times \mathbf{W}\left(\sum _{j=0}^{\infty} \rho^j\mathbf{W}^j\right).
\end{align*}
Furthermore, by properties of derivative of convergent series, we obtain that:
\begin{equation*}
 \frac{\partial\mathbf{A}^{-1}}{\partial\rho} = \sum _{j=1}^{\infty} j\rho^{j-1}\mathbf{W}^j
\end{equation*}
Additionally, using the properties of derivative of a inverse matrix, we obtain that:
\begin{equation*}
 \frac{\partial\mathbf{A}^{-1}}{\partial\rho} =-\mathbf{A}^{-1}\frac{\partial\mathbf{A}}{\partial\rho}\mathbf{A}^{-1} =\mathbf{A}^{-1}\mathbf{WA}^{-1} =\left(\sum_{ j=0}^{\infty} \rho^j\mathbf{W}^j\right) \mathbf{W} \left(\sum _{j=0}^{\infty} \rho^j\mathbf{W}^j\right)
\end{equation*}
Therefore, $\frac{\partial\mathbf{A}^{-1}}{\partial\rho}$ is also absolutely convergent.

Under Assumptions~\ref{as:gmu}-\ref{as::w1}, we obtain that the derivative of $g$ with respect to $\mu$ exists and is continuous. Moreover, strict monotonicity and continuous differentiability imply that $g$ is locally invertible. Hence, by the inverse function theorem, the inverse link function $g^{-1}(\cdot)$ exists and is continuously differentiable on its domain. 
By differentiability of the matrix inverse and by the chain rule, $\mu_i = g^{-1}(\eta_i)$ is also differentiable with respect to $\rho$.

All the above conditions are met to guarantee that the proposed model satisfies the conditions of Theorem 5.8 of \cite[pp.~27]{casella2002point} on exponential families and we obtain that
\begin{itemize}
    \item[R1)] The following functions are known to be continuous and $k$-times differentiable:
\begin{equation}\label{eq::ktimes}
\frac{\partial g^{-1}(\eta_i)}{\partial\eta_i}, \; \frac{\partial^2 g^{- 1}(\eta_i)}{\partial\eta_i^2},\; \frac{\partial g(\mu_i)}{\partial\eta_i} \text{, and } \; \frac{\partial^2 g(\mu_i)}{\partial\eta_i^2}
\end{equation}
\end{itemize}
Thus, using R1) it is concluded that the functions in \eqref{deta} and \eqref{deta3} are continuous and $k$-times differentiable for all $\rho \in (-1,1)$. The value of $k$ is determined by the functional form of $g( \cdot)$ in Equation \eqref{eq:mu}. 

Theorem 2.1 of \citet[pp.~127]{nie2006strong} guarantees that only by using a usual maximum likelihood estimator can obtained consistently estimators of the parameters involved in the mean. Therefore, we will first show that the conditions for applying this theorem are satisfied.
\begin{itemize}
    \item[R2)] Using Theorem 5.8 of \cite[pp.~27]{casella2002point}, Equation \eqref{eq::ktimes}, and the fact of $Y_i$ belongs to exponential family, we obtain that 
    \begin{align*}
    f_{Y_i}(y_i) =\exp\{y_i\mu_i - a(\mu_i)+b(y_i)\} 
\end{align*}
admits all first, second and third derivatives for values of $\rho$, $\phi$ and $\pmb{\beta}$.
\item[R3)] Under Assumptions~\ref{as:gmu}-\ref{as::xp} and a), $\mbox{Var}(\hat{\rho})$ defined in Equation \eqref{eq:varrho1} satisfies that:
\begin{align*}
    0< \mbox{Var}(\hat{\rho}) <\infty
\end{align*}
for all $n$.
\item[R4)] Let the matrices $\tilde{\mathbf{J}}=\tilde{\mathbf{X}}^{T} \mathbf{V}(\pmb{\mu})^{-\frac{1}{2}} \mathbf{V}(\pmb{\mu})^{-\frac{1}{2}}\tilde{\mathbf{X}}$, $\tilde{\mathbf{B}}= \mathbf{N}^{\top}\mathbf{V}(\pmb{\mu})^{-\frac{1}{2}}\mathbf{V}(\pmb{\mu})^{-\frac{1}{2}} \mathbf{N}$, $\mathbf{N} =\mbox{diag}\left\{\frac{\partial \mu_1}{\partial \eta_1}, \ldots, \frac{\partial \mu_n}{\partial \eta_n} \right\}$, and
$\lambda_n(\tilde{\mathbf{J}})$ the smallest eigenvalue of $\tilde{\mathbf{J}}$. Under Assumptions~\ref{as:gmu}-\ref{as::xp}, we obtain that:
\begin{align*}
0&<\lambda_n(\tilde{\mathbf{J}})<\infty\\
    0&<\liminf_{n\to \infty}\lambda_n(\tilde{\mathbf{J}})<\infty
\end{align*}
and hence, $\tilde{\mathbf{J}}^{-1} \tilde{\mathbf{B}} \tilde{\mathbf{J}}^{-1}$ exists.
\end{itemize}
R1), R2) R3), and R4) allow using the Theorem 2.1 of \citet[pp.~127]{nie2006strong} over the estimator of $\rho$ and obtain that:
\begin{itemize}
    \item[R5)] $P\left(\lim_{n\to \infty}\hat {\rho}=\rho\right) = 1$.
\end{itemize}
This means that if we estimate $\rho$  using an estimation equation with an independent correlation matrix, the estimator converge asymptotically to the true value of the parameter $\rho$. This makes our estimators behave analogously to what was demonstrated by Theorem 1 of \citet[pp.~15]{liang1986longitudinal}.
The principal disadvantage of $\hat {\rho}$ and, consequently of $\hat{\pmb{\beta}}$ is that it may not have high efficiency in cases where the autocorrelation is large, i.e, $\rho$ is large. 

Now, in order to apply Theorem 2 of \citet[pp.~15]{liang1986longitudinal}, we must now demonstrate that our proposed correlation structure approximates the true correlation structure of the variables.
Under Assumptions~\ref{as:gmu}-\ref{as::xp}, we obtain that:
\begin{enumerate}
    \item[R6)] Under Assumption~\ref{as::xp}, it is obtain that:
    \begin{align*}
        \frac{1}{n}\mathbf{X}^\top\mathbf{X}
    \end{align*}
    is positive definite for each $n$.
    \item[R7)]  Let
    \begin{align*}
  d_n = -\sum_{i=1}^{n}\left(\frac{\partial\mu_i}{\partial\rho}\right)^2 \mathbb{E}\left[\frac{\partial^2\ell_i(\rho, \hat{\pmb{\beta}}, Y_i, \mathbf{X}, \mathbf{W})}{\partial \mu_i^2} \right] 
\end{align*}
Using R6) and Assumption~\ref{as::w1}, we obtain that  $d_n>0$.
  \item[R8)] Let 
  \begin{align}
    \mathbf{J}&=\tilde{\mathbf{X}}^{T} \mathbf{V}(\pmb{\mu})^{-\frac{1}{2}}\left( \mathbf{A}^\top\mathbf{A}\right) \mathbf{V}(\pmb{\mu})^{-\frac{1}{2}}\tilde{\mathbf{X}}\nonumber\\
    \mathbf{B} &= \mathbf{N}^{\top}\mathbf{V}(\pmb{\mu})^{-\frac{1}{2}}\left( \mathbf{Y}-\hat{\pmb{\mu}}\right) \left( \mathbf{Y}-\hat{\pmb{\mu}}\right)^\top\mathbf{V}(\pmb{\mu})^{-\frac{1}{2}} \mathbf{N}  \nonumber 
\end{align}
Using R1)-R4), and R6) - R8) and Assumptions~\ref{as:gmu}-\ref{as::xp}, we obtain that $\mathbf{J}$ is positive definite. 
\end{enumerate}
R1)-R6) allow using Theorem 2 of \cite{yin2006asymptotic} over $\rho$, and we obtain that
\begin{equation}
    \hat{\rho} \xrightarrow[n\to \infty]{d} \mathcal{N}\left(\rho, d_n\right)
\end{equation}
\begin{itemize}
    \item[R9)] Using Multivariate Delta Method's (Theorem 3.9.4 of \citet[pp.~374]{van1996weak}), it is known that the covariance function of the vector $\mathbf{Y}_n$ can be approximated by the expression:
\begin{equation}\label{deltas}
 \mbox{Cov}(\mathbf{Y}_n)=\mbox{diag}\left(\frac{\partial g^{-1}(\eta_i)}{\eta_i}\right)\left( \mathbf{A}^\top\mathbf {A}\right)^{-1} \mbox{diag}\left(\frac{\partial g^{-1}(\eta_i)}{\eta_i}\right)
\end{equation}
\end{itemize}

Using R2) and R9) the covariance $\text{Cov}(\mathbf{Y}_n)$ is well approximated by a first-order Taylor series, which is verified by the matrix \eqref{deltas}. Finally, R1)-R9) allow applying Theorem 2 of \cite[pp.~15]{liang1986longitudinal} and Theorem 1 of \cite[pp.~903]{pan2001robust}, we obtained that:
\begin{equation*}
\hat{\pmb{\beta}} \xrightarrow[n\to \infty]{d} \mathcal{N}_{p}\left(\pmb{\beta}, \mathbf{J}^{-1} \mathbf{B} \mathbf{J}^{-1} \right)
\end{equation*}
\end{proof}
\subsection{Proof of Corollary \ref{cor1}}\label{proofcor1}
\begin{proof}
  When $\rho=0$ in Equation \eqref{eq:eta_vector}, then $\mathbf{A}=\mathbb{I}_n$, Therefore, the model is equal to a GLM proposed by \cite{mccullagh2019generalized} and \cite{dobson2018introduction}.  
\end{proof}
\subsection{Proof of Lemma \ref{lem1}}\label{prooflema1}
\begin{proof}
By Theorem \ref{te1}, it is known that if the true $\rho=0$, then:
\begin{equation*}
 P\left(\lim_{n\to \infty}\hat{\rho}=0\right) = 1
\end{equation*}
Therefore, the $\pmb{\beta}$ estimator holds that:
\begin{align}
 \lim_{n\to \infty}\mathbf{J}&=\tilde{\mathbf{X}}^{T} \mathbf{V}(\pmb{\mu})^{-\frac{1}{2}}\left( \mathbf{A}^\top\mathbf{A}\right) \mathbf{V}(\pmb{\mu})^{-\frac{1}{2}}\tilde{\mathbf{X}}\nonumber\\
 \lim_{n\to \infty} \mathbf{B} &=\mathbf{N}^{\top}\mathbf{V}(\pmb{\mu})^{-\frac{1}{2}}\left( \mathbf{Y}-\pmb{\mu}\right) \left( \mathbf{Y}-\pmb{\mu}\right)^\top\mathbf{V}(\pmb{\mu})^{-\frac{1}{2}} \mathbf{N} \nonumber
\end{align}
Then, by Theorem 2 of \citet[pp.~903]{pan2001robust}, it is obtained  that $\pmb{\beta}$ converges to the GLM estimator because $\mathbb{E}\left[\left( \mathbf{Y}-\pmb{\mu}\right) \left( \mathbf{Y}-\pmb{\mu} \right)^\top\right] = \mathbb{I}_n$, and it is clear that $\mathbf{A}=\mathbb{I}_n-0\mathbf{W}= \mathbb{I}_n$
\end{proof}
\subsection{Proof of Corollary \ref{cor2}}\label{proofcor2}
\begin{proof}
 Assume that
\[
\mathbf{Y}\sim
\mathcal{N}\!\left(
\mathbf{A}^{-1}\mathbf{X}\pmb{\beta},
\sigma^2(\mathbf{A}^{\top}\mathbf{A})^{-1}
\right),
\]
with
\[
\mathbf{A}=\mathbf{I}-\rho\mathbf{W}.
\]

Under the Gaussian specification, the quasi-likelihood in
Equation~\eqref{eq:qlfunction} coincides with the Gaussian log-likelihood up
to an additive constant. Therefore, the estimating equations used in
Algorithm~\ref{algorithm} reduce to generalized least squares.

For fixed $\rho$, the GEE update for $\pmb{\beta}$ becomes

\[
\hat{\pmb{\beta}}=\left(\tilde{\mathbf X}^{\top}\pmb{\Sigma}^{-1}\tilde{\mathbf X}\right)^{-1}\tilde{\mathbf{X}}^{\top}\pmb{\Sigma}^{-1}\mathbf Y ,
\]
where
\[
\tilde{\mathbf X}
=
\mathbf{A}^{-1}\mathbf X,
\qquad
\pmb{\Sigma}
=
\sigma^2(\mathbf{A}^{\top}\mathbf{A})^{-1}.
\]

Substituting these expressions yields

\[
\hat{\pmb{\beta}}
=
\left(
\mathbf{X}^{\top}\mathbf{X}
\right)^{-1}
\mathbf{X}^{\top}\mathbf{A}\mathbf{Y} .
\]
This is precisely the classical SAR estimator of \cite{anselin1988spatial}. Since Theorem~\ref{te1} establishes consistency and asymptotic normality of the proposed estimator for $\rho$, it follows that the sequence of estimators produced by Algorithm~\ref{algorithm} converges asymptotically to the same limit as the Gaussian SAR estimator. Hence, under the Gaussian specification, the proposed procedure reduces asymptotically to the classical SAR estimator.

\end{proof}
\subsection{Proof of Corollary \ref{cor3}}\label{proofcor3}
\begin{proof}
The result follows by combining Theorem~\ref{te1} with
Theorems~\ref{thm:lz1} and \ref{thm:Pan}. Theorem~\ref{te1}
establishes that the proposed estimator is consistent and
asymptotically normal under Assumptions~\ref{as:gmu}--\ref{as:lee}.
The quasi-likelihood function for $\rho$ proposed by \cite{lambert2010two}
coincides with Equation~\eqref{eq:quasilikelihood} when
$Y_i\sim Po(\lambda_i)$. However, the estimator of
$\pmb{\beta}$ in \cite{lambert2010two} is obtained under an
independence working covariance structure, whereas the proposed
method incorporates the spatial working covariance matrix defined
in Equation~\eqref{eq:covariance}.

Consequently, by Theorem~1 of \cite{liang1986longitudinal} (Theorem~\ref{thm:lz1}),
both estimators are asymptotically unbiased for the same marginal
mean parameter $\pmb{\beta}$. 

\end{proof}
\subsection{Proof of Corollary \ref{cor4}}\label{proofcor4}
\begin{proof}
The estimation function presented in \cite{song2022variable} is constructed as the product of the marginals of each $Y_i$, that is, it is assumed that the dependence is only captured in the mean of the binomial random variable and not in the covariance of the same. Therefore, using reasoning similar to that presented in Corollary \ref{cor3}, the result is obtained.
\end{proof}

\end{document}